\documentclass[12pt]{report}

\usepackage[T1]{fontenc}

\usepackage[svgnames]{xcolor}

\definecolor{lightgrey}{rgb}{0.7,0.7,0.7}
\definecolor{grey}{rgb}{0.5,0.5,0.5}
\definecolor{darkgrey}{rgb}{0.3,0.3,0.3}

\usepackage{multicol}

\usepackage{url}
\usepackage{hyperref}
\usepackage{graphicx}
\usepackage{enumerate}
\usepackage{amssymb}
\usepackage{ifthen}
\newboolean{twocolswitch}

\usepackage{rotating}
\usepackage{array}

\usepackage{lastpage}

\newcounter{PaperCounter}

\usepackage{fancyhdr}

\fancypagestyle{titlepage}{
  \fancyhf{} 
  \fancyfoot[L]{\small \textcolor{gray}{Rebuilt version, \today}}
  \fancyfoot[R]{\small \textcolor{gray}{Peter Sheridan Dodds}}

}

\fancypagestyle{plain}{%
    \lhead[\fancyplain{}{\bfseries\thepage}]{\fancyplain{}{\small\bfseries\nouppercase\rightmark}}
    \rhead[\fancyplain{}{\bfseries\leftmark}]{\fancyplain{}{\small\bfseries\thepage}\textcolor{gray}{/\pageref{LastPage}}}

}

\usepackage{pictex}

\newcommand{\Fb}{\mbox{$\beta F(\beta)$}}

\newcommand{\EF}{E {\widehat{\otimes}}_{\pi} F}
\newcommand{\EstarE}{E^\ast {\widehat{\otimes}}_{\pi} E}
\newcommand{\EstarF}{E^\ast {\widehat{\otimes}}_{\pi} F}
\newcommand{\LB}{{\cal L}^{(\beta)}}
\renewcommand{\L}{{\cal L}}
\renewcommand{\O}{{\cal O}}
\renewcommand{\H}{{\cal H}}
\newcommand{\M}{{\cal M}^{(z,\beta)}}
\newcommand{\R}{{\cal R}^{(z,\beta)}}
\newcommand{\Rk}{{\cal R}_k^{(z,\beta)}}
\newcommand{\Mk}{{\cal M}_{(k)}^{(z,\beta)}}
\newcommand{\MG}{(-1)^k{\cal M}^{(z,\beta+k)}}
\newcommand{\Proj}{{\cal P}_N}
\newcommand{\T}{\cal T}
\newcommand{\Tk}{{\cal T}_k}
\newcommand{\Proof}{{\em Proof}}
\newcommand{\HD}{H_{\infty}(D)}
\newcommand{\half}{\frac{1}{2}}

\newcommand{\tr}{\mbox{trace}}
\newcommand{\envtr}{\mbox{\em trace}}
\renewcommand{\th}{\mbox{\scriptsize th}}
\newcommand{\sumi}{\sum_{\{i\}}}
\newcommand{\cne}{[(1-\epsilon)n]}
\newcommand{\en}{[\epsilon n]}
\newcommand{\cneb}{[\frac{(1-\epsilon)n+1}{2}]}
\newcommand{\enb}{[\frac{\epsilon n}{2}]}
\newcommand{\gk}{\gamma^{(k)}}

\newcommand{\lapprox}{\ \raisebox{-.8 ex}{$\stackrel{\textstyle <}{\sim}$} \ }

\newcommand{\be}{\begin{equation}}
\newcommand{\ee}{\end{equation}}
\newcommand{\bdm}{\begin{displaymath}}
\newcommand{\edm}{\end{displaymath}}

\newtheorem{theorem}{Theorem}[chapter]
\newtheorem{lemma}[theorem]{Lemma}
\newtheorem{defn}[theorem]{Definition}
\newtheorem{corollary}[theorem]{Corollary}
\newtheorem{prop}[theorem]{Proposition}
\newtheorem{rem}[theorem]{Remark}
\newtheorem{con}[theorem]{Conjecture}

\def\cfrac#1#2{\dfrac{\strut#1}{#2}\kern-\nulldelimiterspace}
\def\dfrac#1#2{{\displaystyle{#1\over#2}}}

\usepackage[nottoc,notlot,notlof]{tocbibind}

\usepackage{abstract}

\begin{document}

\pagestyle{empty}

\mbox{}

\vspace{50pt}

\begin{center}
  {
    \Huge

    \textcolor{darkgray}{\textbf{On the Thermodynamic Formalism for the Farey Map
}}

  }

  \bigskip
  \bigskip
  \bigskip

  \includegraphics[height=0.25\textheight]{farey}
  \qquad
  \includegraphics[height=0.25\textheight]{fareypart}

  \bigskip
  \bigskip
  \bigskip

  {
    {\Large \textbf{A  Masters Thesis by Peter Sheridan Dodds}}
    
    {\large Department of Mathematics} \\
    {\large Department of Physics} \\
    {\large University of Melbourne, Victoria, Australia}

    {\large Advisor: Dr. Thomas Prellberg}

    {\normalsize Started January 1994, completed August 1994; Masters awarded 1995.}

    {\normalsize \textcolor{gray}{Rebuilt version, \today}}
  }

\end{center}

\clearpage

\renewcommand{\abstractnamefont}{\normalfont\Huge\bfseries}
\renewcommand{\abstracttextfont}{\normalfont\Huge}
\begin{abstract}
  \large
  The chaotic phenomenon of intermittency is modeled by
a simple map of the unit interval, the Farey map.  
The long term dynamical behaviour of a point under iteration of 
the map is translated into a spin system via symbolic dynamics.
Methods from dynamical systems theory and statistical mechanics 
may then be used to analyse the 
map, respectively the zeta function and the transfer operator.  Intermittency is 
seen to be problematic to analyze due to the presence of an `indifferent 
fixed point'.   Points under iteration of the map move away from this point 
extremely slowly creating pathological convergence times for calculations.
This difficulty is removed by going to an appropriate induced subsystem,
which also leads to an induced zeta function and an induced transfer 
operator.  Results obtained there can be transferred back to the
original system.
The main work is then divided into two sections.  The first demonstrates
a connection between the induced versions of the zeta function and the
transfer operator providing useful results regarding the
analyticity of the zeta function.
The second section contains a detailed analysis of the pressure function 
for the induced system and hence the original by considering bounds
on the radius of convergence of the induced zeta function.
In particular, the asymptotic behaviour of the 
pressure function in the limit $\beta$, the inverse of `temperature', 
tends to negative infinity is determined and the existence and nature of 
a phase transition at $\beta=1$ is also discussed. 

\end{abstract}

\tableofcontents
\listoffigures

\addtocontents{toc}{\protect\enlargethispage{\baselineskip}}

\clearpage

\pagestyle{plain}

\chapter{General Overview}
\label{ch:general}

\section{Introduction}
\label{sec:intro}
The long term behaviour of chaotic systems 
may be investigated through the use of the thermodynamic formalism for dynamical 
systems~(\cite{ruelle:therm}, \cite{bowen:therm} and \cite{beck:thermchaos}). 
Sequences of iterates of the map may be likened to 
one-dimensional spin chains through the use of symbolic dynamics,
\cite{tel:multifracs}.  This converts
the problem of examining the dynamics of iterates of the map 
into the language of statistical mechanics. An important mechanism for the transition
from order to chaos is {\em intermittency}, \cite{schuster:detchaos}.  
One such map that exhibits intermittent behaviour is the Farey map, \cite{feig:farey}. 
The examination of the thermodynamics of the spin system 
`generated' by this map
will be the main focus of this thesis.

Two well known methods for calculating the partition function of this 
thermodynamic `Farey system' are presented: the Ruelle zeta function
and the transfer operator method.  Both are developed heuristically and
are seen to give the same thermodynamics even though they appear to
be quite disparate methods.  It is observed that intermittency provides a
problem for these techniques in the form of an indifferent fixed point.
Points do not separate fast enough near this fixed point and the system
is said to be non-hyperbolic.  More will be said on this towards the end of 
this introductory chapter and it is sufficient to state here
that in general the techniques require hyperbolic systems.
A method of  inducing the system is used to obtain a hyperbolic one that is still
strongly linked to the original.  Correspondingly, an induced zeta
function and an induced transfer operator are developed for maps
of an interval.  Examination of these two well behaved objects then 
yields information on the original system.

The work is then broken into two main sections.  
The first of these sections concerns a proof of the meromorphic qualities
of the induced zeta function.  This is done by first establishing a connection
between trace formulas involving the induced transfer operator and the induced
zeta function itself.  The work requires the introduction of nuclear operators
for Banach spaces developed in the 50's by Alexander Grothendieck and 
standard techniques regarding analytic continuation.

The second part concerns
the finding and analysis of an explicit form of the induced zeta function.
The key is to extract the radius of convergence of the induced zeta function
which yields the pressure function of the system.   Several bounds are produced
along with some exact values.  The existence of a phase transition is observed
and the scaling behaviour discussed; this has already been done in a more general
setting in~\cite{prelslawn:interm}.  The behaviour of the pressure function
for inverse temperature approaching negative infinity is also found
analytically.  Along the way, several results involving periodic
continued fractions, Fibonacci numbers and the golden ratio are also obtained.

\section[Farey Map, Intermittency, Symbolic Dynamics, and Stat Mech]{The Farey Map, Intermittency, Symbolic Dynamics, and Statistical Mechanics}
\label{sec:general}
The process of intermittency is a well recognised route to 
chaos first observed by Pomeau and Manneville~\cite{pomman:interm}. 
Physically, it may characterise 
some measured signal that moves randomly between states of 
unpredictability and regularity.  It is observed that with the tuning of 
some external parameter the period of occurrence and duration of these chaotic
bursts continuously changes.  At one extreme of the parameter range, 
the signal will become totally chaotic and at the other, fully
periodic.

\begin{figure}[tp!]
  \includegraphics[width=\textwidth]{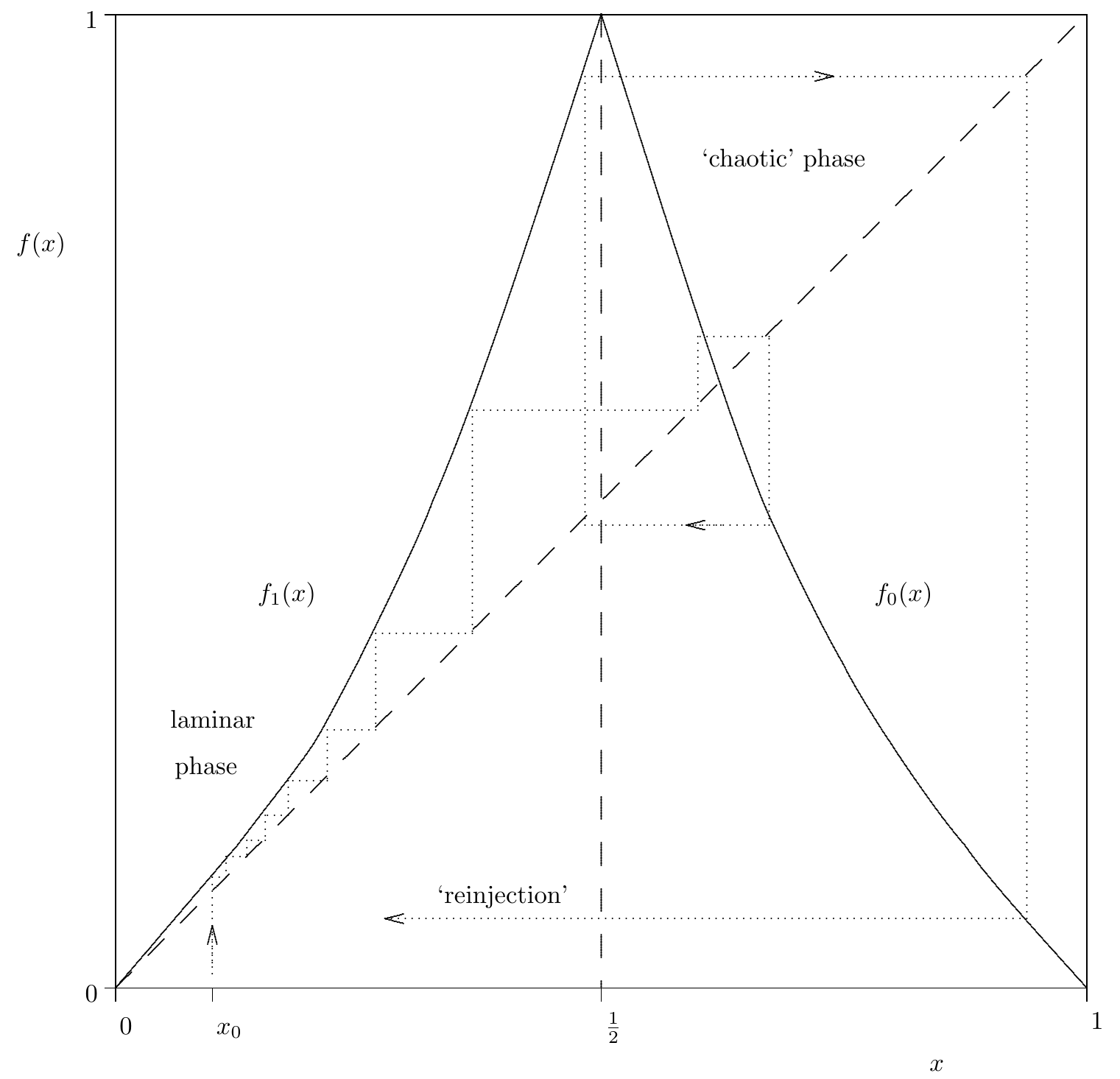}
  \caption{The Farey map}
  \label{fig:farey}
\end{figure}

The process of intermittency may be modelled by simple `toy' maps
of the unit interval.  Apart from being simple to investigate, the 
behaviour of such maps may be qualitatively generalised 
to more complex systems possessing intermittent behaviour --- this
is really just the paradigm of `universality'.  Toy
models have certainly proved helpful in understanding more complicated
systems as in the seminal work of Lorenz's model of the 
atmosphere~\cite{lorenz:air}.  
Some examples of intermittent 
signals are the flow of highway traffic, the electric potential of 
a nerve membrane, the many currents and voltages observed
in electronic components and circuits such as a 
junction diode and the relative velocities of the eddies of 
turbulence~\cite{schuster:detchaos}. 
One particularly simple toy model is provided by the Farey map 
which is otherwise well known
for its number theoretic properties.
The Farey map, $f:I \rightarrow I$ where $I$ 
is the unit interval $[0,1]$ 
is defined as
\be
f(x) = \left\{
\begin{array}{l}
f_1(x) = \frac{x}{1-x}, \ \ \ \ x \in [0,\frac{1}{2}] \\
f_0(x) = \frac{1-x}{x}, \ \ \ \ x \in [\frac{1}{2},1]
\end{array}
\right.
\label{eq:fareydef}
\ee
The inverses of the two branches are written as $F_1$ and $F_0$ and
are given by
\be
\begin{array}{l}
F_1(x) = \frac{x}{1+x}, \ \ \ \ x \in [0,1] \\
F_0(x) = \frac{1}{1+x}, \ \ \ \ x \in [0,1]
\end{array}
\label{eq:invfareydef}
\ee
Figure~(\ref{fig:farey}) displays the Farey map.    

This figure also shows a typical intermittent 
orbit\footnote{Note that all but countably many points in the interval $(0,1)$ 
give rise to `typical' orbits which are those points whose iterates densely fill the
unit interval or more generally the underlying manifold of the system.  
This is because all rationals, and only these points (a countable set),
eventually get mapped to the fixed point
at 0 since the Farey map provides an enumeration of the
rationals by back iteration of 0.  Also, only countably many points
end up stuck at the other fixed point, $\frac{\sqrt{5}-1}{2}$.}, 
i.e. the `motion' of the point $x_0 \in I$ under
iteration of the Farey map.  Initially, the point moves slowly away
from the point $x=0$.  This is equated with the `laminar' or regular phase 
of an intermittent process.  After numerous iterations, it moves to the righthand
side of the unit interval where it is acted upon by the `reinjection' branch
of the map.  Here it may bounce around irregularly ( the chaotic phase )
until it eventually returns to a point near 0 starting the laminar phase again.

\subsection{Long term behaviour of intermittent processes}
\label{subsec:longterm}
Given that the Farey map is simple model of intermittency, the question arises
how best to analyse its dynamics and, indeed, what information about these
dynamics is of interest?
As with many chaotic processes the short term behaviour is inherently
unpredictable and does not always provide tractable problems.  In the long term
however, average behaviour may be detectable and this will be the focus of 
this work.  

Notice that the Farey map has two branches which have been indirectly referred
to as the `laminar' branch, $f_1$, and the `chaotic/reinjection' branch $f_0$.
An important question about a sequence of iterates is how many have been mapped
through each branch?  In other words, given a {\em seed} point $x_0$, how many of
its iterates are, on average, to the left of $x=\half$ ( i.e. undergoing smooth
motion ), and how many are to the right (i.e. undergoing chaotic motion)?  
Translating these ideas into a mathematical setting, it is useful to label
a point by a `1' if it is less than $\half$ and by a `0' if it is greater.
This idea of labelling of orbits is generally referred to as symbolic dynamics.
The movement of a point under iteration can then be represented by a string
of bits.  In figure~(\ref{fig:farey}) for example, iteration of the point $x_0$ 
produces the sequence $\{1,1,1,1,1,1,1,1,0,0,1,0,1,\ldots\}$.  

Naturally, there are many initial points whose 
iterates are encoded in a sequence beginning in the
same way. However, as more and more iterates of $x_0$ are taken, 
fewer points remain that fit its particular sequence.  In this sense, the 
apparent coarseness of this technique is removed by taking the length of the
sequence to infinity.  

It is now possible to make an analogy to a spin 
system\footnote{It has also been observed that the
Farey system is reminiscent of a second-quantized Fermi gas~\cite{feig:transfer}.}.  
In particular, the symbolic dynamics of the Farey map can be equated with a 
one-sided one-dimensional 2-spin system ( that is, a one dimensional lattice of 
particles which may be spin up (1) or spin down (0)).  However, it is not
yet a thermodynamical system as the probabilities of these states and some
notion of temperature have to be introduced.  The former is rather simple:
consider one iteration of the Farey map.  The interval $[0,\half)$ is mapped
through the branch $f_1$ and these points are thus represented 
by the singleton $\{ 1 \}$.  Similarly, all the points in $(\half,1]$ are
represented by $\{0\}$.  So the probability of a random number between 0 and 
1 being in the `up' state is $\half$, which is the Lebesgue measure of $[0,\half)$.
The same idea applies to the `down' state.  Consider then two iterations of the
Farey map.  There are now four possible spin states: $\{1,1\}$,$\{1,0\}$,$\{0,1\}$
and $\{0,0\}$; note that the $n^{\th}$ iterate of the Farey map will provide
$2^n$ states.  The state $\{1,1\}$ corresponds to the interval $[0,\frac{1}{3})$
and the probability of its occurrence is $\frac{1}{3}$ and so on.  Thus, the
probability of a state is given by the Lebesgue measure of the 
interval corresponding to that state.

Furthering the analogy, the idea of considering the symbolic dynamics of a point
in the limit of the number of iterates going to infinity is equivalent to
taking the thermodynamic limit of the corresponding spin system.  Also in this
limit, a finer and finer partition of the unit interval will be constructed
based on the symbolic dynamics.  
Let the $2^n$ lengths at the $n^{\th}$ stage of 
construction of the partition be written as $\ell_i$ where $i=1,\ldots,2^n$.
Note that the probability $p_i$ of a state $i$ in a thermodynamic system may 
be written as $\exp \left(-\beta E_i\right)$ 
where $E_i$ is the energy of the state and
$\beta$ is the inverse of temperature $T$.  Since the probabilities of the
states being considered are also equal to $\ell_i$ the partition function for this
system may be written as 
follows\footnote{ The lengths (probabilities) of the elements
(spin states) of this partition will scale in a consistent way much like
a fractal.  However, the partition will exhibit a spectrum of scaling
exponents and thus the partition may be thought of as a 
multifractal.  Indeed, the theory of multifractals is
intimately linked with the interpretation of dynamical
systems as thermodynamic ones
(see~\cite{tel:multifracs}, ~\cite{feig:transfer})
and~\cite{dodds:honsthesis}. } 
\be
\sum_{n=1}^{2^n}\exp -\beta E_i \equiv \sum_{n=1}^{2^n}\ell_i^{\beta} \sim \exp -n \Fb
\label{eq:part fn1}
\ee
where $F(\beta)$ is the free energy of the system and \Fb\ is the pressure function.

\subsection{Methods for calculating the Pressure Function}
This section presents some non-rigorous, 
heuristic arguments for the construction of 
the zeta function and the transfer operator.  The conclusions will then be 
compared with formal definitions of these objects.
Before going on to these two methods for the evaluation of the partition function,
the partition itself will be presented in more detail.  There are three important ways
to view the partition's construction: forward iteration of the Farey map, back iteration
of the Farey map and the enumeration of the rationals via Farey addition.
Note that the first method relates to the zeta function and
the second to the transfer 
operator while the third demonstrates the number theoretic properties of the map.

Consider the $n^{\th}$ iterate of the Farey map, $f^n$.  This function will 
have $2^n$ branches.  Since these branches are composed of $f_1$ and $f_0$,
each branch will lie precisely above the corresponding element of 
the partition in its $n^{\th}$ stage of construction.
Consider secondly the preimages of $x=1$.  The first preimage is just $x=\half$
which divides the unit interval as per the partition in its first stage.
Next, the preimage of the point $\half$ is $\frac{1}{3}$ and $\frac{2}{3}$.
These three points give the partition in its second stage and so on.

\begin{figure}[tp!]
  \includegraphics[width=\textwidth]{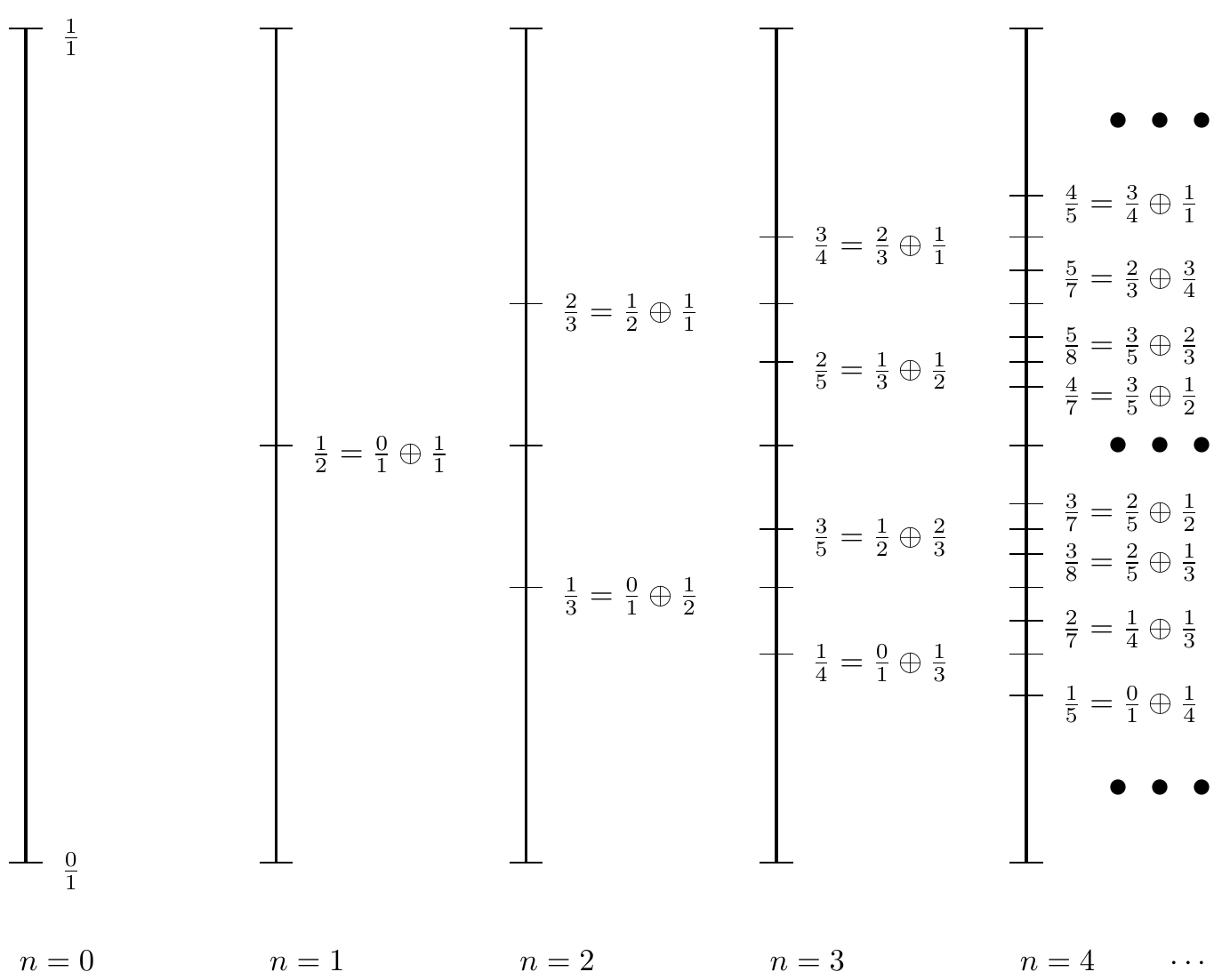}
  \caption{
    Creation of the Partition of the Farey System.
  }
  \label{fig:fareypart}
\end{figure}

Finally, the endpoints of the intervals making up the 
partition may be formed by Farey addition
of the rationals.  The first few steps of this process are demonstrated
in figure~(\ref{fig:fareypart}). 
The operation of the Farey addition of two rational
numbers is represented by the symbol $\oplus$ and is defined by 
\be
\frac{a}{b} \oplus \frac{c}{d} = \frac{a+c}{b+d}
\label{eq:fareyadd}
\ee
So, starting with $0=\frac{0}{1}$ and $1=\frac{1}{1}$, the added
endpoint for the first stage is $\frac{0}{1} \oplus \frac{1}{1} = \frac{1}{2}$.
Now, the additions $\frac{0}{1} \oplus \frac{1}{2} = \frac{1}{3}$ and
$\frac{1}{2} \oplus \frac{1}{1} = \frac{2}{3}$ give the extra endpoints
required for the second stage.
In general, adding consecutive endpoints of the current stage via Farey
addition gives the additional endpoints for the next stage; see
figure~\ref{fig:fareypart}.

\subsubsection{The Ruelle Zeta Function}
One way of approximating the partition is as follows: at the $n^{\th}$ stage
of construction take any point $x_i$ in each
interval of the partition and estimate the width of the interval using the 
slope of $f^n$ at that point; i.e.
\be
\ell_i \sim \left({(f^n)}'(x_i)\right)^{-1} \mbox{ for large $n$}
\label{eq:lengthapprox}
\ee
Intuitively, it can be seen that in the
limit $n$ approaches infinity, the slope of each branch becomes more and
more uniform throughout the interval.  Hence the slope at
any point inside an interval will do
and, for this approximation, the fixed points of $f^n$ may be used for the $x_i$
\footnote{This may cause some difficulties at $x=0$}.  
Thus the partition sum in equation~(\ref{eq:part fn1}) may be estimated by
\be
\sum_{i=1}^{2^n} \ell_i^{\beta} \sim \sum_{f^n x=x} \left({(f^n)}'(x)\right)^{-\beta}
\mbox{ for large $n$}
\label{eq:part fn2}
\ee
It is expected that this approximation and the actual partition will
match up in the thermodynamic limit.

It will be now be useful to introduce the zeta function that is used in
fields of statistical mechanics and dynamical systems.     
It was the work of Artin and Mazur~\cite{artin:periodic} in 1967 that 
first introduced zeta functions into the study of dynamical systems.
Ruelle~\cite{ruelle:zeta} generalised this idea
to systems with interactions by incorporating the probability of the
states of the system into the structure of the zeta function.
Ruelle's definition of the zeta function is as follows
\begin{defn}
\be
\zeta(z,w) = \exp \sum_{n=1}^{\infty}
\frac{z^n}{n} \sum_{f^{n}(x)=x} \prod_{k=0}^{n-1} w (f^{k}x)
\nonumber
\ee
\label{defn:zeta1}
\end{defn}
where $z \in C$ and $w$ is the {\em weight function} of the system.
A more convenient definition of the zeta function for the work
here is provided when $w$ is replaced by $\exp{\phi}$; 
$\phi$ is the {\em interaction} of the system.
\begin{defn}
\be
\zeta(z,\phi) = \exp \sum_{n=1}^{\infty}
\frac{z^n}{n} \sum_{f^{n}(x)=x} \exp \sum_{k=0}^{n-1} \phi (f^{k}x) \nonumber
\ee
\label{defn:zeta2}
\end{defn}
Throughout this work, the second definition will be used. 

It may be shown that the radius of convergence of the zeta 
function is equal to $\exp -\Fb$, see~\cite{ruelle:anosov}.
I.e.,
\be
\lim_{n \rightarrow \infty}
\left| \frac{1}{n} \sum_{f^{n}(x)=x} \exp \sum_{k=0}^{n-1} \phi (f^{k}x)  
\right|^{\frac{1}{n}} = \exp -\Fb
\label{eq:partconverge1}
\ee
Note that $\left(\frac{1}{n}\right)^{\frac{1}{n}} \rightarrow 1$ so this
term is unimportant.  Equations~(\ref{eq:part fn1}) and~(\ref{eq:part fn2}) 
together show that 
\be
\left| \sum_{f^{n}(x)=x} \left({(f^n)}'(x)\right)^{-\beta}
\right|^{\frac{1}{n}} \sim  \exp -\Fb
\label{eq:partconverge2}
\ee
Equations~(\ref{eq:partconverge1}) and~(\ref{eq:partconverge2}) then imply
that for the Farey system a natural choice of the interaction $\phi$ is 
\be
\phi(x) = -\beta \log |f'(x)|
\label{eq:interaction}
\ee
since ${(f^n)}'(x) = \prod_{k=0}^{n-1} f'(f^{k}x)$ by the chain rule.
Thus the interaction of a state\footnote{note that~\cite{prelslawn:interm}
employs this same interaction}, $i$, is the energy of the state, $E_i$, weighted by 
the inverse of the temperature, $\beta$.

Finally, the Farey system may be conveniently and compactly represented by the triplet
\bdm
X = (I,f,\phi)
\edm
I.e., in the form of ( state space, evolution operator, interaction).

\subsubsection{The Transfer Operator}
The transfer operator method is based on the idea of finding an operator
that creates the partition function of the partition at its
$n^{\th}$ stage of construction by $n$ iterations
upon an appropriate initial function.
Since the partition function grows like
$\exp -n\Fb$ it is expected that the largest eigenvalue of such an
operator would be precisely $\exp -\Fb$.  Thus the choice of the initial 
function becomes unimportant as it is now the spectrum of this operator
that contains information about the pressure function.  

\begin{figure}[tp!]
  \centering
  \includegraphics[width=0.8\textwidth]{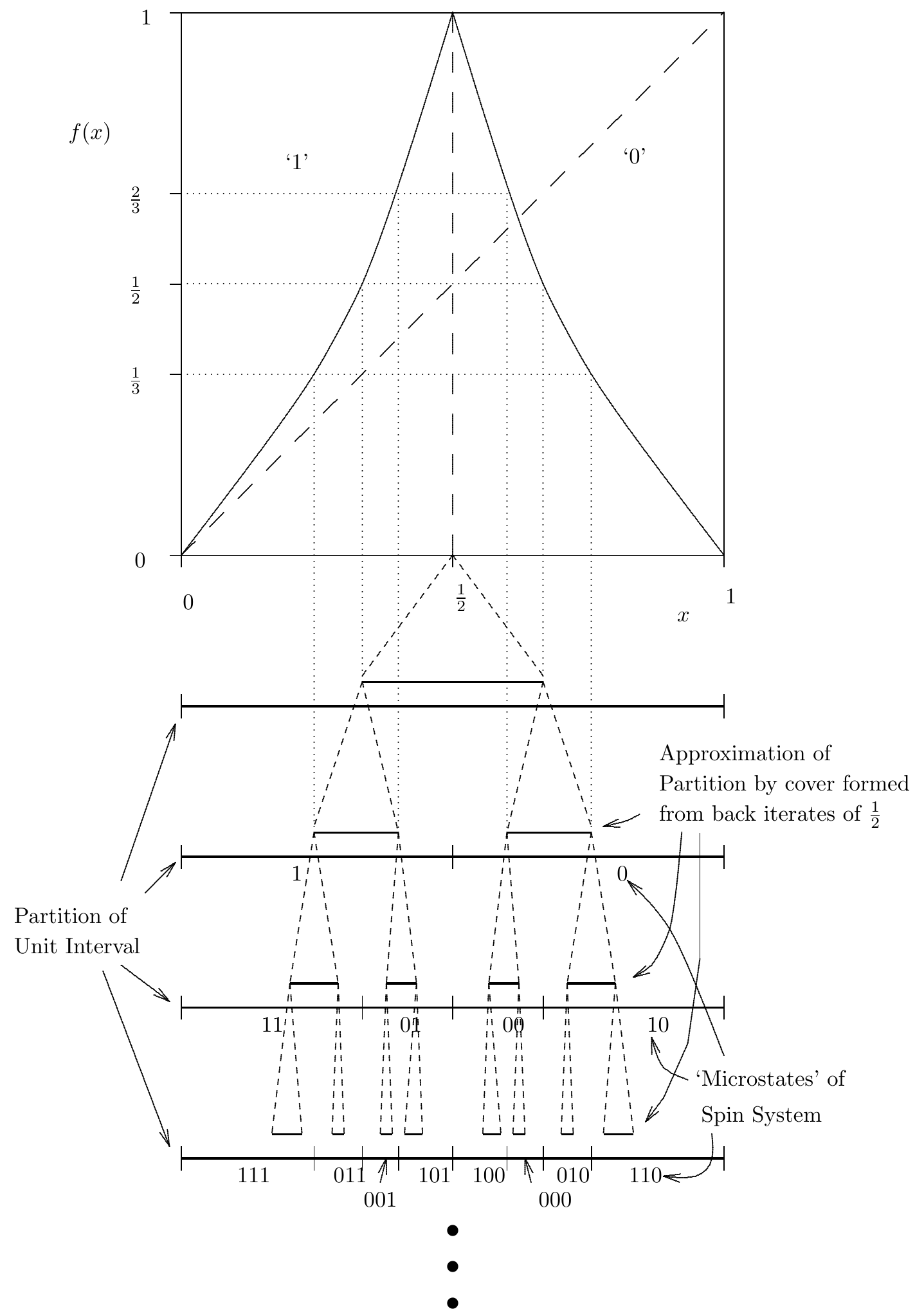}
  \caption{Partition created by back iteration of Farey map}
  \label{fig:partition}
\end{figure}

As with the zeta function method, an approximation for the actual partition
is also employed here, the statistical mechanics of both expected to be the
same~\cite{feig:transfer}. The construction of this second approximation is 
demonstrated in figure~(\ref{fig:partition}) and is described as follows. 
Consider a {\em seed point}, $x^\ast$
in the interval $I=(0,1)$ ( note that $x^\ast$ in the example figure has been chosen 
to be $x=\half$). Back iteration of this point by the Farey map
gives the two points $F_1(x^\ast)$ and $F_0(x^\ast)$.  The interval 
$[F_1(x^\ast),F_0(x^\ast)]$ then serves as an approximation for the unit interval
(i.e. the $0^{\th}$ stage of construction).  The preimages of these two points are,
in order along the unit interval,
$F_1 \circ F_1(x^\ast)$, $F_1 \circ F_0(x^\ast)$, $F_0 \circ F_0(x^\ast)$ and
$F_0 \circ F_1(x^\ast)$.  The two intervals defined by 
$[F_1 \circ F_1(x^\ast),F_1 \circ F_0(x^\ast)]$ and 
$[F_0 \circ F_0(x^\ast),F_0 \circ F_1(x^\ast)]$ then give the 
approximation at the first stage of construction.  Note that the closer $x^\ast$ is
to 0, the more precise the approximation.  In general, the $n^{\th}$ stage of 
construction will be made up of $2^{n+1}$ points of the form
\be
x_i^{(n)} = F_{e_{n}} \circ F_{e_{n-1}} \circ \cdots \circ F_{e_0}(x^\ast)
\label{eq:toppoints1}
\ee
\clearpage
where $i=1,\ldots,2^{n+1}$ and $e_j=0,1$.  As per the symbolic dynamics, each
point may be represented by a sequence of 0's and 1's: $\{e_{n},e_{n-1},\ldots,e_0\}$.
The $x_i^{(n)}$ are paired off as endpoints of the $2^n$ intervals needed to give
an estimation of the true partition with two points being endpoints of the same
interval if their sequence representations match up except for the entry $e_0$
(which must be different).
So the approximation for the true length of the partition, $\ell_i$, is given by the 
difference between these two points and will be denoted by $l_i$:
\be
\ell_i \sim l_i = 
F_{e_{n}} \circ F_{e_{n-1}} \circ \cdots \circ F_{e_1} \circ F_{0}(x^\ast)
-F_{e_{n}} \circ F_{e_{n-1}} \circ \cdots \circ F_{e_1} \circ F_{1}(x^\ast)
\label{eq:toppoints2}
\ee
The $l_i$ may now be themselves approximated in the following way.  Note that
for $x \in (0,1)$, $F_1$ and $F_0$ are both contractive mappings, 
i.e. $|F_1'(x)|,|F_0'(x)| < 1$.  The two endpoints of an interval must approach
each other with increasing $n$.  Thus a taylor approximation may be used for large $n$:
\begin{eqnarray}
\lefteqn{F_{e_{n}} \circ F_{e_{n-1}} \circ \cdots \circ F_{e_1} \circ F_{0}(x^\ast)}
\nonumber \\
 & = & F_{e_{n}} \circ F_{e_{n-1}} \circ \cdots \circ F_{e_1} 
\circ \left( F_{1}(x^\ast) +  (F_{0}(x^\ast) - F_{1}(x^\ast)) \right)
\nonumber \\
 & \approx & F_{e_{n}} \circ F_{e_{n-1}} \circ \cdots \circ F_{e_1} \circ F_{1}(x^\ast)
\nonumber \\ 
 &   & + (F_{0}(x^\ast) - F_{1}(x^\ast))  \left. \frac{d}{dy}
F_{e_{n}} \circ F_{e_{n-1}} \circ \cdots \circ F_{e_1} (y) \right|_{y=F_{1}(x^\ast)}
\nonumber \\
 & = & F_{e_{n}} \circ F_{e_{n-1}} \circ \cdots \circ F_{e_1} \circ F_{1}(x^\ast)
\nonumber \\ 
 &   & + (F_{0}(x^\ast) - F_{1}(x^\ast))  
\prod_{k=1}^{n}
F_{e_{k}}' \left( F_{e_{k-1}} \circ \cdots \circ F_{e_1} \circ F_{1}(x^\ast) \right)
\label{eq:toppoints3}
\end{eqnarray}
Substituting this into equation~(\ref{eq:toppoints2}), the 
partition function, equation~(\ref{eq:part fn1}), becomes
\be
\sum_{i=1}^{2^n} \ell_i^{\beta}
\sim \sum_{i=1}^{2^n} l_i^{\beta}
\sim \sum_{e_n,\ldots,e_1}  \prod_{k=1}^{n} \left|
F_{e_{k}}' \left( F_{e_{k-1}} \circ \cdots \circ F_{e_1} \circ F_{1}(x^\ast) \right)
\right|^{\beta}
\label{eq:part fn3}
\ee
The transfer operator will now appear as a way of calculating the expression
on the righthand side of equation~(\ref{eq:part fn3}).  Consider the following function:
\be
\psi_j^{(\beta)}(x) \equiv \psi^{(\beta)}(x;e_j,e_{j-1},\ldots,e_1) \equiv
\sum_{e_n,\ldots,e_j} \prod_{k=j}^n \left|
F_{e_{k}}' \left( F_{e_{k-1}} \circ \cdots \circ F_{e_1} \circ F_{1}(x) \right)
\right|^{\beta}
\label{eq:transopderiv1}
\ee
Note that $\psi_1^{(\beta)}(x^\ast)$ is actually
the `$n^{\th}$ stage' partition function.
Now set $y = F_{e_{j-1}} \circ \cdots \circ F_{e_1} \circ F_{1}(x)$.
The definition of the $\psi_j^{(\beta)}$ becomes:
\be
\psi_j^{(\beta)}(y) \equiv \psi^{(\beta)}(y;e_j,e_{j-1},\ldots,e_1) \equiv
\sum_{e_n,\ldots,e_j} \prod_{k=j}^n \left|
F_{e_{k}}' \left( F_{e_{k-1}} \circ \cdots \circ F_{j}(y) \right)
\right|^{\beta}
\label{eq:transopderiv2}
\ee
Next, consider the following manipulation of the function $\psi_{j}^{(\beta)}$:
\begin{eqnarray}
\psi_j^{(\beta)}(y) & = & \psi^{(\beta)}(y;e_j,e_{j-1},\ldots,e_1) 
\nonumber \\
 & = & \sum_{e_n,\ldots,e_j} 
\prod_{k=j}^n \left| F_{e_{k}}' \left( F_{e_{k-1}} \circ \cdots \circ F_{j}(y) \right)
\right|^{\beta}
\nonumber \\
 & = & \sum_{e_j} \left| F_{e_j}'(y)\right|^{\beta}
\sum_{e_n,\ldots,e_{j+1}} 
\prod_{k=j+1}^n \left| F_{e_{k}}' \left( F_{e_{k-1}} \circ \cdots \circ F_{j}(y) 
\right) \right|^{\beta}
\nonumber \\
 & = & \sum_{e_j} \left| F_{e_j}'(y)\right|^{\beta}
\psi^{(\beta)}(F_{e_j}(y);e_{j+1},e_{j},\ldots,e_1)
\nonumber \\
 & = & \sum_{e_j} \left| F_{e_j}'(y)\right|^{\beta} \psi_{j+1}^{(\beta)}(F_{e_j}(y))
\label{eq:transopderiv3}
\end{eqnarray}
Thus, a method for producing the partition function by iteration of an 
operator on some initial function has been obtained.
For large $n$, the partition function behaves like
$\exp - n\Fb$.  Thus, it follows from inspection of
equation~(\ref{eq:transopderiv3}) that in the limit of large $n$ and large $n-j$,
that $\psi_j^{(\beta)} \sim \exp(-\Fb)\psi_{j+1}^{(\beta)}$.
The substitution of this into the 
final line of equation~(\ref{eq:transopderiv3}), and also
writing $\lambda(\beta)=\exp(-\Fb)$,
$e_j$ simply as $e$, $x$ for $y$
and $\psi_j^{(\beta)}$ as $\psi$,
then yields the following eigenvalue equation
\be
\lambda(\beta) \psi(x) = 
\sum_{e=0,1} \left| F_{e}'(x)\right|^{\beta} \psi(F_{e}(x))
\label{eq:transopderiv4}
\ee
and also the definition of a transfer operator $\LB$,
\be
\LB \psi(x) = 
\sum_{e=0,1} \left| F_{e}'(x)\right|^{\beta} \psi(F_{e}(x))
\label{eq:transferop}
\ee
By derivation, the largest eigenvalue of this operator is the exponential of 
the pressure function.  More generally, the exponential of the pressure 
function is seen to correspond to the {\em spectral radius} of the 
transfer operator.  The spectral radius of an operator $\O$, $r(\O)$, is defined as the 
supremum over the magnitudes of all the elements of $\sigma(\O)$, the 
spectrum of $\O$ (\cite{ruelle:therm}).  
Noting that $F_{e}'(x)=\frac{1}{f'(F_{e}(x))}$ and setting
$y=F_{e}(x)$, this definition
can be recast in another useful form:
\begin{defn}
\bdm
\LB \circ \psi(x) = \sum_{f(y)=x}\frac{\psi(y)}{|f'(y)|^{\beta}}
\edm
\label{defn:transferop2}
\end{defn}
A formal definition of the transfer operator is given as 
follows~\cite{prelslawn:interm}:
\be
\LB \circ \psi(x) = \sum_{f(y)=x}\psi(y)\exp \phi(y)
\label{eq:transferop1}
\ee
Using the interaction found in the discussion of the zeta
function, equation~(\ref{eq:interaction}), the transfer operator 
for the Farey map given in definition~(\ref{defn:transferop2}) 
is seen to agree with this more formal statement.

Note that for $\beta=1$, the transfer operator is just the Perron-Frobenius operator. 
The corresponding eigenfunction is referred to as the 
{\em invariant density}~\footnote{The nature of the
eigenfunctions $\psi$ for general $\beta$ is not discussed here and
the reader is referred to~\cite{prelslawn:interm} as a starting point.
It is the largest eigenvalue that is the concern of the present work}, $\psi_I$.
This density is actually a probability measure for the long term behaviour of 
the iterates of the map concerned.  In the case of the Farey map, 
$\psi_I = \frac{1}{x}$, (see \cite{feig:transfer}).   However, an attempt to
normalize this function is problematic as it is of infinite measure on $[0,1]$.
The only sensible normalization of $\psi_I$ is actually the Dirac delta function $\delta(x)$.
So, in the long term iterates are expected to be on average at the fixed point 0,
a typical feature of intermittent maps. 
Definition~(\ref{defn:transferop2}) becomes the Perron-Frobenius equation 
$\L \circ \psi(x) = \sum_{f(y)=x}\frac{\psi(y)}{|f'(y)|}$.
The transfer operator is thus also known as the Ruelle-Perron-Frobenius
operator since it was the work of Ruelle that extended the original idea
(\cite{ruelle:therm}).

It is important to impress the fact that the thermodynamics calculated from the zeta
function and the transfer operator methods are the same.  At first it may
appear that they are unrelated.  However, as has been discussed
in this section, they are based on the same thermodynamic system and are
expected to deliver the same results and in particular the same pressure
function.  A much deeper connection between the two methods will be presented 
in the first half of this work.

\subsection{A problem}
\label{subsec:prob}
The above methods of calculating the pressure function are very efficient for
hyperbolic systems. However, for the Farey map they become
somewhat ineffective.  In the case of a simple
map of an interval like the Farey map, the term hyperbolic means the map must be 
{\em uniformly expanding}.   If a map
is uniformly expanding then the iterates of any two nearby points will separate in exponential
fashion; i.e., it is required that on all points of the interval
$|f'| \geq 1 + \epsilon$ where $\epsilon > 0$.
This is not true for the Farey map and, in particular, 
breaks down at the fixed point at 0 which
is referred to as an {\em indifferent fixed point}.  Clearly it is an unstable
fixed point as points eventually move away, something that can be intuited by
the shape of the function around 0, see figure~(\ref{fig:farey}).  Iterates do
not separate exponentially from 0 however and can even be so chosen as to 
take arbitrarily long times to do so, ~\cite{prelslawn:interm}.  As was mentioned 
previously, the invariant density of the Farey map is the Dirac delta 
function $\delta(x)$, further demonstrating the singular nature of intermittency.
Indeed, the generic feature
of such intermittent maps is that their derivative approaches and becomes 1
at $x=0$, creating an indifferent fixed point there.  
In particular, for the Farey map, $f'(x) = (1-x)^{-2} \sim 1+2x$ for $x$ near 0.
The presence of such an indifferent fixed point numerically leads to slow convergence 
to the thermodynamic limit, if at all, and also may bring about singularities in
the pressure function.  This problem is dealt with by the so-called
method of inducing which is discussed in the following chapter.

\chapter{The Induced System}
\label{ch:induce}
In order to remove the pathology of the fixed point at $0$, 
a method of inducing is employed.  Inducing was first used in this setting in a
paper by Prellberg and Slawny, \cite{prelslawn:interm}.  The system
is said to be `induced' onto a subset
$J$ of $I$ creating a new induced system. The main
point of this technique is that an expanding map is produced which can
be analysed via the devices of the zeta function and transfer operator
with none of the problems associated with the indifferent fixed point.
Also, if the interaction of the system is chosen appropriately, very
strong connections between the induced system and the original system
can be demonstrated and utilized forthwith.  These connections will be 
presented at the end of the chapter.  Note that this work applies to 
maps of intervals in general and is in no way particular to the Farey map.

The induced map or the first return map,
$g$, is defined in the following way:   
first consider the function $n:J \rightarrow N$ where
\be
n(x)=n \ \ \mbox{if} \ \ f^k(x) \not\in J \ \ \forall \ \ 
k \in \left\{ 1,\ldots,n-1 \right\}
\ \ \mbox{and} \ \ f^n(x) \in J
\label{eq:n(x)def}
\ee
The induced map\footnote{Note that the number of points in $J$ where no 
finite $n$ exists such that the orbit returns to $J$ is at most countable 
and $g$ may be defined arbitrarily at these points with no added 
problems, see~\cite{prelslawn:interm}.}
is then simply given by $g:J \rightarrow J$ where
\be
g(x) = f^{n(x)}(x)
\label{eq:g(x)def}
\ee
Also, defining $J^c$ to be the complement of $J$ in $I$, i.e. $J^c=I \setminus J$,
consider the following set of points:
\be
K = \{ x \in J^c \mid \ \forall \ n: \ f^n(x) \in J^c   \}
\label{eq:compK}
\ee
These are exactly the points that do not mix with the induced system.
Recall that the Farey system was written as the triplet $X=(I,f,\phi)$.
The method of inducing can now be seen to split this system into two
new non-interacting dynamical systems: an induced system $Y=(J,g,\phi_Y)$ and 
a complementary system $Y^c=(K,f \mid_{K},\phi_{Y^c})$.  The interactions of 
these two systems, $\phi_Y$ and $\phi_{Y^c}$, must be chosen appropriately so
that information regarding the original system is not lost; the work that
follows will determine these functions via an examination of the 
corresponding zeta functions and transfer operators.

\begin{figure}[htbp!]
  \includegraphics[width=\textwidth]{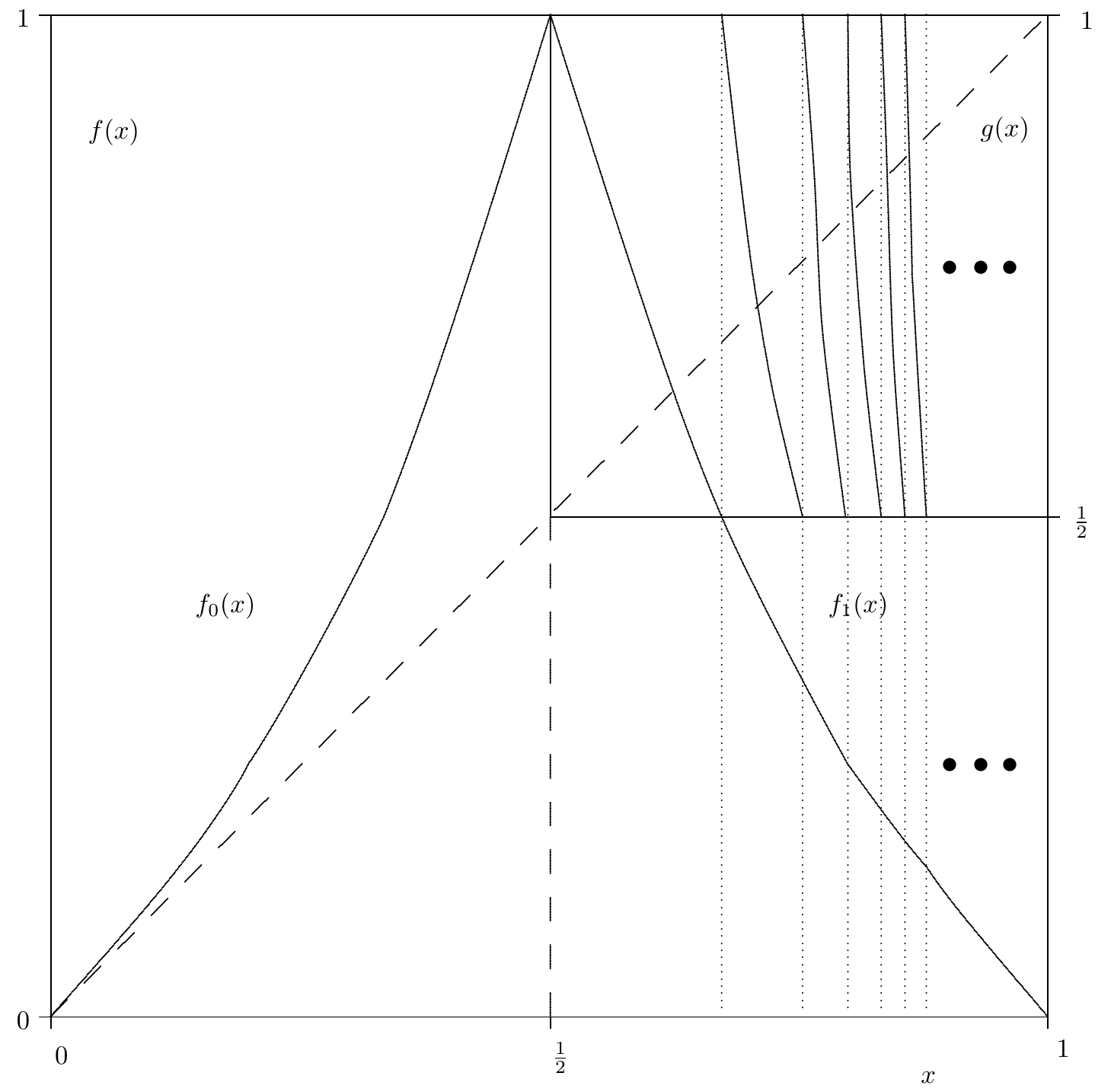}
  \caption{The induced Farey map}
  \label{fig:induced}
\end{figure}

For the Farey map, a good choice for $J$ is the interval $[\frac{1}{2},1]$.
The induced map for this $J$ is given by
\be
g(y) = \left\{ 
\begin{array}{lll}
\frac{1-y}{1+n(y-1)}, & \mbox{for} \ y 
\in (\frac{n}{n+1} ,\frac{n+1}{n+2} ]  &  n=1,2,3,\ldots  \\        
1 & \mbox{for} x= \frac{1}{2} &
\end{array}
\right.
\label{eq:inducedfarey}
\ee 
This new function is displayed along with the Farey map in figure~(\ref{fig:induced}).

\section{The Induced Zeta Function and the Induced Transfer Operator}
\label{sec:zeta}

The concept of the induced system naturally
gives rise to an induced zeta function and an induced transfer operator.
The following is a derivation of the form of the induced zeta function
which, as stated above, will help determine the appropriate
interaction of the the induced system.

The logarithm of the zeta function of the Farey map can be 
decomposed into two parts: another zeta function relating to 
the induced system and a simple term relating to the indifferent 
fixed point at $0$. More generally, the zeta function for any 
system $X$ can be thought of as the product of the zeta functions 
$\zeta_1(z)$ and $\zeta_2(z)$ relating to the systems $Y$ and the 
induced system $Y^c$ respectively.  In fact, they will demonstrate
natural choices for the interactions of these systems, $\phi_{Y}$ and $\phi_{Y^c}$
based on the fact that $\phi(x) = -\beta\log|f'(x)|$ from
equation~(\ref{eq:interaction}).
Explicitly, we have from the 
definition of the zeta function~(\ref{defn:zeta2}):
\begin{eqnarray*}
\lefteqn{\log\zeta_{X}(z,\phi) = \sum_{n=1}^{\infty} \frac{z^n}{n}
\sum_{f^{n}(x)=x} \exp \sum_{k=0}^{n-1} \phi(f^{k}x) }\\ 
 & = & \sum_{n=1}^{\infty} \frac{z^n}{n} 
\left(
\sum_{f^{n}(x)=x  , \exists k \mid f^{k}x \in J}
+ \sum_{f^{n}(x)=x  ,  \not\exists k \mid f^{k}x \in J} 
\right) 
\exp \sum_{k=0}^{n-1} \phi(f^{k}x) \\  
 & = & \sum_{n=1}^{\infty} \frac{z^n}{n} 
\sum_{f^{n}(x)=x , \exists k \mid f^{k}x \in J}
\exp \sum_{k=0}^{n-1} \phi(f^{k}x)
+
\sum_{n=1}^{\infty} \frac{z^n}{n} 
\sum_{f^{n}(x)=x  , \not\exists k \mid f^{k}x \in J} 
\exp \sum_{k=0}^{n-1} \phi(f^{k}x) \\
 & = & \log\zeta_1(z) + \log\zeta_2(z)
\end{eqnarray*}

Now consider the complementary system 
$Y^c=(K,f \mid_K,\phi_{Y^c})$ which was defined 
at the start of this chapter with the interacion $\phi_{Y^c} = \phi$ . 
The fixed points of $f^{n}$ in $K$ 
are by definition precisely those that are part of cycles 
that have no point in $J$. The second term is  
therefore the logarithm of the zeta 
function $\zeta_{Y^c}(z)$.
In the case of the Farey map, $\log \zeta_{Y^c}(z)$ is very simple.
$K$ is just the indifferent fixed point at 0.  Since 
$\phi(f^k 0)=\phi(0)=-\log f'(0)=-\log 1 =0$ the zeta function reduces to
$ \log \zeta_{Y^c}(z) = \sum_{n=1}^{\infty} \frac{z^n}{n}(1) $. This 
series is simply the expansion of $-\log(1-z)$ about $z=0$.  Thus, the 
series for $\zeta_{Y^c}(z)$ can be analytically extended to 
a meromorphic function in the whole of the $z$ plane independent of $\beta$
with a simple pole at $z=1$.
\be
\log \zeta_{Y^c}(z) = \log\frac{1}{1-z} \Rightarrow 
\zeta_{Y^c}(z) = \frac{1}{1-z}
\label{eq:zetacomp}
\ee

Returning to the first term, it can be rewritten as follows:
\begin{eqnarray*}
\lefteqn{\log\zeta_1(z) =  \sum_{n=1}^{\infty} \frac{z^n}{n} 
\sum_{f^{n}(x)=x  ,\exists k \mid f^{k}x \in J} 
\exp \sum_{k=0}^{n-1} \phi(f^{k}x)} \\
& = & \sum_{n=1}^{\infty} \frac{z^n}{n} 
\sum_{f^{n}(x)=x  ,\exists k \mid f^{k}x \in J} 
\exp \sum_{k=0}^{n-1} -\beta\log |f'(f^{k}x)| \\
& = & \sum_{n=1}^{\infty} \frac{1}{n}
\sum_{f^{n}(x)=x  ,\exists k \mid f^{k}x \in J} 
\exp \sum_{k=0}^{n-1} \left( \log z - \beta\log |f'(f^{k}x)| \right)\\
& = & \sum_{n=1}^{\infty} \sum_{m=1}^{n} \frac{1}{m} 
\sum_{\stackrel{\mbox{\scriptsize $f^n(x)=x, x \in J$ s.t.}}
{\mbox{\scriptsize $n$-cycle has $m$ different points in $J$}}} 
\exp \sum_{k=0}^{n-1} \left( \log z - \beta\log |f'(f^{k}x)| \right) \\
& = & \sum_{m=1}^{\infty} \frac{1}{m} \sum_{n=m}^{\infty} 
\sum_{\stackrel{g^{m}(x)=x}{\scriptscriptstyle \sum_{k=0}^{m-1} n(g^k(x)) = n}}
\exp \sum_{k=0}^{n-1} \left( \log z - \beta\log |f'(f^{k}x)| \right) \\
\end{eqnarray*}
Since $g^{m}(x)=x$, the sum in the exponential of the previous
line may be manipulated in the following way:
\begin{eqnarray*}
\lefteqn{\sum_{k=0}^{n-1} \left( \log z - \beta\log |f'(f^{k}x)| \right) }  \\
 & = & n(x) \log z - \beta \log \prod_{k=0}^{n(x)-1} |f'(f^{k}\circ g^{0}(x))|  \\
 & + & n(g^{1}x) \log z - \beta
\log \prod_{k=0}^{n(g^{1}(x))-1} |f'(f^{k}\circ g^{1}(x))|  \\
 & \vdots & \vdots  \\
 & + & n(g^{m-1}x) \log z - \beta \log \prod_{k=0}^{n(g^{m-1}(x))-1} |f'(f^{k}
           \circ g^{m-1}(x))|  \\
 & = & \sum_{k=0}^{m-1} \left( 
n(g^k x) \log z - \beta \log |g'(g^k x)| \right)
\end{eqnarray*}
where the chain rule has been used:
$g'(y)=\frac{d}{dy}f^{n(y)}(y)=\prod_{k=0}^{n(y)-1} f'(f^{k}y)$.

So:
\begin{eqnarray*}
\log \zeta_1(z) & = & \sum_{m=1}^{\infty} \frac{1}{m} \sum_{n=m}^{\infty} 
\sum_{\stackrel{g^{m}(x)=x}{\scriptscriptstyle \sum_{k=0}^{m-1} n(g^k(x)) = n}}
\exp \sum_{k=0}^{m-1} \left( n(g^k x) \log z - \beta \log |g'(g^k x)| \right) \\
& = & \sum_{m=1}^{\infty} \frac{1}{m}  
\sum_{g^{m}(x)=x} 
\exp \sum_{k=0}^{m-1} \left( n(g^k x) \log z - \beta \log |g'(g^k x)| \right) 
\end{eqnarray*}
Now, define a new interaction, which depends explicitly on $z$:
\be
\phi_z(y)= n(y) \log z - \beta \log |g'(y)|
\ee
To demonstrate the link to the induced system, it is useful to 
to introduce a parameter $\sigma$ in the following way:
\be
\log \zeta_1(z) =
\left. \zeta_1(z,\sigma) \right|_{\sigma=1} =   
\left. \sum_{m=1}^{\infty} \frac{\sigma^{m}}{m}  
\sum_{g^{m}(x)=x} 
\exp \sum_{k=0}^{m-1}  \phi_z(g^{k}x)  \right|_{\sigma = 1}
\label{eq:zetawithxi}
\ee
It is clear then that $\zeta_1(z,\sigma)=\zeta_{Y}(\sigma)$ if the induced
system is given the interaction $\phi_Y = \phi_z$ defined above.  The notation
$\zeta_{\mbox{ind}}(z,\beta)$ will also be used to denote this function. Thus, 
the definition of the induced system is given by the triplet
\bdm
Y_z=(J,g,\phi_z)
\edm
where the $z$ subscript has been introduced to emphasize the fact that this
system has an explicit $z$ dependence.
The identifications of $\zeta_1(z)$ and $\zeta_2(z)$
with the zeta functions of the induced and 
the original system may now be stated as a result
\begin{lemma}
The zeta functions of the original, the induced and the complementary system
are related in the following way:
\bdm
\zeta_X(z) \equiv \zeta_{Y_z}(1) \zeta_{Y^c}(z)
\edm
\label{lemma:zetafconn}
\end{lemma}

Using the interaction found in the discussion of the zeta
function and the definition of the transfer operator~(\ref{eq:transferop1}),
the transfer operator 
for the induced Farey map, which will be denoted by $\M$, will have the 
following definition:
\begin{defn}
\bdm
\M \circ \phi(x) = \sum_{f(y)=x}\frac{z^{n(y)}\phi(y)}{|g'(y)|^{\beta}}
\edm
\label{defn:indtransferop}
\end{defn}
Note, that a naive definition of the transfer operator might not include the
$z^{n(y)}$.  This factor provides the link back to the original system as 
it carries the information about how many iterates of the Farey map 
it takes for a point to return to $J$.  Thus, knowledge of time has been retained
but has been removed from the actual dynamics of $g$ making it a simpler map
to study.  

Consider now a decomposition of $\LB$ into $\LB_{1}+\LB_{0}$ where
$\LB_{1}\psi=\LB(\chi_{J^c}\psi)$ and $\LB_{0}\psi=\LB(\chi_{J}\psi)$.
Here, $\chi$ is the characteristic map: $\chi_A x = 1$ if $x \in J$ 
and  $\chi_A x = 0$ if $x \not \in A$ where $A$ is some set.
 
The main connection to the thermodynamics of the original map
as far as the the transfer operator is concerned is the following result
( which will be stated without proof).
\begin{theorem}[\cite{prelslawn:interm}]
Suppose $0 < |z| < 1/r(\LB_1)$.  Then $\frac{1}{z}$ is an eigenvalue
of $\LB$ if and only if 1 is an eigenvalue of $\M$.  In addition, the
geometric multiplicity of the eigenvalue $\frac{1}{z}$ for $\LB$ is the
same as that of 1 for $\M$.
\label{theorem:topconn}
\end{theorem}
Here, $r(\O)$ is the spectral radius of the operator $\O$ and
is given by 
\bdm
r(\O)=\sup_{\lambda \in \sigma(\O)} |\lambda|
\edm
Note that $\sigma(\O)$ denotes the spectrum of the operator $\O$ which 
are all the values $\lambda$ such that $(\lambda I-\O)$ does not have a continuous
inverse, (\cite{rob:topvecsp}). The value of $r(\LB_1)$ for the induced Farey map
is 1 and this is shown in the appendices in section~(\ref{Asec:radL1})
More details on these observations and
further connections regarding the eigenfunctions of $\LB$ and $\M$ can 
be found in~\cite{prelslawn:interm}.

\chapter{The connection between the Induced Transfer Operator and
the Induced Zeta Function}
\label{chap:connection}
This section provides a result linking the Fredholm determinant of the
induced transfer operator to the zeta function.  Using this connection,
results on the meromorphic properties of the zeta function will
then be obtained.
The method applied has been used in a general setting by Ruelle for expanding maps
and Anosov flows in~\cite{ruelle:anosov} 
and for the Gauss map by Mayer in~\cite{mayer:zeta}.  While the concepts
are not new, the present work is a novel application of the ideas
involved.  An extension of the
work done here would hopefully demonstrate the same relationship for
induced transfer operators and the corresponding induced zeta functions
in a more general setting.

\section{Nuclear Operators according to Grothendieck}
\label{sec:nuclear ops}

This section presents some important results from the theory of 
Nuclear Operators which was developed by Grothendieck in 
the 1950's~\cite{groth:nuclear}.  Much of what is presented here comes from
a useful summary of Grothendieck's work provided by Mayer in~\cite{mayer:transop}
and also in~\cite{ergsymdyn}.  The basic notion of the work is to provide 
a way of finding operators with well defined traces in the setting of 
general Banach spaces.

The first step is to associate a linear operator, acting from any Banach space
to another, with a Fredholm kernel.
A brief description of the latter is thus provided here.  
\subsection{Fredholm kernels}
\label{subsec:fredholmkernels}

Consider any two Banach spaces $(E,\| \ \|_E)$ and $(F,\| \ \|_F)$
and their tensor product $E \otimes F$.   Next take the so-called
$\pi$-norm, $\| \ \|_{\pi}$, which is defined as follows:
\be
\|X\|_{\pi} \equiv \inf \sumi \| e_i \|_E \| f_i \|_F
\label{eq:pinorm}
\ee
where the infimum is taken over all finite sets of $\{e_i\} \in E$ 
and $\{f_i \in F\}$ such that $X=\sumi e_i \otimes f_i$.  The completion
of $E \otimes F$ with respect to the $\pi$-norm is the `projective 
topological tensor product' of $E$ and $F$.  This new Banach space will be 
denoted by $\EF$; its elements are referred to as the Fredholm kernels.
Note that by choosing $a_i \neq 0$ and $b_i \neq 0$ 
so that $\|a_i^{-1} e_i \|_E=1$ and $\|b_i^{-1} f_i \|_F=1$
and then setting $\lambda_i=a_i b_i$, each Fredholm kernel has a representation
\be
X=\sumi \lambda_i e_i' \otimes f_i'
\label{eq:repnote}
\ee
such that $\| e_i' \|_E=1$ and $\| f_i' \|_F=1$. 
This also implies that $\sumi \| \lambda_i \| < \infty$. 

\subsection{Nuclear Operators}
\label{subsec:nuclear ops}
Again, consider any two Banach spaces $(E,\| \ \|_E)$ and $(F,\| \ \|_F)$
and now also the space $\EstarF$.  Here, $E^\ast$ is the dual of the Banach space $E$ 
which is defined as the set
of all bounded linear functionals on $E$.  The norm of any $f \in E^\ast$ is
given by
\be
\| f \|_{E^\ast} \equiv \sup_{e \in E; \| e \| \leq 1}|f(e)| 
\label{eq:dualnorm}
\ee
The contraction of any element of $\EstarF$ with an element of $E$ yields
an element of $F$.  In this sense, every element of $\EstarF$ is equivalent to
a bounded linear operator $\L_X:E \rightarrow F$, i.e. 
$\L_X \in B(E,F)$\footnote{$B(E,F)$ is the space of all linear bounded maps of 
$E$ to $F$}.  Considering the representation of $X$ in equation~(\ref{eq:repnote}),
the operation of $\L_X$ may always be written as
\be
\L e \equiv \sumi \lambda_i e_i^\ast(e) f_i \mbox{ \ forall $e \in E$}
\label{eq:loprep}
\ee
where $\| e_i \|_E=1$ and $\| f_i \|_F=1$.  So, the above indicates there is
a natural mapping $\phi:\EstarF \rightarrow B(E,F)$.  It is important at this
point to simply note that $\phi$ is not always an injective mapping.

Now, for any Banach space $E$, the definition of a nuclear operator $\L$ is 
given as follows:
\begin{defn}
Take any $\L:E \rightarrow F$ such that $\L$ is a linear bounded operator and
$E$ is any Banach space.
$\L$ is said to be nuclear if there exists a Fredholm kernel 
$X \in \EF$ with $\L = \L_X$.
\label{defn:nukeop}
\end{defn}

One of the most important features of nuclear operators is the possible
existence of a trace.
Firstly, the trace of a Fredholm kernel $X \in \EstarE$ with the representation
$X=\sumi \lambda_i e_i^\ast \otimes e_i$ is given by
\be
\tr X \equiv \sumi \lambda_i e_i^\ast(e_i)
\label{eq:tracefred}
\ee
where $e_i^\ast \in E^\ast$, $e_i \in E$, $\|e_i^\ast \|_E^\ast=1$, 
$\|e_i\|_E=1$ and $\{\lambda_i\} \in \ell_1$.  While this is known to be 
a well defined function, problems arise when considering an operator 
$\L \in B(E,F)$.  This is due to the fact that $\phi$ is not necessarily
injective --- if it is not injective than there will be ambiguity in trying
to define a trace of $\L$ based on the trace of its Fredholm kernel, 
as it has more than one such kernel.\footnote{Note that a Fredholm kernel has many
representations and hence the fact that two representations are different does not mean
their corresponding kernels are different.   However, two different kernels may
easily provide different traces giving rise to the problem mentioned in the text.}

The search for trace-class operators acting on Banach spaces led Grothendieck
to the following classification of Fredholm kernels and thence nuclear
operators.
\begin{defn}
Take any Fredholm kernel $X \in \EF$.  $X$ is said to be `p-summable' if
$X$ has a representation $X = \sumi \lambda_i e_i \otimes f_i$
such that $\sum_{i=1}^\infty |\lambda_i|^p < \infty$, i.e. $\{\lambda_i\} \in \ell_p$.
The `order' of the Fredholm kernel is the number q which is the infimum of all
p such that $0 < p \leq 1 $ and $X$ is p-summable.  Note that, in general,
$0 \leq q \leq 1$ while $0 < p \leq 1$.
\label{defn:psummfkernel}
\end{defn}

\begin{defn}
Take $\L:E \rightarrow F$, a nuclear operator.  $\L$ is p-summable
if there is a p-summable $X \in \EF$ such that $\L = \L_X$.  The order of 
$\L$ is the same as the order of $X$.
\label{defn:psummnukeop}
\end{defn}

\begin{theorem}
Consider a nuclear operator $\L:E \rightarrow E$ of order p such 
that $0 \leq p \leq \frac{2}{3}$.  Then $\L$ is of trace class:
\bdm
\mbox{\em trace} \L = \sumi \nu_i
\edm
where $\nu_i$ are the eigenvalues of $\L$ counted according to their
algebraic multiplicity.  The Fredholm determinant $\det (1-\xi\L)$ is an
entire function of $\xi$ given by the formula 
\bdm
\det(1-\xi\L)=\prod_i(1-\nu_i\xi)=\exp \mbox{\em trace} \log (1-\xi\L)
\edm
Also, if $\L=\L(\sigma)$ and $\sigma \rightarrow \L(\sigma)$ is a holomorphic
function of $\sigma$ in some domain $D$ then the function 
$\det(1-\L(\sigma))$ is holomorphic in $D$.
\label{theorem:nuke trace}
\end{theorem}

Another important result concerns the composition of a nuclear operator
with bounded operators:
\begin{lemma}
Consider $\L:E \rightarrow F$, a nuclear operator of order q, $\O_1 \in L(F,G)$ 
and $\O_2 \in L(G,E)$.  Then the composition mapping 
$\O_1 \cdot \L \cdot \O_2:G \rightarrow G$ is also a nuclear operator of order q.  
\label{lemma:nukecomp}
\end{lemma}

\subsection{Nuclear Spaces}
\label{subsec:nuke spaces}
A nuclear space is defined by the fact that any bounded linear operator 
mapping it to any Banach space is a nuclear operator.  Nuclear
spaces are actually a class of Fr\'{e}chet spaces.  A Fr\'{e}chet
space is a convex topological metric space that admits a metric such that
its topology is reproduced by the metric and it is complete with respect to
this same measure of distance. Note that not every Fr\'{e}chet
space is normable so this is not a direct generalization of Banach spaces.
In fact, it is known that any infinite dimensional Banach space is not nuclear.
One particularly useful nuclear space is $\H(D)$ where $D$ is some open
set in $C^n$. Indeed, this space will aid in the proof that $\M$ is 
a nuclear operator.  $\H(D)$ is the space of holomorphic functions on $D$
which are continuous on $\overline{D}$ together with the seminorms $\| \ \|_K$,
where $K$ is compact in $D$ and 
\be
\| f \|_K =\sup_{\xi \in K}|f(\xi)|
\label{eq:seminormdef}
\ee
Also, for certain special Banach spaces, every nuclear operator $\L$ is of order 0
and considering, theorem~(\ref{theorem:nuke trace}), must be therefore of trace class.
One such example is any Banach space of holomorphic functions over a domain
$D$ in $C^n$\label{order0}.

\subsection{Compact Operators}
\label{subsec:compact ops}
Compact operators are defined as per~\cite{rob:topvecsp} which will be the main 
reference for this section:
\begin{defn}[(\cite{rob:topvecsp}), VIII, \S 1, p. 143]
Let $\L$ be a linear operator such that 
$\L:E \rightarrow F$ where $E$ and $F$ are convex
vector spaces.  $\L$ is said to be {\em compact} if for any neighbourhood of 
the origin $U \in E$,
there exists a compact set $K \in F$ such that $\L(U) \subseteq K$.
\label{defn:compactops}
\end{defn}
Compact operators \label{compact}
have many nice properties and, with regards to spectrum,
behave in many ways like finite dimensional matrix operators.  The following
results and observations are for a compact operator that maps a convex
Hausdorff space $E$ into itself.
The operator's spectrum, apart from possibly 0\footnote{If $E$ is infinite dimensional,
0 always belongs to the spectrum}, comprises entirely of eigenvalues
and the eigenspace associated with each eigenvalue is finite dimensional.
These eigenvalues are also either a finite set of values or a sequence
which is convergent to zero.
In particular, one result which will be important for later on is 
recorded here as a lemma:  
\begin{lemma}[(\cite{rob:topvecsp}), VIII, \S 1, Corollary 2, p. 147]
If $\L$ is a compact linear operator mapping a convex Hausdorff space $E$
into itself and $W=\lambda I -\L$, where $0 \neq \lambda \in C$, then
$\lambda$ is not an eigenvalue of $\L$ iff $W$ is bijective.
\label{lemma:compact1}
\end{lemma}
Finally, note that any operator that has a countable set of eigenvalues
converging to 0 is a compact operator.  Thus all nuclear operators are
compact but the reverse is not true since a sequence whose terms
converge to zero is by no means an absolutely convergent series.

\subsection{Nuclearity of $\M$}
\label{subsec:comnucM}
Using the definition of the induced system,
the action of the induced transfer operator is defined as follows:
\begin{defn}
\bdm
\M \circ \phi(\xi) = \sum_{n=1}^{\infty} \frac{z^n}{(1+n\xi)^{2\beta}}
\phi\left( 1-\frac{\xi}{1+n\xi}\right)
\edm
\label{defn:indtop}
\end{defn}
where the composition operators will be denoted by $G_n$, 
i.e. $G_n(\xi)=1-\frac{\xi}{1+n\xi}$.
For the induced Farey map, $\xi \in [\half,1]$.  In order to analyse this 
operator more fully, however, it is natural to extend the domain into the 
complex plane where the full weight of complex analysis can be brought to
bear on the problem.  Note that each $G_n$ has a simple pole at $\xi=-\frac{1}{n}$.
It is thus advisable to stay away from these points and, in particular, away
from $\xi=0$.  A suitable domain for the work here, which is by no means the 
best or only choice, is given by
\be
D = \{ \xi: |\xi-1| < \frac{3}{4} \}
\label{eq:domain}
\ee
Next consider the domain $D'= \{ \xi: |\xi-1| < \frac{2}{3} \} \subset D$.
\begin{lemma}
\bdm
\overline{G_n(D)} \subseteq \overline{D'}
\edm
where $\overline{A}$ denotes the set of the complex
conjugates of the elements of $A$.
\label{lemma:Gncontract}
\end{lemma}
\Proof. \ \
Expressing $G_n$ in the form
\be
G_n(\xi)=\frac{1+(n-1)\xi}{1+n\xi}
\label{eq:Gnmobius1}
\ee
it is seen to be a M\"{o}bius or a linear fractional transformation, 
see~\cite{rudin:anal}.  It is well known that these transformations
map discs to either a disc or a halfspace in the complex plane.  Since
$G_n$ is bounded on $D'$ it cannot map it to a half space and so $G_n(D')$
must be an open disc.  Note that
\be
\overline{G_n(\xi)}=\frac{\overline{1+(n-1)\xi}}{\overline{1+n\xi}}
=\frac{1+(n-1)\overline{\xi}}{1+n\overline{\xi}}=G_n(\overline{\xi})
\label{eq:Gnmobius2}
\ee
where $\overline{\xi}$ denotes the complex conjugate of $\xi$.
This means that since $D'$ is symmetric about the real axis, the disc
$G_n(D')$ must also have its centre lying on the real axis.  Further, note that the 
boundary of $D'$, $\partial D'$ is mapped to the boundary of 
$G_n(D')$, $\partial G_n(D')$.  Therefore, the two intersections of $\partial D'$
with the real axis are mapped to the two intersections of $\partial G_n(D')$ with
the real axis.  Thus, it is enough to find $G_n(\frac{1}{3})$ and $G_n(\frac{5}{3})$
to totally specify the region $G_n(D)$.
Their absolute difference will give the diameter of $G_n(D)$ and their mean
the position of its centre.
The following is then observed:
\be
G_n(\frac{1}{4})=1-\frac{1}{n+4} \mbox{ and } G_n(\frac{7}{4})=1-\frac{7}{7n+4}
\label{eq:Gnmobius3}
\ee
It is clear then that each circle lies inside $\overline{D'}$ as $G_n(\frac{7}{4})$
and $G_n(\frac{1}{4})$ always lie in $[\frac{1}{3},\frac{2}{3}]$.  Since all points
inside $D$ are mapped inside each circle, the proof is complete.
\hfill
$\Box$

Let $\HD$ be the Banach space comprising of all functions holomorphic on the
domain $D$ and continuous on the closure of $D$.  It is clear from 
lemma~(\ref{lemma:Gncontract}) that $\M$ maps elements of $\HD$ to functions
on $D$.  Since the $G_n$ are holomorphic functions on $\overline{D}$, the
composition of a holomorphic function
$\phi \in \HD$ and $G_n$ is another holomorphic function.
Providing the weighted sum over all such compositions is itself bounded, the
function $\M(\phi)$ will be an element of $\HD$.  So it is clear then
that $\M(\HD) \subset \HD$.
\begin{lemma}
$\M:\H(D) \rightarrow \HD$ is a nuclear operator for $|z|<1$ and
$\beta \in C$.
\label{lemma:M nuke op}
\end{lemma}
\Proof. \ \
$\M$ is a bounded operator for $|z| < 1$ and $\beta \in C$ since
for any $\phi \in \HD$ ( $\|\phi\|_{H_\infty(D)} \neq 0$ )
\begin{eqnarray}
\|\M \circ \phi(\xi)\|_{H_\infty(D)} & = &
\sup_{\xi \in D} |\M \circ \phi(\xi)|
\nonumber \\
 & \leq &
\sup_{\xi \in D} \sum_{n=1}^{\infty}
|z|^n |(1+n\xi)^{-2\beta}|
\left| \phi(1-\frac{\xi}{1+n\xi}) \right|
\nonumber \\
 & \leq & \|\phi(\xi) \|_{H_\infty(D)}  \sup_{\xi \in D}
\sum_{n=1}^{\infty} |z|^n |(1+n\xi)^{-2\beta} |
\end{eqnarray}
Via the ratio test, the sum on the right always converges for $|z| < 1$ 
irrespective of the value of $\beta$.   Since $\H(D)$ is a nuclear space
it follows from the discussion in section~(\ref{subsec:nuke spaces}) that
$\M$ is a nuclear operator.  

\hfill
$\Box$

\begin{corollary}
$\M:\HD \rightarrow \HD$ is a nuclear operator of order 0 for $|z|<1$ and
$\beta \in C$.
\label{cor:M nuke op}
\end{corollary}
\Proof. \ \
Composing $\M$ with the injective function $i:\HD \rightarrow \H(D)$
and noting lemma~(\ref{lemma:nukecomp}), it follows that 
$\M:\HD \rightarrow \HD$ is a nuclear operator. 
The last remark of section~(\ref{subsec:nuke spaces})
then gives that $\M$ must be a nuclear operator of order 0.

\hfill
$\Box$

Nuclearity of $\M$ for $|z|=1$ will be proven later with appropriate
restrictions on $\beta$.  Note that the above
also shows that $\M:\HD \rightarrow \HD$ is
a compact operator for $|z|<1$ and $\beta \in C$.

\section{The trace of $\M$}
\label{sec:traceM}
\begin{prop}
The trace of the induced transfer operator can be expressed as
\bdm
\mbox{\em trace} \M = \sum_{n=1}^\infty \mbox{\em trace} \M_n
= \sum_{n=1}^\infty 
\frac{z^n([\overline{n}])^{-2\beta}}{1+([\overline{n}])^2} 
\edm
where $[\overline{n}]$ is the continued fraction
\bdm
\cfrac{1}{n+\cfrac{1}{n+\cfrac{1}{n+\cdots}}}
\edm
\label{prop:trace of Mzbet}
\end{prop}
\Proof. \ \
The following analysis is for $z$ inside the unit disc and 
arbitrary $\beta$.
Consider the operator $\M$ in the form of a sum of operators: 
$\M=\sum_{n=1}^\infty \M_n$ where 
$\M_n\circ\phi(\xi)=\frac{z^n}{(1+n\xi)^{2\beta}}\phi(1-\frac{\xi}{1+n\xi})
=z^n (-G_n'(\xi))^{\beta}\phi(G_n(\xi))$.
The point of this is that $\tr \M = \sum_{n=1}^\infty \tr \M_n$ since 
trace is a linear function on operators and 
the problem can thus be broken down into 
one of determining the trace of the simpler operators $\M_n$.  
The method used to determine the spectrum of
$\M_n$ and hence $\tr \M_n$ is based on the work of 
Kamowitz found in~\cite{kamw:compop} and~\cite{kamw:endspectra}. The same
idea is also used for the Gauss map by Mayer~\cite{mayer:zeta}.

Recall that $G_n$ has exactly one fixed point $\xi_n^\ast$ in $D$,
see page~\pageref{uniquefp}, which was found to be
\be
\xi_n^\ast=[1,\overline{n}]=\frac{n-2+\sqrt{n^2+4}}{2n}
\label{eq:fpdef2}
\ee
The eigenvalue equation for each $\M_n$ is 
\be
\M_n\circ\phi(\xi)=\lambda_{n} \phi(\xi)=z^n (-G_n'(\xi))^{\beta}\phi(G_n(\xi))
\label{eq:eigM_n}
\ee
At the fixed point $\xi_n^\ast$, $G_n(\xi_n^\ast)=x_n^\ast$ and so 
equation~(\ref{eq:eigM_n}) becomes
\be
\lambda_{n} \phi(\xi_n^\ast)=z^n (-G_n'(\xi_n^\ast))^{\beta}\phi(xi_n^\ast)
\label{eq:eigM_nfp}
\ee
Providing $\phi(\xi_n^\ast) \neq 0$, equation~(\ref{eq:eigM_nfp}) shows
that
\be
\lambda_n=\lambda_{n,0}=z^n (-G_n'(\xi_n^\ast))^{\beta}
\ee
In the case that $\phi(\xi_n^\ast)=0$, the differentiation of
equation~(\ref{eq:eigM_n}) with respect to $\xi$ 
yields another possible value of $\lambda_n$, $\lambda_{n,1}$:
\be
\lambda_{n,1} \phi'(\xi_n^\ast) = z^n \left(
2\beta(-G_n'(\xi))^{\beta-1}(-G_n''(\xi))\phi(G_n(\xi))
+(-G_n'(\xi))^{\beta}\phi'(G_n(\xi))G_n'(\xi)  
\right)
\label{eq:eigM_ndiff}
\ee
Once again, at the fixed point $\xi_n^\ast$, this equation simplifies.  The first
term on the righthand side of equation~(\ref{eq:eigM_ndiff})
disappears as $\phi(G_n(\xi_n^\ast))=\phi(\xi_n^\ast)=0$ by assumption.
This leads to the equation
\be
\lambda_{n,1} \phi'(\xi_n^\ast) = 
z^n (-1)^1 (-G_n'(\xi_n^\ast))^{\beta+1}\phi'(\xi_n^\ast)
\label{eq:eigM_ndifffp}
\ee
Again, providing $\phi'(\xi_n^\ast) \neq 0$, equation~(\ref{eq:eigM_ndifffp})
shows that
\be
\lambda_{n,1}=z^n (-1)^1 (-G_n'(\xi_n^\ast))^{\beta+1}
\ee
Clearly, for the $k^{\mbox{th}}$ differentiation, only the term that has
$\frac{d}{d^k\xi}\phi(\xi)|_{\xi=\xi_n^\ast}$ may be non-zero, 
as the assumption has been mad that $\frac{d}{d^l\xi}\phi(\xi)|_{\xi=\xi_n^\ast}=0$
for all $l \in \{0,1,\ldots,k-1\}$.
So this process immediately generalises to show that each
\be
\lambda_{n,j}=z^n (-1)^k (-G_n'(\xi_n^\ast))^{\beta+k} 
= \frac{z^n (-1)^k}{(1+n\xi_n^\ast))^{\beta+k}}
\ee
is in the spectrum of $\M_n$, where $k \in \{0,1,2,\ldots\}$.
Finally, the only other possiblity for the eigenvalue 
equation~(\ref{eq:eigM_n}) is that $\lambda_n = 0$.  From the remark
in a footnote to section~(\ref{subsec:compact ops}),  0 is actually part of the
spectrum and hence an eigenvalue since $\HD$ is infinite dimensional.
So the spectrum of $\M_n$, denoted by
$\sigma(\M_n)$ has been found to be at most this set of values. I.e.
\be
\sigma(\M_n) \subset \{0\} \bigcup 
\left\{z^n (-1)^k (-G_n'(\xi_n^\ast))^{\beta+k}; k \in \{0,1,2,\ldots\ \} \right\}
\label{eq:specM_n}
\ee

The next step is to show that these numbers are indeed eigenvalues of
$\M_n$.
One way of doing this is to show that 
$(\xi-\xi_n^\ast)^k$ is not in the range of
the operator $(\lambda_{n,k}-\M_n)$.  Lemma~(\ref{lemma:compact1}) then
implies that $\lambda_{n,k}$ must be an eigenvalue since otherwise 
$(\lambda_{n,k}-\M_n)$ would be bijective.
A result from Kamowitz, lemma 2 in~\cite{kamw:endspectra}, is followed
here as a means of proving this fact.
\begin{lemma}
\be
\sigma(\M_n) =
\{z^n (-1)^k (-G_n'(\xi_n^\ast))^{\beta+k}; k \in \{0,1,2,\ldots\} \}
\ee
In addition, these eigenvalues have algebraic multiplicity of one.
\label{lemma:Kameig}
\end{lemma}
\Proof \ \ 
The observation that $(\xi-\xi_n^\ast)^k 
\not\in R(\lambda_{n,k}\M)$ may be
shown by way of a contradiction.  Assume that there exists a
function $\phi \in \HD$ such that 
$(\lambda_{n,k}-\M)\phi(\xi)=(\xi-\xi_n^\ast)^k$.  Inserting the
expressions for $\lambda_{n,k}$ and $\M$ this assumption becomes
\be
z^n (-1)^k (-G_n'(\xi_n^\ast))^{\beta+k}\phi(\xi)
-z^n (-G_n'(\xi))^{\beta}\phi(G_n(\xi))
=(\xi-\xi_n^\ast)^k
\label{eq:kameig1}
\ee
If $k=0$ and $\xi=\xi_n^\ast$, equation~(\ref{eq:kameig1})
reduces to the following:
\be
z^n (-G_n'(\xi_n^\ast))^{\beta}\phi(\xi_n^\ast)
-z^n (-G_n'(\xi_n^\ast))^{\beta}\phi(\xi_n^\ast)=1
\label{eq:kameig2}
\ee
where the fact that $G_n(\xi_n^\ast)=\xi_n^\ast$ has been used.
The left hand side of this is obviously 0 which means the 
assumption is incorrect for $k=0$.
In the case $k>0$, a similar contradiction occurs.
When $\xi$ is set to $\xi_n^\ast$, equation~(\ref{eq:kameig1})
simplifies with the righthand side equalling 0.
\be
z^n (-1)^k (-G_n'(\xi_n^\ast))^{\beta+k}\phi(\xi_n^\ast)
-z^n (-G_n'(\xi_n^\ast))^{\beta}\phi(\xi_n^\ast)=0
\label{eq:kameig3}
\ee  
It follows immediately that $\phi(\xi_n^\ast)$ must be 0. 
Differentiating~(\ref{eq:kameig1}):
\begin{eqnarray}
\lefteqn{z^n (-1)^k (-G_n'(\xi_n^\ast))^{\beta+k}\phi'(\xi)
-z^n (-1)^1(-G_n'(\xi))^{\beta+1}\phi'(G_n(\xi)) }
\nonumber \\
 & & -z^n (-G_n''(\xi))\beta(-G_n'(\xi))^{\beta-1}\phi(G_n(\xi))
=k(\xi-\xi_n^\ast)^{k-1}
\label{eq:kameig4}
\end{eqnarray}
If $k=1$, the righthand side of equation~(\ref{eq:kameig4})
equals one.  Setting $\xi=\xi_n^\ast$ and recalling that
$\phi(\xi_n^\ast)=0$, the lefthand side becomes 0, again a
contradiction.  If $k>1$ however and $\xi=\xi_n^\ast$,
equation~(\ref{eq:kameig4}) becomes
\be
z^n (-1)^k (-G_n'(\xi_n^\ast))^{\beta+k}\phi'(\xi_n^\ast)
-z^n (-1)^1(-G_n'(\xi_n^\ast))^{\beta+1}\phi'(\xi_n^\ast) =0
\label{eq:kameig5}
\ee
which indicates that $\phi'(\xi_n^\ast)=0$.

In general, for all $j<k$ the $j^{\mbox{th}}$ differentiation
of equation~(\ref{eq:kameig1}) is
\begin{eqnarray}
\lefteqn{z^n (-1)^k (-G_n'(\xi_n^\ast))^{\beta+k}\frac{d^j}{d\xi^j}\phi(\xi)
-z^n (-1)^j (-G_n'(\xi))^{\beta+j}\frac{d^j}{d\xi^j}\phi(G_n(\xi)) } \\
 & + & (\mbox{terms involving 
$\frac{d^l}{d\xi^l}\phi(\xi)$ where $l<j$}) 
 = k(k-1)\cdots(k-j+1)(\xi-\xi_n^\ast)^{(k-j)}
\nonumber
\label{eq:kameig6}
\end{eqnarray}
Again the righthand side vanishes when $\xi=\xi_n^\ast$.
If it is assumed that for all $l<j$, 
$\frac{d^l}{d\xi^l}\phi(\xi_n^\ast)=0$, then the term in
brackets disappears.  Equation~(\ref{eq:kameig6}) reduces to 
\be
z^n (-1)^k (-G_n'(\xi_n^\ast))^{\beta+k}
\frac{d^j}{d\xi^j}\phi(\xi_n^\ast)
-z^n (-1)^j (-G_n'(\xi_n^\ast))^{\beta+j}
\frac{d^j}{d\xi^j}\phi(\xi_n^\ast)
= 0
\label{eq:kameig7}
\ee
Since $k \neq j$, $\frac{d^j}{d\xi^j}\phi(\xi_n^\ast)$
must also be equal to 0, i.e. $\frac{d^l}{d\xi^l}\phi(\xi_n^\ast)=0$
for all $l<j+1$.  Since this was shown to be true for $j=0$,
then, by induction, $\frac{d^j}{d\xi^j}\phi(\xi_n^\ast)=0$ for all
$j=0,1,\ldots,k-1$.  Looking now at the $k^{\mbox{th}}$ differentiation
of equation~(\ref{eq:kameig1}) 
\begin{eqnarray}
k! & = & z^n (-1)^k (-G_n'(\xi_n^\ast))^{\beta+k}\frac{d^k}{d\xi^k}\phi(\xi)
-z^n (-1)^k (-G_n'(\xi))^{\beta+k}\frac{d^k}{d\xi^k}\phi(G_n(\xi)) \cr
 &  & \mbox{} + (\mbox{terms involving $\frac{d^j}{d\xi^j}\phi(\xi)$ where $j<k$}) 
\label{eq:kameig8}
\end{eqnarray}
Now, setting $\xi$ to $\xi_n^\ast$,
the terms in the brackets vanish as has been discussed and the
first two terms of the equation cancel each other.  So the righthand
side of equation~(\ref{eq:kameig8}) equals 0 but as the 
lefthand side is $k!$, the assumption that 
$(\xi-\xi_n^\ast)^k$ is in the range
of $(\lambda_{n,k}-\M_n)$ is clearly false.

Finally, it remains to be shown that each eigenvalue pertains to a one
dimensional eigenspace.  This will be done by demonstrating that the 
eigenfunctions can be determined iteratively. It was shown in the above that
for any eigenfunction, $\phi_{n,k}$, corresponding to $\lambda_{n,k}$, the 
first $k-1$ derivatives evaluated at $\xi_n^\ast$ vanished.  The $k^{\th}$
derivative, equation~(\ref{eq:kameig8}), reveals no information about
the value of $\frac{d^k}{d\xi^k}\phi_{n,k}(\xi_n^\ast)$ except that it must not
be equal to zero.  Thus $\frac{d^k}{d\xi^k}\phi_{n,k}(\xi_n^\ast)$ is free to
be any complex constant.
Now, consider the ${k+1}^{\th}$ derivative of the eigenvalue equation evaluated at
$\xi = \xi_n^\ast$:
\begin{eqnarray}
0 & = & z^n (-1)^k(-G_n'(\xi_n^\ast))^{\beta+k}
\frac{d^{k+1}}{d\xi^{k+1}}\phi_{n,k}(\xi_n^\ast) \cr
 & - & z^n (-1)^{k+1} (-G_n'(\xi_n^\ast))^{\beta+k+1}
\frac{d^{k+1}}{d\xi^{k+1}}\phi_{n,k}(G_n(\xi_n^\ast)) \cr
 & - & z^n (-1)^k (-G_n''(\xi_n^\ast))(\beta+k)(-G_n'(\xi_n^\ast))^{\beta+k-1}
\frac{d^k}{d\xi^k}\phi_{n,k}(G_n(\xi_n^\ast))  
\label{eq:kameig9}
\end{eqnarray}
Clearly, $\frac{d^{k+1}}{d\xi^{k+1}}\phi_{n,k}(\xi_n^\ast)$ is proportional to and 
is uniquely determined by the value of 
$\frac{d^{k}}{d\xi^{k}}\phi_{n,k}(\xi_n^\ast)$.  In general, the value of 
$\frac{d^{k+l}}{d\xi^{k+l}}\phi_{n,k}(\xi_n^\ast)$ will be partly proportional to the
values of $\frac{d^{k+j}}{d\xi^{k+j}}\phi_{n,k}(\xi_n^\ast)$ where $0 \leq j \leq l-1$
and therefore ultimately exactly proportional to the value of 
$\frac{d^{k}}{d\xi^{k}}\phi_{n,k}(\xi_n^\ast)$.  Thus the eigenfunction is 
determined uniquely by the choice of value for 
$\frac{d^{k}}{d\xi^{k}}\phi_{n,k}(\xi_n^\ast)$ and is indeed proportional to
this value.   This means that up to a multiplicative constant, there is 
one eigenfunction for each $\lambda_{n,k}$, 
i.e. each eigenvalue is of algebraic multiplicity 1.

\hfill $\Box$

The trace of $\M_n$ is then simply the sum over all of the $\lambda_{n,k}$.
Since $|G_n'| < 1$, (see page~\pageref{uniquefp}), this is a convergent
geometric series:
\begin{eqnarray}
\tr \M_n & = & \sum_{k=0}^\infty \lambda_{n,k}  
\nonumber \\
 & = & \sum_{k=0}^\infty \frac{z^n (-1)^k}{(1+n\xi_n^\ast)^{2\beta+2k}}  
\nonumber \\
 & = & \frac{z^n}{(1+n\xi_n^\ast)^{2\beta}} 
\sum_{k=0}^\infty \left(\frac{-1}{(1+n\xi_n^\ast)^2}\right)^k
\nonumber \\
 & = & \frac{z^n(1+n\xi_n^\ast)^{2-2\beta}}
{1+(1+n\xi_n^\ast)^2} 
\label{eq:trM_n}
\end{eqnarray}
The quantity $(1+n\xi_n^\ast)$ simplifies as follows.
From the definition of $\xi_n^\ast$ in~(\ref{eq:fpdef2}), it may be written as
\be
\xi_n^\ast=\frac{1}{1+[\overline{n}]}
\label{eq:fpas1over1+n}
\ee
where $[\overline{n}]$ is the continued fraction $[n,n,n,\ldots]=[n;n,n,\ldots]^{-1}
=[\overline{n;}]^{-1}$.
So, using the above representation of $\xi_n^\ast$, $1+n\xi_n^\ast)$ becomes:
\begin{eqnarray}
1+n\xi_n^\ast & = & 1+n\frac{1}{1+[\overline{n}]}
 = \frac{1+\frac{1}{[\overline{n}]}}{1+[\overline{n}]}
 = \frac{1}{[\overline{n}]} \cr
 & = & \frac{(1+[\overline{n}])}{(1+[\overline{n}])}
 = \frac{1}{[\overline{n}]}=[\overline{n;}]
\label{eq:tracesimp}
\end{eqnarray}
The formula for the trace of $\M_n$, equation~(\ref{eq:trM_n}), may then be written
as
\be
\tr \M_n = \frac{z^n([\overline{n}])^{-2\beta}}
{1+([\overline{n}])^2} 
=\frac{z^n([n;\overline{n}])^{2-2\beta}}{1+([n;\overline{n}])^2}
\label{eq:trM_n1}
\ee
Returning to the object of this section, the trace of $\M$ itself is
then found to be:
\be
\tr \M = \sum_{n=1}^\infty \tr \M_n
= \sum_{n=1}^\infty 
\frac{z^n([\overline{n}])^{-2\beta}}{1+([\overline{n}])^2} 
\label{eq:trM}
\ee

\hfill
$\Box$

\section{A Generalized Induced Transfer Operator, $\Mk$}
\label{sec:tracegen}
At this point, it is useful to introduce a generalization of the induced transfer
operator, $\Mk$.  The idea here follows from the work of Ruelle
in~\cite{ruelle:anosov} and Mayer in~\cite{mayer:zeta}.
Ruelle shows that the $\zeta$-function for an expanding map is given in terms
of the Fredholm determinants of the generalized transfer operator.
\be
\zeta(z,\beta)=\prod_{k=0}^N [\det(1-z{\cal L}_k^{\beta})]^{(-1)^{k+1}}
\label{eq:ruezgen}
\ee
where ${\cal L}_k^{\beta}$ is the generalized transfer operator of the 
system and N is the dimension of the compact manifold on which the system is defined.
Note that for the Farey map and its induced version 
the action takes place on closed intervals, so this
dimension $N$ is 1.  Therefore, the induced zeta function may be expected to be
the quotient of the Fredholm determinants of $\M_1$ and $\M_0=\M$.
The next few sections aim to show that this result is indeed true for the
induced system.  

This generalized induced transfer operator is defined for $k=0,1,2,\ldots$ and is 
given by
\begin{defn}
\be
\Mk \circ \phi(\xi) = 
\MG \circ \phi(\xi) = 
(-1)^k \sum_{n=1}^\infty \frac{z^n}{(1+n\xi)^{2\beta+2k}} \phi(1-\frac{\xi}{1+n\xi})
\ee
\label{defn:genindtopdef}
\end{defn}
Notice that to obtain $\Mk$ from $\M$ all that is needed is a multiplicative
factor of $(-1)^k$ and the linear shift $\beta \rightarrow \beta+k$.  
The trace of $\Mk$ is therefore easily found from the trace of $\M$ given
in~(\ref{eq:trM})
\be
\tr \Mk = \tr \MG =
(-1)^k \sum_{n=1}^\infty 
\frac{z^n([\overline{n}])^{-2\beta-2k}}{1+([\overline{n}])^2} 
\label{eq:trgenM}
\ee
These generalised operators are also nuclear operators for all $|z|<1$ and
$\beta \in C$.  This follows immediately from lemma~(\ref{lemma:M nuke op}) due to
the direct correspondence between $\Mk$ and $\M$.

\section{The trace of $(\M)^N$}
\label{sec:ntrace}
Here the trace of $(\M)^N$ is determined.  Also, the trace
of $(\M_1)^N$ is calculated in the search for a 
connection between these generalized 
induced transfer operators and the induced $\zeta$-function of $\M$.
The method used is similar to the previous sections so some details will
be spared.

\begin{prop}
The trace of the induced transfer operator composed with itself $N$ times
is given by
\bdm
\envtr ({\M})^N = \sum_{i_1=1}^\infty \sum_{i_2=1}^\infty \cdots
\sum_{i_N=1}^\infty \frac{\prod_{l=1}^N z^{i_l}
[\overline{i_l,i_{l-1},\ldots,i_1,i_N\ldots,i_{l+1}}]^{2\beta}}
{1-(-1)^N \prod_{l=1}^N
[\overline{i_l,i_{l-1},\ldots,i_1,i_N\ldots,i_{l+1}}]^2}
\edm
\label{prop:trace of M^N}
\end{prop}
\Proof. \ \
It is useful to once more express $\M$ in the form 
$\M=\sum_{n=1}^\infty \M_n$.  $(\M)^N$ may itself then be written as
a sum of simpler operators:
\be
(\M)^N = \sum_{i_1=1}^\infty \sum_{i_2=1}^\infty \cdots
\sum_{i_N=1}^\infty \M_{i_1} \M_{i_2}  \cdots \M_{i_N}
\label{eq:Mnsum}
\ee
and its trace is then given by
\be
\tr (\M)^N = \sum_{i_1=1}^\infty \sum_{i_2=1}^\infty \cdots
\sum_{i_N=1}^\infty \tr \left( \M_{i_1} \M_{i_2}  \cdots \M_{i_N} \right)
\label{eq:Mntrace}
\ee
Attention may now be focussed on the composition of operators on the
righthand side of equation~(\ref{eq:Mntrace}) which will be denoted
by $\M_{I_N}$ where $I_N=\{i_1,i_2,\ldots,i_N \}$.
To find an explicit form for $\M_{I_N}$, consider 
$\M_{i_{N-1}} \circ \M_{i_N} \circ \phi(\xi)$
and the representation $\M_n\phi(\xi)=z^n (-G_n'(\xi))^{\beta}\phi(G_n(\xi))$.
\begin{eqnarray}
\lefteqn{  \M_{i_{N-1}} \circ \M_{i_N} \circ \phi(\xi) =
\M_{i_{N-1}} \circ z^{i_N} (-G_{i_N}'(\xi))^{\beta}\phi(G_{i_N}(\xi))  }
\nonumber \\
 & = & z^{i_{N-1}+i_N} (-G_{i_{N-1}}'(\xi))^{\beta}
(-G_{i_N}'(G_{i_{N-1}}(\xi)))^{\beta}\phi(G_{i_N}(G_{i_{N-1}}(\xi)))
\label{eq:MnIstart}
\end{eqnarray}
Continuing on in the same way by applying in order the operators $\M_{i_{N-2}}$,
$\M_{i_{N-3}}$ and so on, the following expression for $\M_{I_N}$ is obtained:
\be
\M_{I_N} \circ \phi(\xi)=
\prod_{l=1}^N 
z^{i_l}
\left[ -G_{i_l}'(G_{i_{l-1}} G_{i_{l-2}} \cdots G_{i_1}(\xi))
\right]^{\beta} 
\phi(G_{i_N}G_{i_{N-1}} \cdots G_{i_1}(\xi))
\label{eq:MnI}
\ee
Notice that the function $\phi$ in equation~(\ref{eq:MnI}) has as its argument
a composition of the inverse branches of the induced map $g$.  This is 
exactly the object whose fixed point was needed to find the 
exact expression for the induced zeta function, see section~(\ref{sec:indzf}).
There it was shown that $G_{i_N}G_{i_{N-1}} \cdots G_{i_1}$ had 2 fixed points
and only one inside $D$, see page~\pageref{compmapguff}.  
On examination of equation~(\ref{eq:compmapfp}),
the fixed point of $G_{i_N}G_{i_{N-1}} \cdots G_{i_1}$, denoted by
$\xi_{I_N}^\ast$ is seen to be:
\be
\xi_{I_N}^\ast=[1,\overline{i_N,i_{N-1},\ldots,i_1}]
\label{eq:newcompmapfp}
\ee
Now, the eigenvalue equation for $\M_{I_N}$ is the following
\begin{eqnarray}
\lefteqn{\lambda \phi(\xi) =
(-1)^N \prod_{l=1}^N z^{i_l}
\left[G_{i_l}'(G_{i_{l-1}} G_{i_{l-2}} \cdots G_{i_1}(\xi)) \right]^{\beta} 
\phi(G_{i_N}G_{i_{N-1}} \cdots G_{i_1}(\xi)) }
\nonumber \\
 & = &  \prod_{l=1}^N z^{i_l}
\left[(-1)^N \frac{d}{d\xi}(G_{i_N} G_{i_{N-1}} \cdots G_{i_1}(\xi)) \right]^{\beta} 
\phi(G_{i_N}G_{i_{N-1}} \cdots G_{i_1}(\xi)) 
\label{eq:MINeig}
\end{eqnarray}
Consider the product term in equation~(\ref{eq:MINeig}), 
$\prod_{l=1}^N z^{i_l}
\left[-G_{i_l}'(G_{i_{l-1}} G_{i_{l-2}} \cdots G_{i_l}(\xi)) \right]^{\beta} $.
Differentiating both sides of the equation $g_n(G_n(\xi))=\xi$ gives the result
\be
G_n'(\xi)=\frac{1}{g_n'(G_n(\xi)}
\label{eq:diffinv}
\ee
Upon substitution of this result, the product term becomes
\be
\prod_{l=1}^N z^{i_l}
\left[-G_{i_l}'(G_{i_{l-1}} G_{i_{l-2}} \cdots G_{i_1}(\xi)) \right]^{\beta} 
=\prod_{l=1}^N z^{i_l}
\left[-g_{i_l}'(G_{i_l} G_{i_{l-1}} \cdots G_{i_1}(\xi)) \right]^{-\beta} 
\ee
Also, with the observation that 
\be
G_{i_l} G_{i_{l-1}} \cdots G_{i_1}(\xi)=
[1,\overline{i_l,i_{l-1},\ldots,i_1,i_N\ldots,i_{l+1}}]
\ee
equation~(\ref{eq:MINeig}) further simplifies when $\xi$ is set to $\xi_{I_N}^\ast$, 
the unique fixed point.
\be
\lambda \phi(\xi_{I_N}^\ast) =
\prod_{l=1}^N z^{i_l}
\left[-g_{i_l}'([1,\overline{i_l,i_{l-1},\ldots,i_1,i_N\ldots,i_{l+1}}]) 
\right]^{-\beta} 
\phi(\xi_{I_N}^\ast)
\label{eq:MINeig2}
\ee
Using lemma~(\ref{lemma:product}) equation~(\ref{eq:MINeig2}) becomes
\be
\lambda \phi(\xi_{I_N}^\ast) =
\prod_{l=1}^N z^{i_l}
[\overline{i_l,i_{l-1},\ldots,i_1,i_N\ldots,i_{l+1}}]^{2\beta} 
\phi(\xi_{I_N}^\ast)
\label{eq:MINeig3}
\ee
So, providing $\phi(\xi_{I_N}^\ast) \neq 0$, equation~(\ref{eq:MINeig3}) yields that
\be
\lambda=\lambda_{I_N,0}=
\prod_{l=1}^N z^{i_l}
[\overline{i_l,i_{l-1},\ldots,i_1,i_N\ldots,i_{l+1}}]^{2\beta}
\ee
is a possible eigenvalue.  

Now, exactly the same procedure used in the previous section may be applied
here.  At the unique fixed point of $G_{i_N}G_{i_{N-1}} \cdots G_{i_1}$
equation~(\ref{eq:MINeig3}) and the $l^{\mbox{th}}$ differentiation 
of equation~(\ref{eq:MINeig}) show that if 
$\frac{d^j}{d\xi^j}\phi(\xi)|_{\xi=\xi_{I_N}^\ast}=0 \ \forall \ j=0,1,\ldots,k-1$
and $\frac{d^k}{d\xi^k}\phi(\xi)|_{\xi=\xi_{I_N}^\ast} \neq 0$ then
\be
\lambda=\lambda_{I_N,k}=
(-1)^{Nk} \prod_{l=1}^N z^{i_l}
[\overline{i_l,i_{l-1},\ldots,i_1,i_N\ldots,i_{l+1}}]^{2\beta+2k} 
\ee
where $k=0,1,2,\ldots$.

The fact that these values are actually eigenvalues needs to be then
shown; the approach will be the same as that used in
the previous section for $M_n$ and not as much detail will be shown.
In particular, the following work will demonstrate that 
$(\lambda_{I_N,k}-\M_{I_N})$ is not a bijective mapping.

\begin{lemma}
\be
\sigma(\M_{I_N}) = (-1)^{Nk} \prod_{l=1}^N z^{i_l}
[\overline{i_l,i_{l-1},\ldots,i_1,i_N\ldots,i_{l+1}}]^{2\beta+2k};
 k \in \{0,1,2,\ldots\} \}
\ee
These eigenvalues have algebraic multiplicity of one.
\label{lemma:Kameigb}
\end{lemma}
\Proof \ \ 
Assume that there exists a $\phi \in \HD$ such that 
\be
(\lambda_{I_N,k}-\M_{I_N})\phi(\xi)=(\xi-\xi_{I_N}^\ast)^k
\label{eq:Kameigb1}
\ee
Written out in its full glory, the above becomes
\begin{eqnarray}
\lefteqn{ (-1)^{Nk} \prod_{l=1}^N z^{i_l}
[\overline{i_l,i_{l-1},\ldots,i_1,i_N\ldots,i_{l+1}}]^{2\beta+2k} \phi(\xi) } \\
 & - & \prod_{l=1}^N z^{i_l}
\left[(-1)^N \frac{d}{d\xi}(G_{i_N} G_{i_{N-1}} \cdots G_{i_1}(\xi)) \right]^{\beta} 
\phi(G_{i_N}G_{i_{N-1}} \cdots G_{i_1}(\xi)) 
=(\xi-\xi_{I_N}^\ast)^k \nonumber
\label{eq:Kameigb2}
\end{eqnarray}
As before, assume that for any $0<j<k$ that for all $l<j$, 
$\frac{d^l}{d\xi^l} \phi(\xi_{I_N}^\ast)=0$.  Differentiating 
equation~(\ref{eq:Kameigb2}) $j$ times:
\begin{eqnarray}
 &  & (-1)^{Nk} \prod_{l=1}^N z^{i_l}
[\overline{i_l,i_{l-1},\ldots,i_1,i_N\ldots,i_{l+1}}]^{2\beta+2k} 
\frac{d^j}{d\xi^j} \phi(\xi)
\nonumber \\
 & - & \prod_{l=1}^N z^{i_l}
\left[(-1)^{N} 
\frac{d}{d\xi}(G_{i_N} G_{i_{N-1}} \cdots G_{i_1}(\xi)) \right]^{\beta+j} 
\frac{d^j}{d\xi^j} \phi(G_{i_N}G_{i_{N-1}} \cdots G_{i_1}(\xi)) 
\nonumber \\
 & + & (\mbox{terms involving $\frac{d^l}{d\xi^l} \phi(\xi)$ where $l<j$})
\nonumber \\
 & = & k(k-1)\cdots(k-j+1)(\xi-\xi_{I_N}^\ast)^{k-j}
\label{eq:Kameigb3}
\end{eqnarray}
and then setting $\xi$ to $\xi_{I_N}^\ast$ yields the following equation:
\begin{eqnarray}
 &   & (-1)^{Nk} \prod_{l=1}^N z^{i_l}
[\overline{i_l,i_{l-1},\ldots,i_1,i_N\ldots,i_{l+1}}]^{2\beta+2k} 
\frac{d^j}{d\xi^j} \phi(\xi_{I_N}^\ast)
\nonumber \\
 & - &  (-1)^{Nj} \prod_{l=1}^N z^{i_l}
[\overline{i_l,i_{l-1},\ldots,i_1,i_N\ldots,i_{l+1}}]^{2\beta+2j} 
\frac{d^j}{d\xi^j} \phi(\xi_{I_N}^\ast)  =  0
\label{eq:Kameigb4}
\end{eqnarray}
Thus, since $j<k$, the two factors multiplying $\frac{d^j}{d\xi^j}
\phi(\xi_{I_N}^\ast)$ are different and $\frac{d^j}{d\xi^j} \phi(\xi_{I_N}^\ast)$ 
must be itself 0.  Therefore $\frac{d^l}{d\xi^l} \phi(\xi_{I_N}^\ast) = 0$
for all $l<j+1$.  Since this is true for $j=1$ it must be true by induction for all
$0<j<k$.  Consider finally the $k^{\mbox{th}}$ differentiation of 
equation~(\ref{eq:Kameigb2}): 
\begin{eqnarray}
 &  & (-1)^{Nk} \prod_{l=1}^N z^{i_l}
[\overline{i_l,i_{l-1},\ldots,i_1,i_N\ldots,i_{l+1}}]^{2\beta+2k} 
\frac{d^k}{d\xi^k} \phi(\xi)
\nonumber \\
 & - & \prod_{l=1}^N z^{i_l}
\left[(-1)^{N} \frac{d}{d\xi}(G_{i_N} G_{i_{N-1}} 
\cdots G_{i_1}(\xi)) \right]^{\beta+k} 
\frac{d^k}{d\xi^k} \phi(G_{i_N}G_{i_{N-1}} \cdots G_{i_1}(\xi)) 
\nonumber \\
 & + & (\mbox{terms involving $\frac{d^l}{d\xi^l} \phi(\xi)$ where $l<k$})
\nonumber \\
 & = & k!
\label{eq:Kameigb5}
\end{eqnarray}
At $\xi=\xi_{I_N}^\ast$ the first two terms on 
the left cancel each other, while the term in brackets
has been shown to be 0.  Since the righthand side is not equal to 0, the 
assumption has led to a contradiction.
Therefore, $(\xi-\xi_{I_N}^\ast)^k$ is not in the range of 
$(\lambda_{I_N,k}-\M_{I_N})$.  Thus $(\lambda_{I_N,k}-\M_{I_N})$ is not
bijective and lemma~(\ref{lemma:compact1}) implies that $\lambda_{I_N,k}$ must be
an eigenvalue of $M_{I_N}$.

In showing that each eigenvalue is of algebraic multiplicity one, 
exactly the same argument used in lemma~(\ref{lemma:Kameig}) may be 
employed here.  It is seen that the eigenfunction is uniquely determined by
the choice of the value of $\frac{d^k}{d\xi^k} \phi(\xi_n^\ast)$ and is in 
fact proportional to it.  Thus, up to a multiplicative constant, each
eigenvalue possesses one eigenfunction.  This completes the proof of the
lemma.

\hfill
$\Box$

The trace for each $\M_{I_N}$ is thus given by
\begin{eqnarray}
\lefteqn{ \tr \M_{I_N} = \sum_{k=0}^\infty \lambda_{I_N,k} } \nonumber \\
 & = &\sum_{k=0}^\infty (-1)^{Nk} \prod_{l=1}^N z^{i_l}
[\overline{i_l,i_{l-1},\ldots,i_1,i_N\ldots,i_{l+1}}]^{2\beta+2k} 
\nonumber \\
 & = & \prod_{l=1}^N z^{i_l}
[\overline{i_l,i_{l-1},\ldots,i_1,i_N\ldots,i_{l+1}}]^{2\beta} 
\sum_{k=0}^\infty  
\left( (-1)^N \prod_{l=1}^N 
[\overline{i_l,i_{l-1},\ldots,i_1,i_N\ldots,i_{l+1}}]^2 
\right)^k 
\nonumber \\
 & = &\frac{\prod_{l=1}^N z^{i_l}
[\overline{i_l,i_{l-1},\ldots,i_1,i_N\ldots,i_{l+1}}]^{2\beta}}
{1-(-1)^N \prod_{l=1}^N[\overline{i_l,i_{l-1},\ldots,i_1,i_N\ldots,i_{l+1}}]^2}
\label{eq:trMIN}
\end{eqnarray}

Substituting the result of~(\ref{eq:trMIN}) into equation~(\ref{eq:Mntrace})
the following expression for the trace of ${\M}^N$ is obtained:
\be
\tr {\M}^N = \sum_{i_1=1}^\infty \sum_{i_2=1}^\infty \cdots
\sum_{i_N=1}^\infty \frac{\prod_{l=1}^N z^{i_l}
[\overline{i_l,i_{l-1},\ldots,i_1,i_N\ldots,i_{l+1}}]^{2\beta}}
{1-(-1)^N \prod_{l=1}^N
[\overline{i_l,i_{l-1},\ldots,i_1,i_N\ldots,i_{l+1}}]^2}
\label{eq:Mntraceexp}
\ee

\hfill
$\Box$

The trace of $N$ compositions of the generalized induced 
transfer operator with itself, ${\Mk}^N$, is then given by 
setting $\beta \rightarrow \beta+k$ in the trace 
formula~(\ref{eq:Mntraceexp}) and multiplying by $(-1)^kN$.
\begin{eqnarray}
\lefteqn{ \tr {\Mk}^N =\tr {\MG}^N = 
(-1)^{kN} \tr {{\cal M}^{(z,\beta+k)}}^N }
\nonumber \\
 & = & (-1)^{kN} 
\sum_{i_1=1}^\infty \sum_{i_2=1}^\infty \cdots
\sum_{i_N=1}^\infty \frac{\prod_{l=1}^N z^{i_l}
[\overline{i_l,i_{l-1},\ldots,i_1,i_N\ldots,i_{l+1}}]^{2\beta+2k}}
{1-(-1)^N\prod_{l=1}^N
[\overline{i_l,i_{l-1},\ldots,i_1,i_N\ldots,i_{l+1}}]^2}
\label{eq:Mntrgen}
\end{eqnarray}
and, in particular, for the first generalized operator, $k=1$:
\be
\tr {\M_{(1)}}^N = (-1)^{N} 
\sum_{i_1=1}^\infty \sum_{i_2=1}^\infty \cdots
\sum_{i_N=1}^\infty \frac{\prod_{l=1}^N z^{i_l}
[\overline{i_l,i_{l-1},\ldots,i_1,i_N\ldots,i_{l+1}}]^{2\beta+2}}
{1-(-1)^N \prod_{l=1}^N
[\overline{i_l,i_{l-1},\ldots,i_1,i_N\ldots,i_{l+1}}]^2}
\label{eq:Mntrgen1}
\ee
In examining the various representation of the induced zeta function given 
in equations~(\ref{eq:fzetaind1}), \ref{eq:fzetaind2},
\ref{eq:fzetaind3}, \ref{eq:fzetaind4} and~(\ref{eq:fzetaind5}), 
its form is seen to closely resemble the trace formulas
given in~(\ref{eq:Mntraceexp}) and~(\ref{eq:Mntrgen1}).  In analogy
with the work of Ruelle in~\cite{ruelle:anosov}, the logarithm of
the induced zeta function is found to be equal to be a sum over $N$ of
the difference between $\tr {\M}^N$ and $\tr {\M_{(1)}}^N$ divided by $N$.
This difference is given by:
\begin{eqnarray}
\lefteqn{\tr {\M}^N-\tr {\M_{(1)}}^N}
\nonumber \\
 & = & \tr {\M_{(0)}}^N-\tr {\M_{(1)}}^N 
\nonumber \\
 & = & \sum_{i_1=1}^\infty \sum_{i_2=1}^\infty \cdots
\sum_{i_N=1}^\infty \frac{\prod_{l=1}^N z^{i_l}
[\overline{i_l,i_{l-1},\ldots,i_1,i_N\ldots,i_{l+1}}]^{2\beta}}
{1-(-1)^N \prod_{l=1}^N
[\overline{i_l,i_{l-1},\ldots,i_1,i_N\ldots,i_{l+1}}]^2}
\nonumber \\
 &   & - (-1)^{N} \sum_{i_1=1}^\infty \sum_{i_2=1}^\infty \cdots
\sum_{i_N=1}^\infty \frac{\prod_{l=1}^N z^{i_l}
[\overline{i_l,i_{l-1},\ldots,i_1,i_N\ldots,i_{l+1}}]^{2\beta+2}}
{1-(-1)^N \prod_{l=1}^N
[\overline{i_l,i_{l-1},\ldots,i_1,i_N\ldots,i_{l+1}}]^2}
\nonumber \\
 & = & 
\sum_{i_1=1}^\infty \sum_{i_2=1}^\infty \cdots
\sum_{i_N=1}^\infty \prod_{l=1}^N z^{i_l}
[\overline{i_l,i_{l-1},\ldots,i_1,i_N\ldots,i_{l+1}}]^{2\beta}
\nonumber \\
 &   & \times
\frac{1-(-1)^N \prod_{l=1}^N
[\overline{i_l,i_{l-1},\ldots,i_1,i_N\ldots,i_{l+1}}]^2}
{1-(-1)^N \prod_{l=1}^N
[\overline{i_l,i_{l-1},\ldots,i_1,i_N\ldots,i_{l+1}}]^2}
\nonumber \\
 & = &
\sum_{i_1=1}^\infty \sum_{i_2=1}^\infty \cdots
\sum_{i_N=1}^\infty \prod_{l=1}^N z^{i_l}
[\overline{i_l,i_{l-1},\ldots,i_1,i_N\ldots,i_{l+1}}]^{2\beta}
\label{eq:Mndiff}
\end{eqnarray}
For the sake of generality, 
a further parameter $\sigma$ can be introduced as per 
equation~(\ref{eq:zetawithxi}) at the end of the 
derivation of the induced zeta function.  
The sum over $N$ of these differences with a weighting $\frac{\sigma^N}{N}$ 
is then 
\begin{eqnarray}
\lefteqn{\sum_{N=1}^\infty \frac{\sigma^N}{N} 
\left( \tr {\M}^N-\tr {\M_{(1)}}^N \right) }
\nonumber \\
 & = & \sum_{N=1}^\infty \frac{\sigma^N}{N}
\sum_{i_1=1}^\infty \sum_{i_2=1}^\infty \cdots
\sum_{i_N=1}^\infty \prod_{l=1}^N z^{i_l}
[\overline{i_l,i_{l-1},\ldots,i_1,i_N\ldots,i_{l+1}}]^{2\beta}
\nonumber \\
 & = & \sum_{N=1}^\infty \frac{\sigma^N}{N}
\sum_{n=N}^\infty \sum_{ \{i_k\}_1^N; \sum_{k=1}^N i_k=n} 
\prod_{l=1}^N z^{i_l}
[\overline{i_l,i_{l-1},\ldots,i_1,i_N\ldots,i_{l+1}}]^{2\beta}
\nonumber \\
 & = & \sum_{n=1}^\infty z^n \sum_{N=1}^n
\sum_{ \{i_k\}_1^N; \sum_{k=1}^N i_k=n}  \frac{\sigma^N}{N}
\prod_{l=1}^N 
[\overline{i_l,i_{l-1},\ldots,i_1,i_N\ldots,i_{l+1}}]^{2\beta}
\nonumber \\
 & = & \sum_{n=1}^\infty z^n \sum_{m=1}^n 
\sum_{ \{i_k\}_1^m; \sum_{k=1}^m i_k=n} \frac{\sigma^m}{m} 
\prod_{l=1}^m 
[\overline{i_l,i_{l-1},\ldots,i_1,i_m\ldots,i_{l+1}}]^{2\beta}
\label{eq:sumMndiff}
\end{eqnarray}
where the same method of enumeration used in arriving at 
expression~(\ref{eq:fzetaind2}) for the induced $\zeta$-function has been
employed and the dummy indices have been relabelled to match up with
those of the previous section.  Note, that the order of the entries
in the continued fraction in the last line of equation~(\ref{eq:sumMndiff})
may be reversed.  The reasoning for this is that for 
every sequence $\{i_k\}_1^m$ that appears in the summation, 
so too does the reverse one $\{i_k\}_m^1$, trivially so if they are
the same.  Therefore, in reversing the order of the entries, 
the product term of every sequence is still summed over. 
\begin{eqnarray}
\lefteqn{ \sum_{N=1}^\infty \frac{\sigma^N}{N} 
\left( \tr {\M}^N-\tr {\M_{(1)}}^N \right)   }
\nonumber \\
 & = & \sum_{n=1}^\infty z^n \sum_{m=1}^n 
\sum_{ \{i_k\}_1^m; \sum_{k=1}^m i_k=n} \frac{\sigma^m}{m} 
\prod_{l=1}^m 
[\overline{i_l,i_{l+1},\ldots,i_m,i_1\ldots,i_{l-1}}]^{2\beta}
\label{eq:sumMndiff2}
\end{eqnarray}
So, for $\sigma=1$, this final expression is exactly the same as that
of the logarithm of the induced zeta function found in~(\ref{eq:fzetaind5}).
Therefore, the following relationship has been shown to be true:
\be
\zeta_{\mbox{\em ind}}(z,\beta)=\exp \sum_{n=1}^\infty \frac{1}{n} 
\left( \tr {\M}^n-\tr {\M_{(1)}}^n \right) 
\label{eq:zftopconn1}
\ee
A very important observation is that these sums of traces are 
Fredholm determinants since, by definition:  
\be
\det (1-\sigma \Mk) = \exp\left( -\sum_{n=1}^\infty 
\frac{\sigma^n}{n} \tr {\Mk}^n \right)
\label{eq:fredM}
\ee
Thus, there is the following theorem:
\begin{theorem}
The induced zeta function and the induced
transfer operators are related in the following way:
\bdm
\zeta_{\mbox{\em ind}}(z,\beta)=\frac{\det(1-\M_{(1)})}{\det(1-\M)}
=\frac{\det(1-\M_{(1)})}{\det(1-\M_{(0)})}
\label{eq:zftopconn2}
\edm
\label{theorem:zftopconn}
\end{theorem}
\Proof. \ \ 
The result follows immediately from equations~(\ref{eq:zftopconn1}) 
and~(\ref{eq:fredM}).
\hfill
$\Box$

As $\M$ and $\M_{(1)}$ are nuclear operators,
it is known that these determinants are entire functions
of $\sigma$ and are of order 0 for $|z| < 1$ and $\beta \in C$.
A stronger and more important 
result regarding the analyticity of the Fredholm determinants
of the $\Mk$ with respect to each of $\beta$ and $z$ 
within certain regions of $(z,\beta)$ space, will be shown in a 
later section.   
The analyticity properties of the zeta function will
then easily follow from relationship~(\ref{eq:zftopconn2}).

\section{Meromorphy of the map $(z,\beta) \rightarrow \M$}
\label{sec:merozbeta->M}

Recall that $D$ was defined as the set $\{ \xi \in C:|\xi-1|<\frac{3}{4}\}$.
\begin{theorem}
The map $(z,\beta) \rightarrow \M$ is a holomorphic function in $z$ 
for all $|z|<1$, $\beta\ \mbox{fixed} \in C$ and a holomorphic function in $\beta$
for all $\beta \in C$, $|z|\ \mbox{fixed} \leq 1$, 
$|z| \neq 1$.  ${\cal M}^{(1\beta)}$ is 
a meromorphic function of $\beta$ in the whole of the $\beta$ plane with
simple poles located at $\beta=\frac{1-k}{2}$, $k=0,1,2,\ldots$
possessing as residues the operators
$\Tk$ where 
\bdm
\Tk \circ \phi(\xi) = (-1)^k \phi^{(k)}(1)/2k!
\edm
$\M$ is a nuclear operator of order 0 for all $|z| \leq 1$, $\beta \in C$.
\label{theorem:meroM}
\end{theorem}

\Proof.
By lemma~(\ref{lemma:M nuke op}), $\M$ is a bounded operator for
$|z| < 1$ and $\beta \in C$. Corollary~(\ref{cor:M nuke op}) shows that it
is therefore a nuclear operator of order 0 for these values of $z$ and $\beta$.
Hence the first part of the thereom has already been shown. 
Recall that specifically lemma~(\ref{lemma:M nuke op}) showed that 
\be
\|\M \circ \phi(\xi)\|_{H_\infty(D)}
\leq \|\phi(\xi) \|_{H_\infty(D)}  \sup_{\xi \in D}
\sum_{n=1}^{\infty} |z|^n |(1+n\xi)^{-2\beta} |
\ee
Thus, for $|z|=1$,
the terms in the above sum  behave in the limit $n \rightarrow \infty$ like
$|(n\xi)^{-2\beta}|=|\xi^{-2\beta}n^{-2\beta}|=\mbox{constant} \times
|n^{-2\Re(\beta)}n^{-2i\Im(\beta)}|
=\mbox{constant} \times|n^{-2\Re(\beta)}\exp-(2i\Im(\beta)\log n)|
=\mbox{constant} \times|n^{-2\Re(\beta)}|$.  So this is a bounded sum 
providing $\Re(\beta) > \half$. 
The work of Mayer, see references~\cite{mayer:therm} and ~\cite{ergsymdyn},
on the thermodynamic formalism for
the Gauss map provides a method for examining the rest of the $\beta$-plane
when $z=1$ and a similar approach is followed here.

To investigate the analyticity of $\M$ with respect to $\beta$
for $|z| = 1$ firstly
consider any $\phi \in \HD$.  Trivially, $\phi$ is holomorphic in the disc
$D'=\{\xi:|\xi-1| \leq \frac{2}{3}\}$ since $D' \subset D$.
Hence, $\phi$ has a well defined taylor series around $\xi=1$ that is uniformly
convergent in $D'$.  $\phi$ may be broken up into two parts
as follows. Let $\phi=(\phi-\phi_N)+\phi_N$ where
\be
\phi_N(\xi)=\phi(\xi)-\sum_{k=0}^N \frac{\phi^{(k)}(1)}{k!} (\xi-1)^k
\label{phi_N}
\ee
Note that $\phi_N \in \HD$ since it differs from $\phi$ by a polynomial of
finite order $N$.  Each $\phi$ is be uniformly convergent in $D'$
and so $\phi_N$ satisfies (see p.? in~\cite{rudin:anal})
\be
|\phi_N(\xi)| \leq C|\xi-1|^{N+1}
\label{phi_Nbound}
\ee
for $\xi \in D'$.
 
$\phi_N$ may be thought of as a projection of $\phi$ onto $H^N_\infty(D)$:
this is the subspace of $\HD$ that contains all the functions that behave like
$(\xi-1)^M$ where $M$ is some integer greater than $N$, around $\xi=1$.  
The projection map will be denoted by 
$\Proj: \HD \rightarrow H^N_\infty(D)$ with the definition:
\be
\Proj  \circ \phi = \phi_N
\label{proj}
\ee
$\M \circ \phi$ may now be rewritten for $|z| \leq 1$ and $\beta > \half$
\be
\M \circ \phi = \M \circ (\phi - \phi_N) + \M \circ \phi_N
\ee
The first term is easily evaluated:
\begin{eqnarray}
\M \left( \sum_{k=0}^N \frac{\phi^{(k)}(1)}{k!} (\xi-1)^k \right) & = &
\sum_{k=0}^N (-1)^k \frac{\phi^{(k)}(1)}{k!}
\sum_{n=1}^{\infty} \frac{z^n}{(1+n\xi)^{2\beta}}
\left( \frac{\xi}{1+n\xi}\right)^k \\ \nonumber
 & = & \sum_{k=0}^N (-1)^k \frac{\phi^{(k)}(1)}{k!}
\frac{z}{\xi^{2\beta}} \Phi(z,2\beta+k,\frac{1}{\xi}+1)
\end{eqnarray}
where $\Phi(z,s,\nu)$ is the so-called Lerch 
transcendent~\cite{lerch} defined by
\be
\Phi(z,s,\nu)=\sum_{n=0}^\infty \frac{z^n}{(\nu + n)^s}
\ee
for $|z| \leq 1$ and $\nu > 1$.

\begin{figure}[tp!]
  \centering
  \includegraphics[width=0.5\textwidth]{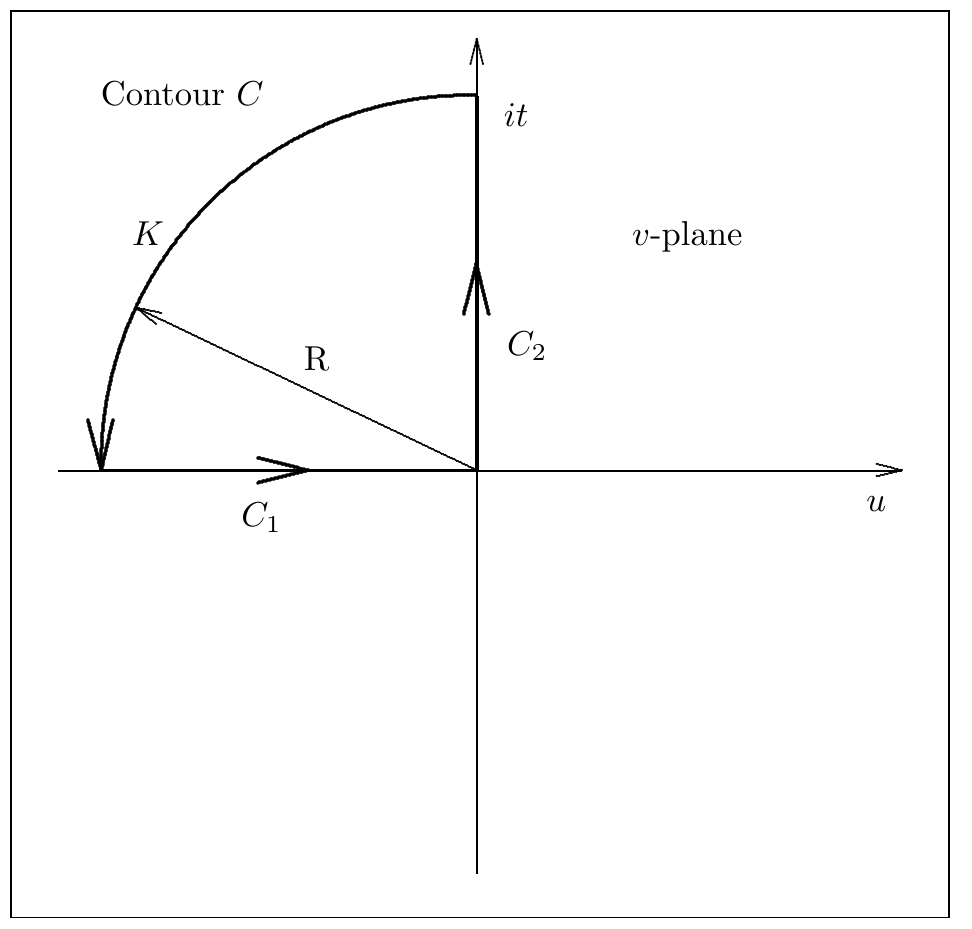}
  \caption{Contour for evaluation of integral}
  \label{fig:contour}
\end{figure}

This can be thought of as the action
of the operators $\Rk , k=0,1,\ldots,N$, acting on $\phi$ where
\be
\Rk \circ \phi(\xi) = \frac{ (-1)^k \phi^{(k)}(1)}{k!}
\frac{z}{\xi^{2\beta}} \Phi(z,2\beta+k,\frac{1}{\xi}+1)
\label{eq:Rop}
\ee 
$\M$ may now be decomposed in the following way:
\be
\M = \sum_{k=0}^N \Rk + \M \circ \Proj
\label{Mzbdecomp}
\ee
These pieces will be examined individually. 
The map $(z,\beta) \rightarrow \Rk$:
the $\Rk$'s are relatively simple operators in that they map $\HD$ to 
a one dimensional subspace of itself.  I.e., the $\Rk$'s are rank 1 operators,
map any $\phi(\xi)$ to a
multiple of the function $\frac{z}{\xi^{2\beta}} \Phi(z,2\beta+k,\frac{1}{\xi}+1)$ .
The singularity structure of the map $(z,\beta) \rightarrow \Rk$ 
will be determined by that of the
analytic continuation of $\Phi$ found in~\cite{lerch}.
For $|z|=1$ there are two possibilites:

\noindent
$\bullet$ Case 1: $z=e^{i\theta}, 0<\theta<2\pi$
\be
\Phi(e^{i\theta},s,\nu)=\frac{1}{2\nu^s}
+\int_0^\infty \frac{e^{it\theta}}{(\nu+t)^s} dt
-2\int_0^\infty  \sin[it\theta - s \tan^{-1}(t/\nu)]
\frac{dt}{(\nu^2+t^2)^{\frac{s}{2}}(e^{2\pi t}-1)}
\label{eq:Phicon}
\ee
for $\Re (\nu) > 0$. Note that $\nu$ corresponds to $\frac{1}{\xi}+1$.  
This is a M\"{o}ebius transformation of the domain $D$ and maps it to
the half plane $\{\xi:\Re(\xi)>3/2 \}$.  Hence $\Re(\frac{1}{\xi}+1)>0$.
The first term on the right hand side of equation~(\ref{eq:Phicon}) is 
clearly holomorphic in $\beta$ in the whole $\beta$-plane.  The third
is also well defined for all $s$: the integrand dies away exponentially and 
behaves like $\frac{i\theta-s/\nu}{2\pi\nu^s} + O(t)$ around $t=0$.
The second term is a possible source of `trouble'. Substituting $s$ and
$\nu$ for $2\beta+k$ and $\frac{1}{\xi}+1$ respectively, the integral becomes:
\be
\int_0^\infty \frac{e^{it \theta} dt}{{(\frac{1}{\xi}+1+t)}^{2\beta+k}}
\label{eq:intterm2}
\ee
Bearing a close resemblance to the Gamma function as it does, it is no
surprise that the above may be expressed in terms of the incomplete
Gamma function, $\Gamma(p,q)$, which is defined as
\be
\Gamma(p,q) \equiv \int_q^\infty e^{-t}t^{p-1}dt
\ee
and can be analytically continued to complex values of $p$ and $q$.
Indeed, for fixed $q \neq 0$, $\Gamma(p,q)$ is an entire 
function of $p$,(see~\cite{grad:int}), which will be the important factor
in determining the analyticity of this second term.
Firstly the integral must be manipulated to a more tractable form. 
Consider the complex variable $v=u+it$.  The integral can now be 
thought of as an integration of the function 
$f(v;\theta,2\beta+k,\xi)=
\frac{(i)^{2\beta+k-1}e^{v\theta}}{(v+i(\frac{1}{\xi}+1))^{2\beta+k}}$ 
along the positive imaginary axis beginning at $v=0$.  
Next consider the contour $C$ shown in figure~(\ref{fig:contour}).

For fixed $\xi$, the pole of $f(v;\theta,2\beta+k,\xi)$ occurs
at $v = -i(\frac{1}{\xi}+1)$.  These points are outside of the contour $C$ as
the set $\{\xi' \in C:\xi'=-i(\frac{1}{\xi}+1):\xi \in D\} 
\equiv \{\xi' \in C:\Im (\xi')<-3/2\}$.
Thus by Cauchy's theorem,  the integral around the contour is 0.  Also, the 
integral around $K$ vanishes in the limit $R \rightarrow 0$.
Putting this together,
\begin{eqnarray}
\lefteqn{\int_0^\infty \frac{e^{it \theta} dt}{{(\frac{1}{\xi}+1+t)}^{2\beta+k}}} 
\nonumber \\
 & = & \int_{C_2} f(v;\theta,2\beta+k,\xi) dv
\nonumber \\
 & = & -\int_{C_1} f(v;\theta,2\beta+k,\xi) dv
\nonumber \\
 & = & -(-i)^{2\beta+k-1}\int_{-\infty}^0 \frac{e^{u\theta}du}
{(u+i(\frac{1}{\xi}+1))^{2\beta+k}}
\nonumber \\
 & = & (-i)^{2\beta+k-1}\int_0^\infty\frac{e^{-u\theta}du}
{(u-i(\frac{1}{\xi}+1))^{2\beta+k}}
\nonumber \\
 & = & (-i\theta)^{2\beta+k-1} e^{-i(\frac{1}{\xi}+1)\theta} 
\Gamma(2\beta+k+1,-i(\frac{1}{\xi}+1)\theta)
\end{eqnarray}
where the final evaluation of the integral is taken from~\cite{grad:int}
subject to the conditions $|\arg(-i(\frac{1}{\xi}+1))|<\pi$ and $\Re(\theta)>0$.
The second criterion is trivially met while for the first,
$\Im (-i(\frac{1}{\xi}+1))<-3/2$ always, 
so this is also satisfied.  As mentioned previously, 
$\Gamma(2\beta+k+1,-i\nu\theta)$ is
entire in $2\beta+k+1$, hence in $\beta$, for all $-i\nu\theta \neq 0$.  
As both $\theta$ and $\nu$ are never 0, it has now been shown 
that the map $(z,\beta) \rightarrow
\Rk$ is entire in $\beta$ for all $z$ such that $|z| \leq 1, z \neq 1$.
Finally, the map $(z,\beta) \rightarrow \sum_{k=0}^N \Rk$ is also entire
in $\beta$ for the same values of $z$.

\noindent
$\bullet$ Case 2: $z=1$ 

Here, the Lerch transcendent reduces to a simpler
generalization of Riemann's zeta function, the Hurwitz zeta function~\cite{lerch}.
\be
\Phi(1,s,\nu)=\zeta(s,\nu)=\sum_{n=0}^\infty \frac{1}{(\nu + n)^s}
\ee
Hermite's representation of $\zeta(s,\nu)$ gives the analytic continuation
for $\Re(\nu) > 0$:
\be
\zeta(s,\nu)=\frac{1}{2\nu^s}
+ \frac{\nu^{1-s}}{s-1}
+ 2\int_0^\infty \frac{\sin[s \tan^{-1}(t/\nu)]}{(\nu^2+t^2)^{\frac{s}{2}}} 
\frac{dt}{e^{2\pi t}-1}
\label{eq:hurzetacon}
\ee
The above indicates that the $\Phi(1,2\beta+k,\frac{1}{\xi}+1)$ is a
meromorphic function of $\beta$ in the whole of the $\beta$-plane with
one simple pole at $\beta=(1-k)/2$ with residue 1.   The map
$(z,\beta) \rightarrow \Rk$ for $z=1$ is in turn a meromorphic function of
$\beta$ in the whole of the $\beta$-plane with a simple pole at $\beta=1$
with residue the operator $\Tk:\HD \rightarrow \HD$ given by
\be
\Tk \circ \phi(\xi) = \frac{(-1)^k \phi^{(k)}(1)}{2k!}
\ee
Note that $0 \equiv \Tk \circ \T_l \circ \phi$ for all $k \geq 1,l \geq 0$
$\Rightarrow$ i.e., the $\Tk$'s are nilpotent for $k \geq 1$.  In total, the map
$(z,\beta) \rightarrow \sum_{k=0}^\infty \Rk$ for $z=1$ is a meromorphic
function of $\beta$ in the whole of the $\beta$-plane with simple poles
at $\beta=(1-k)/2, k \in \{ 0,1,\ldots,N\}$ having residue the operators
$\Tk$ defined above.

Note that, trivially, the operator $\sum_{k=0}^N \Rk$ is nuclear of order 0
since it is of finite rank.  Of course, this only 
applies for $\beta$ away from the above points when $z=1$.

Now consider the map $(z,\beta) \rightarrow \M \circ \Proj$.
The reason for introducing the disc $D'$ is that $\M \circ \Phi(\xi)$
is a sum of weightings of $\Phi \circ G_n(\xi)$ and, as mentioned before,
the $G_n$ are contraction mappings with
$G_n(D) \subset D_0=\{\xi:|\xi-1| \leq \frac{2}{3} \} 
\subset D' \ \forall n$.
The bound on $\phi_N(\xi)$ given in 
equation~(\ref{phi_Nbound}) for $\xi \in D' \subset D_0$ 
leads to the result that $\M \circ \phi_N$ is bounded for $\beta>-N/2$
(The choice of $\frac{2}{3}$ is rather arbitrary, needing only to be 
between $\frac{4}{11}$ and $\frac{3}{4}$).  Recalling that this is for $|z|=1$, this
result may be demonstrated as follows:
\begin{eqnarray}
\left\| \M \circ \Proj \circ \phi(\xi) \right\|_{\HD} & = & 
\sup_{\xi \in \HD} \left| \sum_{n=1}^\infty
\frac{z^n}{(1+n\xi)^{2\beta}}\phi(1-\frac{\xi}{1+n\xi}) \right|
\nonumber \\
 & \leq & \sup_{\xi \in \HD} \sum_{n=1}^{\infty}
\left| \frac{z^n}{(1+n\xi)^{2\beta}} \right| 
C \left| \frac{-\xi}{1+n\xi} \right|^{N+1}
\nonumber \\
 & = & \sup_{\xi \in \HD}\frac{C}{|\xi^{2\beta}|} \sum_{n=1}^\infty 
\left| \frac{1}{1/\xi+n}\right|^{2\beta+N+1}
\end{eqnarray}
For fixed $\xi \in D$, the terms in the series behave like
$(n\xi)^{-2\beta-N-1}=\mbox{constant} \times n^{-2\beta-N-1}$ and
therefore converges providing $2\beta+N+1>1$, i.e. if $\beta>-N/2$.
Thus, the map $(z,\beta) \rightarrow \M \circ \Proj$ is a holomorphic
function of $\beta$ for all $|\beta| < N/2$ and for fixed $z$, $|z| = 1$.

Using the same arguments as lemma~(\ref{lemma:M nuke op}) and 
corollary~(\ref{cor:M nuke op}) showed for $\M$, since 
$\M: $ is a bounded operator, it is then a nuclear operator of order 0.
The composition of it with $\Proj$, another bounded operator acting on
a Banach space, gives that $\M \circ \Proj$ is a nuclear operator of 
order 0 by lemma~(\ref{lemma:nukecomp}).

Since all of the above is true for any $N$, $\M$ is a meromorphic function
of $\beta$ for $z=1$.  It has simple poles at $\beta=(1-k)/2$ with residues
the operators $\T$ defined above.  For all other $z$ such that $|z| \leq 1$,
$\M$ is entire in $\beta$.  As was shown in the first part of the proof,
$\M$ is also a holomorphic function of $z$ for all $\beta \in C$ and $z$, $|z|<1$.
The preceding has also demonstrated that $\M$ is a nuclear operator away from 
these specified singular points. This completes the proof of 
theorem~(\ref{theorem:meroM}).
\hfill
$\Box$ 

It appears feasible that the above theorem may be improved to cater for
an analytic extension of the map $(z,\beta) \rightarrow \M$ to the
whole of $(z,\beta)$ space with a cut along the positive $z$ axis
beginning at $z = 1$.  The cut would arise from the presence of such a cut in the
analytic continuation of the Lerch transcendent.  

All of the above applies to the generalized induced transfer operator since
this is basically the induced transfer operator with a linear $\beta$ shift.
\begin{corollary}
The map $(z,\beta) \rightarrow \Mk$ is a holomorphic function in $z$ 
for all $|z|<1$, $\beta\ \mbox{fixed} \in C$ and a 
holomorphic function in $\beta$
for all $\beta \in C$, $|z|\ \mbox{fixed} \leq 1$, $|z| \neq 1$.  
${\cal M}_{(k)}^{(1\beta)}$ is 
a meromorphic function of $\beta$ in the whole of the $\beta$ plane with
simple poles located at $\beta=\frac{1-k-j}{2}$, $j=0,1,2,\ldots$
possessing as residues the operators
$\Tk$ where 
\bdm
\Tk \circ \phi(\xi) = (-1)^{k+j} \phi^{(k+j)}(1)/2(k+j)!
\edm
$\Mk$ is a nuclear operator of order 0 for all $|z| \leq 1$,
$\beta \in C$.
\label{cor:meroM}
\end{corollary}

\section{Meromorphy of the Trace of $\M$}
\label{sec:trace}
An explicit expression for the trace of $\M$ was determined in
section~(\ref{sec:traceM}).  This result, equation~(\ref{eq:trM}), 
is reproduced here: 
\be
\tr \M = \sum_{n=1}^\infty \tr \M_n
= \sum_{n=1}^\infty 
\frac{z^n([\overline{n}])^{-2\beta}}{1+([\overline{n}])^2} 
\label{eq:trM1}
\ee
To uncover the behaviour of the analytic continuation of this
function will require recourse to the techniques of the 
previous section. 
\begin{theorem}
The function $(z,\beta) \rightarrow \tr \M$ is,
in its analytic continuation, a holomorphic
function of $z$ for $|z| < 1$ and for fixed $\beta \in C$ and 
is also a holomorphic function of $\beta$ for $\beta \in C$
and for fixed $z$, $|z| \leq 1$, $z \neq 1$.  It is a meromorphic
function of $\beta$ when $z=1$ and has one simple pole
at $beta=\half$ with residue $\half$.
\label{theorem:tracemero}
\end{theorem}

\Proof \ \
Using the decomposition of $\M$ from the previous section, the trace 
of $\M$ may be written as follows:
\be
\tr \M =\sum_{k=0}^n \tr \Rk + \tr \M \circ \Proj
\label{eq:trdecomp}
\ee
This holds for $|z| \leq 1$ and $\beta>\half$ since the
expression for these operators are well defined there.

The traces of the $\Rk$ are simple to compute since these operators
map $\HD$ to a one dimensional subspace of of itself.
Therefore, each $\Rk$ has precisely one eigenvalue.  By inspection of 
equation~(\ref{eq:Rop}), the definition of $\Rk$, the 
corresponding eigenfunction to the lone eigenvalue is given by
\be
\phi_k(\xi) =
\xi^{-2\beta} \Phi(z,2\beta+k,\frac{1}{\xi}+1)
\label{eq:Ropeigf}
\ee
The eigenvalue is then seen to be 
\be
\lambda_k = \frac{(-1)^k z}{k!}  \phi_k^{(k)}(1)= \frac{(-1)^k z}{k!} 
\frac{d^k}{d\xi^k} \left.
\xi^{-2\beta} \Phi(z,2\beta+k,\frac{1}{\xi}+1) \right|_{\xi=1}
\label{eq:Ropeig1}
\ee
This is not a straightforward calculation being complicated
by the factor $\xi^{-2\beta}$ and the unhelpfulness of the
argument $(\frac{1}{\xi}+1)$ in the Lerch transcendent function.
However, it will be possible to demonstrate the alleged 
analyticity properties of the trace without a completely
explicit formula for the eigenvalues.  However, an attempt 
to present a more detailed calculation will be reserved
for the ``Miscellaneous'' section of the appendices, 
see section~(\ref{Asec:tracedetail}).

The product rule, $\frac{d}{ds} u(s)v(s)=u'(s)v(s)+u(s)v'(s)$,
generalises easily (in analogy to a binomial expansion or Pascal's
triangle) to the following result :
\be
\frac{d^k}{ds^k} u(s)v(s) =
\sum_{l=0}^k {k \choose l}
\frac{d^{(k-l)}}{ds^{(k-l)}}u(s) \frac{d^l}{ds^l}v(s)
\label{eq:prodrule}
\ee
Using this in the expression for $\lambda_k$, 
equation~(\ref{eq:Ropeig1}) becomes
\begin{eqnarray}
\lambda_k & = & \frac{(-1)^k z}{k!}  \frac{d^k}{d\xi^k} \left.
\xi^{-2\beta} \Phi(z,2\beta+k,\frac{1}{\xi}+1) \right|_{\xi=1}
\nonumber \\
 & = & \frac{(-1)^k z}{k!}  \left.
\sum_{l=0}^k {k \choose l}
\frac{d^{(k-l)}}{d\xi^{(k-l)}} \xi^{-2\beta} 
\frac{d^l}{d\xi^l} \Phi(z,2\beta+k,\frac{1}{\xi}+1) 
\right|_{\xi=1}
\label{eq:Ropeig2}
\end{eqnarray}
The first part of the product in the summand may be evaluated:
\begin{eqnarray}
\frac{d^{(k-l)}}{d\xi^{(k-l)}} \xi^{-2\beta}
 & = & \left. (-1)^{(k-l)} 2\beta(2\beta+1)(2\beta+2)\ldots
(2\beta+k-l-1)\,\xi^{-2\beta-k+l} \right|_{\xi=1}
\nonumber \\
 & = & (-1)^{k-l} 2\beta(2\beta+1)(2\beta+2)\ldots
(2\beta+k-l-1)
\label{eq:diff1}
\end{eqnarray}
Also, an expression for the second part of the product is based on 
the differentiation of the Lerch transcendent
which follows simply from its definition:
\be
\frac{d}{d\nu} \Phi(z,s,\nu)
= \frac{d}{d\nu} \sum_{n=0}^\infty \frac{z^n}{(n+\nu)^s}
= -s \sum_{n=0}^\infty \frac{z^n}{(n+\nu)^(s+1)}
= -s \Phi(z,s+1,\nu)
\label{eq:lerchdiff}
\ee
Recall from the previous section, $\Phi(z,s,\nu)$ was found to
have the following properties:
\begin{itemize}
\item $\Phi(z,s,\nu)$ is holomorphic in $z$ for $|z|<1$ and $\beta \in C$;
$\Re(\nu)>0$.
\item $\Phi(z,s,\nu)$ is holomorphic in $\beta$ for $\beta \in C$ and
$|z| \leq 1$, $z \neq 1$; $\Re(\nu)>0$.
\item $\Phi(z,s,\nu)$ is meromorphic in $\beta$ for $\beta \in C$ and 
$z=1$ and has a simple pole at $s=1$ with residue 1; $\Re(\nu)>0$.
\end{itemize}
Notice that the derivative of the Lerch transcendent in 
equation~(\ref{eq:diff1}) is therefore a holomorphic function of $s$ for
$z=1$ since $\Phi(1,s+1,\nu)$ has a simple pole at $s=0$ which is removed 
by the multiplying factor of $s$. In summary, $\frac{d}{d\nu}\Phi(z,s,\nu)$ is 
\begin{itemize}
\item holomorphic in $z$ for $|z|<1$ and $\beta \in C$; $\Re(\nu)>0$.
\item holomorphic in $\beta$ for $\beta \in C$ and
$|z| \leq 1$; $\Re(\nu)>0$.
\end{itemize}
With this observation, it is useful to rewrite the expression, \ref{eq:Ropeig2}
for $\lambda_k$ for $k>0$ as follows
\begin{eqnarray}
\lambda_k & = & \frac{(-1)^k z}{k!} \left. \left( \frac{d^k}{d\xi^k} 
\xi^{-2\beta} \right) \Phi(z,2\beta+k,\frac{1}{\xi}+1) \right|_{\xi=1}
\nonumber \\
 &  & + \frac{(-1)^k z}{k!}  \left.
\sum_{l=1}^k {k \choose l}
\frac{d^{(k-l)}}{d\xi^{(k-l)}} \xi^{-2\beta} 
\frac{d^l}{d\xi^l} \Phi(z,2\beta+k,\frac{1}{\xi}+1) 
\right|_{\xi=1}
\label{eq:Ropeig3}
\end{eqnarray}
Note that $k=0$ is not included in this part of the discussion.  It cannot be
broken up into two non-zero pieces as above.  The second term on the right
of equation~(\ref{eq:Ropeig3}) is holomorphic in $\beta$ and $z$ separately since
it contains only differentiations of the Lerch transcendent.  A more detailed
analysis of this is provided in the appendices under the heading of 
``Miscellaneous''. The first term contains an `untouched' Lerch transcendent 
and so is worth more inspection.
Using equation~(\ref{eq:diff1}) for $l=0$, the first term evaluated at $\xi=1$ may
be written as
\be
\frac{(-1)^k z}{k!} (-1)^{k-l} 2\beta(2\beta+1)(2\beta+2) \ldots (2\beta+k-1)
\Phi(z,2\beta+k,2) 
\label{eq:Ropeigt1}
\ee
Once again this is a holomorphic function of $\beta$ for $z=1$ since the
$(2\beta+k-1)$ factor cancels the simple pole of $\Phi(1,2\beta+k,2)$ at
$\beta=(1-k)/2$.  Note that this is true only for $k>0$ as no such factor appears
for the case $k=0$ because the $\xi^{-\beta}$ term is not differentiated.  
Thus, for the mapping $(z,\beta) \rightarrow \tr \Rk$ the following has been shown
for $k>0$:
\begin{itemize}
\item The function $(z,\beta) \rightarrow \tr \Rk$ is holomorphic 
in $z$ for $|z|<1$ and $\beta \in C$.
\item The function $(z,\beta) \rightarrow \tr \Rk$ is holomorphic
in $\beta$ for $\beta \in C$ and $|z| \leq 1$.
\end{itemize}
For the case $k=0$: the expression for $\tr \R_0 = \lambda_0$ is given by
inserting $k=0$ into equation~(\ref{eq:Ropeig1}):
\be
\tr \R_0 = \frac{z}{\xi^{-2\beta}} \left.
\Phi(z,2\beta+k,\frac{1}{\xi}+1) \right|_{\xi=1}
= z\Phi(z,2\beta,2)
\label{eq:Roptr0}
\ee
Here, there are no saving graces in the form of nice factors appearing to
cancel the pole of the Lerch transcendent.  Recall that for $z=1$, 
$\Phi(1,s,\nu)$ is just the Hurwitz $\zeta$-function.  
But for $\nu=2$, the Hurwitz $\zeta$-function itself reduces to the
Riemann $\zeta$-function, or, more precisely, $\zeta_R(s)-1$.  
The meromorphic qualities are still the same as $\zeta(s)$ extends
to the entire complex $s$-plane with one simple pole at $s=1$ of
residue 1, see~\cite{lerch}.  Therefore
\be
\tr \R_0 = \zeta_R(2\beta)-1
\label{eq:Roptr0a}
\ee
and it now follows that 
\begin{itemize}
\item The function $(z,\beta) \rightarrow \tr \R_0$ is holomorphic 
in $z$ for $|z|<1$ and $\beta \in C$.
\item The function $(z,\beta) \rightarrow \tr \R_0$ is holomorphic
in $\beta$ for $\beta \in C$ and $|z| \leq 1$, $z \neq 1$.
\item The function $(z,\beta) \rightarrow \tr \R_0$ is meromorphic
in $\beta$ for $\beta \in C$ and $z=1$.  It has a 
simple pole at $\beta=\half$ with residue $\half$.
\end{itemize}

The trace of $\M$ composed with the projection operator $\Proj$ is now
examined.  The projection operator has already been shown to be a 
nuclear operator as well as a holomorphic function of $\beta$ for 
$\Re \beta > -N/2$.   As this analysis is for arbitrary $N$ and, it must
be true for for all $N$ and hence for all $\beta \in C$.
Therefore, by theorem~(\ref{theorem:nuke trace}),
its trace is well defined and has no singular behaviour.  This trace is thus 
holomorphic for $\beta$ in the whole of the 
$\beta$-plane for fixed $z \leq 1$ and holomorphic in $z$ for $|z|<1$
and $\beta \in C$.
Since, adding a finite number of holomorphic (meromorphic) functions
together gives another holomorphic (meromorphic) function, the
properties of the trace of $\M$ follow readily from those of its
decomposition. 

In total, the trace of the $\M$ can be categorized as follows:
\begin{itemize}
\item The function $(z,\beta) \rightarrow \tr \M$ is holomorphic 
in $z$ for $|z|<1$ and $\beta \in C$.
\item The function $(z,\beta) \rightarrow \tr \M$ is holomorphic
in $\beta$ for $\beta \in C$ and $|z| \leq 1$, $z \neq 1$.
\item The function $(z,\beta) \rightarrow \tr \M$ is meromorphic
in $\beta$ for $\beta \in C$ and $z=1$.  It has a 
simple pole at $\beta=\half$ with residue $\half$.
\end{itemize}
This completes the proof. 
\hfill
$\Box$

As before there is the simple corollary regarding the generalized induced
transfer operator:
\begin{corollary}
The function $(z,\beta) \rightarrow \tr \Mk$ is,
in its analytic continuation, a holomorphic
function of $z$ for $|z| < 1$ and for fixed $\beta \in C$ and 
is also a holomorphic function of $\beta$ for $\beta \in C$
and for fixed $z$, $|z| \leq 1$, $z \neq 1$.  It is a meromorphic
function of $\beta$ when $z=1$ and has one simple pole
at $beta=\half-k$ with residue $\half$.
\label{cor:tracemero}
\end{corollary}

\section{Analyticity Properties of the Determinant of $(1-\sigma\M)$}
\label{determinant}

The following theorem provides the information about the 
analyticity of the map from $(z,\beta)$-plane to the
Fredholm determinant $\det(1-\M)$.
\begin{theorem}
The function $(z,\beta) \rightarrow \det(1-\M)$ is, in its
analytic continuation, a holomorphic function of $z$ for
$|z|<1$ and for $\beta \in C$ and it is a holomorphic
function of $\beta$ for $\beta \in C$ and for 
$|z| \leq 1$, $z \neq 1$.
It is a meromorphic function of $\beta$ in the whole $\beta$-plane
when $z=1$ with simple poles at $\beta_k=(1-k)/2, \ k=0,1,2,\ldots$.
\label{theorem:detmero}
\end{theorem}
\Proof. \ \
Theorem~(\ref{theorem:nuke trace}) due to Grothendieck shows that 
wherever $\M$ is holomorphic in either $\beta$ or $z$, the function
$\det(1-\M)$ is also holomorphic.  It is clear that where $\M$ is 
singular, $\det(1-\M)$ will also be singular.  Therefore, this
determinant will have the same structure as $\M$ which is demonstrated
in theorem~(\ref{theorem:meroM}) and the proof is finished.
\hfill $\Box$

Once again, the same argument applies for the generalized induced
zeta function:
\begin{corollary}
The function $(z,\beta) \rightarrow \det(1-\Mk)$ is, in its
analytic continuation, a holomorphic function of $z$ for
$|z|<1$ and for $\beta \in C$ and it is a holomorphic
function of $\beta$ for $\beta \in C$ and for 
$|z| \leq 1$, $z \neq 1$.
It is a meromorphic function of $\beta$ in the whole $\beta$-plane
when $z=1$ with simple poles at $\beta_j=(1-j-k)/2, \ j=0,1,2,\ldots$.
\label{cor:detmero}
\end{corollary}

Finally, there is the following theorem regarding the analtyicity
properties of the induced zeta function.
\begin{theorem}
The induced zeta function for the Farey map is a meromorphic 
function of $\beta$ for all $\beta \in C$ and for fixed $z$ such that $|z| \leq 1$.
It is also a meromorphic function of $z$ for all $z$ such that $|z|<1$
and for fixed $\beta \in C$.
\label{theorem:zetamero}
\end{theorem}
\Proof. \ \ 
The theorem follows directly from theorem~(\ref{theorem:zftopconn})
with theorem~(\ref{theorem:detmero}) and corollary~(\ref{cor:detmero}).
\hfill
$\Box$

\begin{corollary}
The zeta function for the Farey map is  a meromorphic 
function of $\beta$ for all $\beta \in C$ and for fixed $z$ such that $|z| \leq 1$
and $z \neq 1$.
It is also a meromorphic function of $z$ for all $z$ such that $|z|<1$
and for fixed $\beta \in C$.
\label{cor:zetamero}
\end{corollary}
\Proof. \ \
This follows immediately from theorem~(\ref{theorem:detmero}) using the relation 
\bdm
\zeta(z,\beta)=\frac{\zeta_{\mbox{ind}}(z,\beta)}{1-z}
\edm
\hfill
$\Box$

\section{Concluding Remarks}
\label{conclusion1}
This chapter has provided strong results on the overall singularity structure
of the analytic continuation of the induced zeta function for the Farey map.
A very strong link between the Fredholm determinant of a simple
generalization of the induced transfer operator and the induced zeta function
has been demonstrated.  This elegant connection reinforces the observation that
the two techniques provide the same thermodynamics for the Farey system.
Indeed, it appears reasonable that this connection would apply in more
general settings, just as it has been shown to do for normal hyperbolic systems
by Ruelle in~\cite{ruelle:anosov}.

It would be interesting to see if the work could be extended to values of $z$ 
outside the unit disc.  However, it is not entirely useful to do this as the
connection between the induced transfer operator and the original transfer
operator, theorem~(\ref{theorem:topconn}), applies only for values of $|z|$
less than 1.

\chapter{Presentation and Analysis of the Induced Zeta Function of the Farey map}
\label{chap:farey}

\section{The Induced Zeta Function for the Farey map}
\label{sec:indzf}

The form for the induced zeta function was determined in the
previous chapter and is repeated here:
\begin{eqnarray}
\log \zeta_{\mbox{ind}}(z,\beta) & = &
\sum_{n=1}^{\infty} \frac{1}{m}\sum_{g^m(x)=x} 
\exp \sum_{k=0}^{m-1} \phi_z (g^k(x))
\nonumber \\
 & = & \sum_{m=1}^{\infty} \frac{1}{m}\sum_{g^m(x)=x} 
\prod_{l=0}^{m-1} z^{n(g^l(x))} |g'(g^k(x))|^{-\beta}
\label{eq:zetaind}
\end{eqnarray}
The main result of this first section is the following theorem regarding the
presentation of an explicit expression for the power series expansion of 
$\log \zeta_{\mbox{ind}}(z,\beta)$.
\begin{theorem}
The power series expansion about $z=0$ of the logarithm of the induced zeta function
of the Farey map is given by
\bdm
\log \zeta_{\mbox{ind}}(z,\beta)= \sum_{n=1}^\infty
 z^n \sum_{m=1}^{n} \frac{1}{m} 
\sum_{ \{i_k\}_1^m; \sum_{k=1}^m i_k=n} 
\prod_{l=1}^{m}
\left| [\overline{i_l,i_{l+1},\ldots,i_m,i_1,\ldots,i_{l-1}}] \right|^{2\beta}
\edm
where $[\overline{i_1,i_2,\ldots,i_m}]$ is the 
periodic continued fraction\footnote{A general outline of continued fractions
and some relevant results are presented in the appendices.  It is suggested that
the reader refer to this section for clarification on notations and definitions. }
\bdm
\cfrac{1}{i_1+\cfrac{1}{i_2+\cfrac{1}{\cdots+
\cfrac{1}{i_m+\cfrac{1}{i_1+\cfrac{1}{1+\cdots}}}}}}
\edm
\label{theorem:zetaind}
\label{theorem:fzetaind5}
\end{theorem}
\Proof \ \
For the Farey map, recall that the induced map $g:[1/2,1) \rightarrow
[1/2,1]$ was given by
\be
g(x) = \left\{ \begin{array}{lll} 
g_n(x)  \  & \forall \ x \in (\frac{n}{n+1},\frac{n+1}{n+2}], 
&  \ \ n=1,2,\ldots  \\1  & \mbox{for} \  x=\half & 
\end{array} \right.
\label{eq:finddef}
\ee
where $g_n$ were defined by
\be
g_n(x)=\frac{1-x}{1-n(1-x)}
\label{eq:gndef}
\ee
The inverse of $G$ cut down on the interval $[\half,1)$ is the 
countable union of the inverse branches $G_n$.
I.e., $G \equiv \bigcup_{n=1}^\infty G_n \equiv \bigcup_{n=1}^\infty 
g_n^{-1}: [\frac{1}{2},1)\rightarrow
(\frac{n}{n+1},\frac{n+1}{n+2}]$ was found to be
\be
G_n(x)=1-\frac{x}{1+nx}
\label{eq:Gndef}
\ee
The end point 1 has only one inverse:$G(1)=\half$.
The first task is to determine the position of the fixed points of $g^m$.
One way of doing this is to consider the inverse problem of $G^n(x)=x$.
To begin with, consider a branch of the inverse map, $G_n$.
$G_n$ may be rewritten as a kind of operator on continued fractions; this
new form will also help to expose the fixed points of the induced map.
\be
G_n(x)=\cfrac{1}{1+\cfrac{1}{n-1+\cfrac{1}{x}}}
\label{eq:Gncontfrac}
\ee
The equation $G_n(x)=x$ has two solutions as this is really just 
a quadratic equation:
\be
x=1-\frac{x}{1+nx} \Rightarrow x+nx^2=1+nx-x \Rightarrow
nx^2+(2-n)x-1=0
\ee
The solutions can be expressed as:
\be
x_{n\pm}=\frac{n-2 \pm \sqrt{(n-2)^2+4n}}{2n}
=\frac{n-2\pm \sqrt{n^2+4}}{2n}
\label{eq:fpgen}
\ee
Notice that the solutions are always real and that $x_{n+}>0$
and $x_{n-}<0$.  Thus, in the domain of functions of relevance,
namely $J=[\half,1]$ and more generally $D$, the $G_n$ have
only one real fixed point at $x_n^\ast=\frac{n-2 + \sqrt{n^2+4}}{2n}$.
Since $n^2+4$ is never a perfect square, the fixed points
$g_{n\pm}$ are `quadratic surds' --- a quadratic surd is a solution
to a quadratic equation $ax^2+bx+c=0$ such that $b^2-4ac$ is
not a perfect square.
Moreover, there is the following theorem
\begin{theorem} 
Any periodic continued fraction is a quadratic surd and, conversely,
every quadratic surd has a periodic continued fraction expansion.  I.e.,
there is a 1-1 correspondence between the quadratic surds and periodic
continued fractions.
\label{theorem:pcf-qsurds}
\end{theorem} 
\Proof.
See~\cite{rockett:contfracs}.
\hfill 
$\Box$
  
Whether by direct manipulation of the expression for $x_n^\ast$ given
in equation~(\ref{eq:fpgen}) or
more simply by inspection of equation~(\ref{eq:Gncontfrac}), it follows that
$x_n^\ast$ can also be written as:
\be
x_n^\ast = [1,\overline{n}]=
\cfrac{1}{1+\cfrac{1}{n+\cfrac{1}{n+\cfrac{1}{n+\cdots}}}} 
\ee
which is trivially a periodic continued fraction having period 1.

The above may be generalised to the function 
$G^m:[\half,1]\rightarrow[\half,1]$.  Whereas $G$ was made up of the
pieces $G_n$, $G^m$ is made up of the composition maps
$G_{\{i_k\}_{k=1}^m}=G_{i_1} \circ G_{i_2} \circ \cdots \circ G_{i_m}$.  
\be
G_{\{i_k\}_{k=1}^m}(x)=
\cfrac{1}{1+\cfrac{1}{i_1+\cfrac{1}{i_2+\cfrac{1}{\cdots+\cfrac{1}{i_m-1+\cfrac{1}{x}}}}}}
\label{eq:compmap}
\ee
Any continued fraction can be expressed in terms of its $\mbox{k}^{\mbox{th}}$
complete quotient $\mu_k$, (see appendix).  Using relationship~(\ref{Aeq:cfquot})
\be
G_{\{i_k\}_{k=1}^m}(x)=
\frac{A_{m+1}\mu_{m+2}+A_m}{B_{m+1}\mu_{m+2}+B_m}
=\frac{A_{m+1}x+A_m}{B_{m+1}x+B_m}
\ee
Hence, the equation $G_{\{i_k\}_{k=1}^m}(x^\ast)=x^\ast$ reduces to finding the
solutions to a quadratic equation:
\be
p(x)=B_{m+1}x^2+(B_m-A_{m+1})x-A_m=0
\ee
Notice that $p(0)=-A_m<0$. Further, $p(-1)=(B_{m+1}-B_m) +(A_{m+1}-A_m)>0$ since
the sequence $\{A_k\ + B_k\}$ has positive terms which are strictly increasing
for $k \geq 1$, (see appendix, equations~(\ref{Aeq:cfrec2})).  One solution must
therefore be real and lie between -1 and 0, outside of $D$.
The other solution is real and positive and lies between 0 and 1.  This may be seen on
inspection of the fixed point which is 
the periodic continued fraction easily identified via equation~(\ref{eq:compmap}):
\label{compmapguff}
\be
x_{I_m}^\ast=[1,\overline{i_1,i_2,\ldots,i_m}]
\label{eq:compmapfp}
\ee
Once again, these maps have only one fixed point in the region of interest, $D$.
Thus, the fixed points for $G^m$, and therefore of $g^m$ have been found.
The derivative of $G_n$ is given by
\be
G_n'(x)=\frac{-1}{(1+nx)^2}
\label{eq:Gn'def}
\ee 
Clearly, for $x \in [\half,1]$, $|G_n'(x)| < 1$ always and the same
is true for $|G_n'(\xi)|$ for $\xi \in D$.
Also of note at this point is that a lemma of Ruelle~\cite{ruelle:anosov}
indicates that $G_{\{i_k\}_{k=1}^m}$
has exactly one fixed point in $D$ and the modulus of its derivative at 
that point is less than 1 \label{uniquefp}.  

Returning to the form of the induced zeta function, the
exact values of the fixed point may now
be substituted into equation~(\ref{eq:zetaind})
\be
\log \zeta_{\mbox{ind}}(z)=
\sum_{m=1}^{\infty} \frac{1}{m}
\sum_{i_1=1}^\infty \sum_{i_2=1}^\infty \cdots \sum_{i_m=1}^\infty
\prod_{l=0}^{m-1} z^{n(g^l([1,\overline{i_1,i_2,\ldots,i_m}]))}
|g'(g^l([1,\overline{i_1,i_2,\ldots,i_m}]))|^{-\beta}
\label{eq:fzetaind1}
\ee

One useful way of enumerating these composition maps
is to consider `ordered integer partitions' of 
the numbers $\sum_{k=1}^{k=m} i_k = m,m+1,m+2,\ldots$.  Note that
ordering of the partition is important.  
For example, in the case of $m=3$, the sequences
$\{2,1,1\}$, $\{1,2,1\}$ and $\{1,1,2\}$ each represent distinct branches of the
mapping $g^3$, namely $g_2g_1g_1$, $g_1g_2g_1$
and $g_1g_1g_2$ respectively.  Evidently, the fixed points of 
$g^n$ may be enumerated in the same way.  This type of partition should not be 
confused with the `classical' partition of integers worked on by people
such as Hardy and Ramanujan (see~\cite{andrews:part}), 
where order is not important.
\be
\log \zeta_{\mbox{ind}}(z)=
\sum_{m=1}^{\infty} \frac{1}{m} \sum_{n=m}^\infty
\sum_{ \{i_k\}_1^m; \sum_{k=1}^m i_k=n} 
\prod_{l=0}^{m-1} z^{n(g^l([1,\overline{i_1,i_2,\ldots,i_m}]))}
|g'(g^l([1,\overline{i_1,i_2,\ldots,i_m}]))|^{-\beta}
\label{eq:fzetaind2}
\ee
where the sum over $g^m(x)=x$ has been replaced by a sum over ordered integer
partitions of $n$ with $n$ running from $m$ to $\infty$.  The product on the
righthand side of~(\ref{eq:fzetaind2}) may still be considerably simplified.
Let
\be
x^\ast=[1,\overline{i_1,i_2,\ldots,i_m}]=\cfrac{1}{1+\cfrac{1}{i_1+f}}
\ee
where $0<f = [\overline{i_2,i_3,\ldots,i_m,i_1}]<1$.
Using $x^\ast=\frac{i_1+f}{i_1+1+f}$, the following bounds can be verified:
\be
\frac{i_1+f}{i_1+1+f}-\frac{i_1}{i_1+1}=
1-\frac{1}{i_1+1+f}-1+\frac{1}{i_1+1}=
\frac{f}{(i_1+1)(i_1+1+f)}>0
\ee
Similarly:
\be
\frac{i_1+f}{i_1+1+f}-\frac{i_1+1}{i_1+2}=
\frac{f-1}{(i_1+2)(i_1+1+f)}<0
\ee
This means that $\frac{i_1}{i_1+1} < x^\ast < \frac{i_1+1}{i_1+2}$ and
therefore $g(x^\ast)=g_{i_1}(x^\ast)$.  Also, $n(x)=n$ 
for $x \in (\frac{n}{n+1},\frac{n+1}{n+2}]$ so $n(x^\ast)=i_1$.  
The action of $g$ on each of the fixed points is just to remove the
first integer of the periodic part of the continued fraction expansion.
This can be seen by noting that $g$ `undoes' whatever the inverse does to
a point in equation~(\ref{eq:Gncontfrac}) 
or simply via the continued fraction representation of $g_n$:
\be
g_n(x)=\cfrac{1}{1-n+\cfrac{1}{-1+\cfrac{1}{x}}}
\label{eq:gncontfrac}
\ee
All this translates to the fact that
\be
g^l(x^\ast)=[1,\overline{i_{l+1},i_{l+2},\ldots,i_m,i_1,\ldots,i_l}]
\ee
and
\be
n(g^l(x^\ast))=i_{l+1}
\ee

The expression~(\ref{eq:fzetaind2}) for the induced zeta function,
may now be rewritten as
\be
\log \zeta_{\mbox{ind}}(z)=
\sum_{m=1}^{\infty} \frac{1}{m} \sum_{n=m}^\infty
\sum_{ \{i_k\}_1^m; \sum_{k=1}^m i_k=n} 
\prod_{l=0}^{m-1} z^{i_{l+1}}
|g_{l+1}'([1,\overline{i_{l+1},i_{l+2},\ldots,i_m,i_1,\ldots,i_l}])|^{-\beta}
\label{eq:fzetaind3}
\ee
This representation also simplifies the exponent of the product of $z^{i_{l+1}}$'s:
\be
\prod_{l=0}^{m-1} z^{i_{l+1}}
=z^{\sum_{l=0}^{m-1} i_{l+1}}=z^{\sum_{l=1}^{m} i_{l}}=z^n
\label{eq:zpower}
\ee
The term $\prod_{l=0}^{m-1}
|g_{l+1}'([1,\overline{i_{l+1},i_{l+2},\ldots,i_m,i_1,\ldots,i_l}])|^{-\beta}$
is now examined in more detail.  Firstly, from definition~(\ref{eq:gndef}),
$g_n': (\frac{n}{n+1},\frac{n+1}{n+2}] \rightarrow [\half,1)$ is given by
\be
g_n'(x)=\frac{-1}{(1-n+nx)^2}
\label{eq:gn'def}
\ee
Thus, the product can be written with $l \rightarrow l-1$
\begin{eqnarray}
\lefteqn{\prod_{l=1}^{m}
|g_{i_l}'([1,\overline{i_l,i_{l+1},\ldots,i_m,i_1,\ldots,i_{l-1}}])|^{-\beta}} 
\nonumber \\
 & = & \prod_{l=1}^{m}
(1-i_l+i_l\cdot[1,\overline{i_l,i_{l+1},\ldots,i_m,i_1,\ldots,i_{l-1}}])^{2\beta}
\label{eq:zetaguts}
\end{eqnarray}
The above is further simplified via the following result.
\begin{lemma}
\label{lemma:product}
The product 
\be
\prod_{l=1}^{m}
(1-i_l+i_l\cdot[1,\overline{i_l,i_{l+1},\ldots,i_m,i_1,\ldots,i_{l-1}}])
\nonumber
\ee
reduces to
\be
\prod_{l=1}^{m}
[\overline{i_l,i_{l+1},\ldots,i_m,i_1,\ldots,i_{l-1}}]
\nonumber
\ee
\end{lemma}
\Proof.  The $\mbox{l}^{\mbox{th}}$ complete quotients $\mu_l$ of 
the fixed points provide a nice way of showing this fact; 
(Note that $\mu_l=\mu_{l+jm}$ for $l=2,3,\ldots$ and
$j=0,1,\ldots$).  By definition  
\begin{eqnarray}
\mu_0 & = & x^\ast = [1,\overline{i_1,i_2,\ldots,i_m}] \\
\mu_1 & = & [1;\overline{i_1,i_2,\ldots,i_m}] \\
\mu_l & = & [\overline{i_{l-1};i_l,\ldots,i_m,i_1,\ldots,i_{l-2}}]
\ \ \forall \ l > 1
\end{eqnarray} 
Each fixed point in the cycle of $\mu_0$ may then be 
expressed in terms of the $\mu_l$ in the following way:
\be
[1,\overline{i_l,i_{l+1},\ldots,i_m,i_1,\ldots,i_{l-1}}]
=\cfrac{1}{1+\cfrac{1}{i_l+\cfrac{1}{\mu_{l+2}}}}
=\frac{i_l+\cfrac{1}{\mu_{l+2}}}{1+i_l+\cfrac{1}{\mu_{l+2}}}
\label{eq:mufprep}
\ee
Now consider the quantity on the righthand side of 
relation~(\ref{eq:zetaguts}), ignoring the power $2\beta$. 
This quantity can be written as in terms of 
the $\mu_l$ using~(\ref{eq:mufprep})
\begin{eqnarray}
\lefteqn{\prod_{l=1}^{m}
\left(1-i_l+i_l \cdot 
\frac{i_l+\frac{1}{\mu_{l+2}}}{1+i_l+\frac{1}{\mu_{l+2}}}\right) }
\nonumber \\
 & = & \prod_{l=1}^{m}
\frac
{(1-i_l)(1+i_l+\frac{1}{\mu_{l+2}})+i_l(i_l+\frac{1}{\mu_{l+2}})}
{1+i_l+\frac{1}{\mu_{l+1}}}   
\nonumber \\
 & = & \prod_{l=1}^{m}
\frac
{(1+i_l+\frac{1}{\mu_{l+2}})-i_l-i_l(i_l+\frac{1}{\mu_{l+2}})
+i_l(i_l+\frac{1}{\mu_{l+2}})}
{1+i_l+\frac{1}{\mu_{l+2}}}
\nonumber \\
 & = & \prod_{l=1}^{m}
\frac
{1+\frac{1}{\mu_{l+2}}}
{1+i_l+\frac{1}{\mu_{l+2}}}
 = \prod_{l=1}^{m}
\frac
{1+\frac{1}{\mu_{l+2}}}
{1+\mu_{l+1}}
 = \prod_{l=1}^{m}
\frac
{1+\frac{1}{\mu_{l+2}}}
{1+\frac{1}{\mu_{l+1}}}
\prod_{l=1}^{m}
\frac{1}{\mu_{l+1}}      \cr
 & = & 
\frac
{(1+\frac{1}{\mu_{3}})}
{(1+\frac{1}{\mu_{2}})}
\frac
{(1+\frac{1}{\mu_{4}})}
{(1+\frac{1}{\mu_{3}})}
\frac
{(1+\frac{1}{\mu_{5}})}
{(1+\frac{1}{\mu_{4}})}
\cdots
\frac
{(1+\frac{1}{\mu_{m}})}
{(1+\frac{1}{\mu_{m-1}})}
\frac
{(1+\frac{1}{\mu_{1}})}
{(1+\frac{1}{\mu_{m}})}
\frac
{(1+\frac{1}{\mu_{2}})}
{(1+\frac{1}{\mu_{1}})}
\prod_{l=1}^{m}
\frac{1}{\mu_{l+1}}
\nonumber \\
 & = & \prod_{l=1}^{m} \frac{1}{\mu_{l+1}} = \prod_{l=1}^{m}
[\overline{i_l,i_{l+1},\ldots,i_m,i_1,\ldots,i_{l-1}}]
\end{eqnarray}
where the fact that $\mu_{m+1}=\mu_1$ and $\mu_{m+2}=\mu_2$ have been used.
\hfill $\Box$

Substituting this result into equation~(\ref{eq:fzetaind3}) along with
the identification~(\ref{eq:zpower}),
the expression for the logarithm of the induced zeta function is now
\be
\log \zeta_{\mbox{ind}}(z)=
\sum_{m=1}^{\infty} \frac{1}{m} \sum_{n=m}^\infty z^n
\sum_{ \{i_k\}_1^m; \sum_{k=1}^m i_k=n} 
\prod_{l=1}^{m} 
\left| [\overline{i_l,i_{l+1},\ldots,i_m,i_1,\ldots,i_{l-1}}] \right|^{2\beta}
\label{eq:fzetaind4}
\ee
Interchanging the summands, 
$\sum_{m=1}^{\infty}\sum_{n=m}^\infty \equiv \sum_{n=1}^\infty\sum_{m=1}^{n}$,
the zeta function reduces to the elegant form
\be
\log \zeta_{\mbox{ind}}(z,\beta)= \sum_{n=1}^\infty
 z^n \sum_{m=1}^{n} \frac{1}{m} 
\sum_{ \{i_k\}_1^m; \sum_{k=1}^m i_k=n} 
\prod_{l=1}^{m}
\left| [\overline{i_l,i_{l+1},\ldots,i_m,i_1,\ldots,i_{l-1}}] \right|^{2\beta}
\label{eq:fzetaind5}
\ee
which completes the proof of theorem~(\ref{theorem:fzetaind5}).
\hfill
$\Box$

Note that this is now a power series in $z$.  Recall that 
the radius of convergence of the power series for $\zeta(z,\beta)$, which
is not necessarily the same as that of the power series for $\log(\zeta(z,\beta)$,
gives the pressure function for the Farey map.  The focus of this
work is next turned to examining the radius of convergence of $\log(\zeta(z,\beta)$
directly via equation~(\ref{eq:fzetaind5}) with conclusions about the 
pressure function being drawn from these observations.

\section{Bounds on the Pressure function of the Farey map}
\label{sec:bounds}

Bounds on the pressure function for the Farey map, -\Fb\ , may now
be obtained by an analysis of the radius of convergence of the 
induced zeta function.  A good picture of the pressure function can
be developed this way, in the absence of an explicit expression, including
the identification of a phase transition and the scaling behaviour near
the critical point.

\begin{figure}[htb]
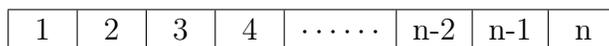

\begin{center}
\begin{tabular}{|c|c|c|c|c|c|c|c|}
\hline
\ 1 \ & \ 2 \ & \ 3 \ & \ 4 \ & $\cdots\cdots$ & n-2 & n-1 & \ n \ \\
\hline
\end{tabular}
\end{center}
\caption{A piece of wood}
\label{fig:wood}
\end{figure}

To begin with, an important ingredient for this work is an answer to the question: how 
many ordered partitions of $n$ into $m$ parts are there?
\begin{lemma}

The number of ordered partitions of an integer $n$ into
$m$ integers is equal to 
${n-1 \choose m-1}$. 
\label{lemma:numpart}
\end{lemma}
\Proof. \ \
One way to think of this question is to imagine a plank of wood
that is $n$ units in length with the numbers 1 to $n$ written evenly
on one side and rulings made between, see figure~(\ref{fig:wood}). 
The number of ordered partitions is equivalent to how many physically
different ways the length of wood can be chopped up 
into $m$ pieces of integral lengths.  Two partitions are `physically
different' if and only if for one partition the wood is cut 
between at least one pair of the inscribed
integers but is not cut between the same pair for the other partition.
\begin{figure}[htb]
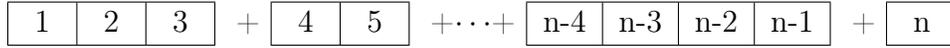

\begin{center}
\begin{tabular}{|c|c|c|}
\hline
\ 1 \ & \ 2 \ & \ 3 \ \\
\hline
\end{tabular}
\ +
\begin{tabular}{|c|c|}
\hline
\ 4 \ & \ 5 \ \\
\hline
\end{tabular}
\ +$\cdots$+
\begin{tabular}{|c|c|c|c|}
\hline
n-4 & n-3 & n-2 & n-1 \\
\hline
\end{tabular}
\ +
\begin{tabular}{|c|}
\hline
\ n \ \\
\hline
\end{tabular}
\end{center}
\caption{A chopped up piece of wood}
\label{fig:cutwood}
\end{figure}
The number of cuts required is $m-1$ and there are $n-1$
places to make these cuts.  Thus, there are $n-1$ places from which
to choose $m-1$.  Mathematically, this means there are 
${n-1 \choose m-1}$ 
ordered partitions of $n$ into $m$ integers. 

\hfill $\Box$

This result will be very useful in the estimates that will be developed
in the rest of this chapter.
The simplest place to start with is the case $\beta=0$.  Here, the 
pressure function is just the topological entropy.  Since the Farey map
has two branches, the pressure function is simply $\log 2$ when $\beta=0$.
Nevertheless, for reasons of pedagogy, it is a useful exercise to calculate the 
pressure function from the zeta function and the induced zeta function for
$\beta=0$ as similar methods will be employed in obtaining other bounds later on.

\subsection{The pressure function at $\beta=0$}
\label{subsec:pfbeta0}
\begin{prop}
The pressure function for the Farey map 
evaluated at $\beta=0$ is equal to $\log 2$.  I.e.,
\bdm
\left. -\Fb \right|_{\beta=0} \equiv \log 2
\edm
\label{prop:beta=0}
\end{prop}
\Proof. \ \
Consider the zeta function for the Farey map for $\beta=0$, (refer
to definition~(\ref{defn:zeta2})).  Recalling that $\phi(x)=-\beta\log|f'(x)|$ is
the definition of the interaction, the zeta function reduces as
follows:
\begin{eqnarray}
\zeta(z,0) & = & \exp \sum_{n=1}^{\infty}
\frac{z^n}{n} \sum_{f^{n}(x)=x} \exp \sum_{k=0}^{n-1} \phi(f^{k}x) 
\left. \right|_{\beta=0}
\nonumber \\
 & = & \exp \sum_{n=1}^{\infty}
\frac{z^n}{n} \sum_{f^{n}(x)=x} \exp \sum_{k=0}^{n-1} -\beta\log|f'(f^{k}x)|
\left. \right|_{\beta=0}
\nonumber \\
 & = & \exp \sum_{n=1}^{\infty} \frac{z^n}{n} \sum_{f^{n}(x)=x} 1
\label{eq:zfbeta0a}
\end{eqnarray}
The sum on the right is equal to the number of fixed points of  $f^n$.
Since the Farey map has 2 branches, $f^2$ will have 4 branches, $f^3$ will
have 8 and so on.  Thus, the number of fixed points is $2^n$.
So, continuing on from the end of equation~(\ref{eq:zfbeta0a}):
\be
\zeta(z,0) = \exp \sum_{n=1}^{\infty} \frac{z^n}{n} 2^n
= \exp \log \frac{1}{1-2z} = \frac{1}{1-2z}
\label{eq:zfbeta0b}
\ee
where 
the standard Taylor series expansion, $-\log(1-x)=\sum_{n=1}^\infty \frac{x^n}{n}$,
has been used in obtaining the last line.
It is clear that the $\zeta$-function has only a simple pole at $z=\half$.
Therefore the radius of convergence is $\half$.  This corresponds to 
$\exp \Fb$ and so the pressure function, $-\Fb$, at $\beta=0$ takes on the
value $\log 2$.
Note that the radius of convergence of a power series 
$\sum_{n=0}^\infty a_n z^n$, is given directly by the calculation
$\frac{1}{\rho_c} = \lim_{n \rightarrow \infty} \sup |a_n|^{\frac{1}{n}}$.
This method of calculation will be 
most useful when an explicit expression for the $\zeta$-function cannot be 
found (i.e. when $\beta \neq 0$!). 

The other way mentioned of finding the pressure function for 
$\beta=0$ was to tackle the induced zeta function.  Using $\beta=0$ in
equation~(\ref{eq:fzetaind5}) the induced zeta function becomes 
\be
\zeta_{\mbox{ind}}(z)= \exp \sum_{n=1}^\infty
 z^n \sum_{m=1}^{n} \frac{1}{m} 
\sum_{ \{i_k\}_1^m; \sum_{k=1}^m i_k=n}   1
\label{eq:fzetaindbeta0a}
\ee
The sum on the rightmost of equation~(\ref{eq:fzetaindbeta0a}) is equal to
the number of ordered partitions of $n$ into $m$ parts.  
Thus, using lemma~(\ref{lemma:numpart}),
the expression for the induced zeta function simplifies to
\be
\zeta_{\mbox{ind}}(z)= \exp \sum_{n=1}^\infty
 z^n \sum_{m=1}^{n} \frac{1}{m} { n-1 \choose m-1 }
\label{eq:fzetaindbeta0b}
\ee
This calculation may now be seen to revolve around the problem of determining the
quantity $\sum_{m=1}^{n} \frac{1}{m} { n-1 \choose m-1 }$.  One way of evaluating
sums like $\sum_{m=0}^n {n \choose m} m^k$, where $k$ is an integer, is to 
replace $m^k$ by $\frac{d^k}{d\lambda^k} e^{m\lambda}\left.\right|_{\lambda=0}$
(or a multiple integral if $k$ is negative and $m \neq 0$).
Pulling the differential(s) (integral(s)) out of the sum, leaves the form of a simple binomial
expansion of $(1+e^{\lambda})^n$.  For example, when $k>0$,
\be
\sum_{m=0}^n {n \choose m} m^k
=\left.\frac{d^k}{d\lambda^k}\sum_{m=0}^n {n \choose m} e^{m\lambda}\right|_{\lambda=0}
=\left.\frac{d^k}{d\lambda^k}(1+e^{\lambda})^n\right|_{\lambda=0}
\label{eq:egsumcalc}
\ee   
However, there is a much simpler way for the particular case $k=-1$
\begin{lemma}
\be
\sum_{m=1}^{n} \frac{1}{m} { n-1 \choose m-1 } = \frac{1}{n}(2^n-1)
\nonumber
\ee
\label{lemma:sumcalc1}
\end{lemma}
\Proof. \ \
\be
\sum_{m=1}^{n} \frac{1}{m} { n-1 \choose m-1 }
= \frac{1}{n}\sum_{m=1}^{n}\frac{n}{m} {n-1 \choose m-1}
= \frac{1}{n}\sum_{m=1}^{n} {n \choose m}
= \frac{1}{n} (2^n-1)
\label{eq:sumcalc1proof}
\ee

\hfill $\Box$

Using this small lemma in equation~(\ref{eq:fzetaindbeta0b}), the 
induced zeta function evaluated at $\beta=0$ can be found as follows:
\begin{eqnarray}
\zeta_{\mbox{ind}}(z) & = & \exp \sum_{n=1}^\infty
 z^n \sum_{m=1}^{n} \frac{1}{m} { n-1 \choose m-1 } 
\nonumber \\
 & = & \exp \sum_{n=1}^\infty z^n \frac{1}{n} (2^n-1)
\nonumber \\
 & = & \exp 
\left( \sum_{n=1}^\infty z^n \frac{2^n}{n} - \sum_{n=1}^\infty z^n \frac{1}{n} \right)
\nonumber \\
 & = & \exp 
\left( \log\frac{1}{1-2z} - \log\frac{1}{1-z} \right)
\nonumber \\
 & = & \frac{1-z}{1-2z}
\label{eq:fzetaindbeta0c}
\end{eqnarray}
The actual zeta function is the product of the induced zeta function
and the complementary zeta function, $\zeta_{Y^c}$.
Recall that $\zeta_{Y^c}$ for the Farey map was found to be $(1-z)^{-1}$,
see equation~(\ref{eq:zetacomp}).  Thus
\be
\zeta(z,0)=\zeta_{Y^c}(z,0)\cdot\zeta_{\mbox{ind}}(z,0)
=\frac{1}{1-z}\frac{1-z}{1-2z}=\frac{1}{1-2z}
\ee
and therefore this second method agrees with the first.

\hfill
$\Box$

\subsection{The pressure function for $\beta \geq 1$}
\label{subsec:pfbeta>=1}

\begin{prop}
The pressure function for the Farey map is monotonically decreasing for
all $\beta$ and is identically 0 for all $\beta \geq 1$.   
\label{prop:pfbeta>=1}
\end{prop}
\Proof. \ \ 
The first step will be to show that the pressure function is always greater
than or equal to zero.  It will then be 
shown to be monotonically decreasing with
$\beta$ and actually equal to zero for $\beta=1$. It is then 
immediately implied that -\Fb\ must be equal to 0 for all $\beta \geq 1$.

In what follows, bounds on the pressure function will naturally come from
bounds on the coefficients of $z$ in the induced zeta function.  These 
coefficients will be denoted by $a_n$; i.e.
$\log \zeta_{\mbox{ind}}(z,\beta)=\sum_{n=1
}^\infty a_n z^n$.  Explicitly, the $a_n$
are given by equation~(\ref{eq:fzetaind5}):
\be
a_n = \sum_{m=1}^{n} b_{(n,m)} = \sum_{m=1}^{n} \frac{1}{m} 
\sum_{ \{i_k\}_1^m; \sum_{k=1}^m i_k=n}  \prod_{l=1}^{m}
\left( [\overline{i_l,i_{l+1},\ldots,i_m,i_1,\ldots,i_{l-1}}] \right)^{2\beta}
\label{eq:fzetaindcoeff}
\ee
where the $b_{(n,m)}$ have been introduced to represent the terms of the sum over $m$.
Note that each $a_n$ is a sum of positive terms.  Two of the $b_{(n,m)}$ are
immediately calculable for all $n$: these are the terms for $m=1$ and $m=n$.

First, consider the term $b_{(n,1)}$.
Trivially, there is only one ordered partition of an integer $n$ into 1 part.
The part must, of course, be itself $n$.  Therefore, the period one 
continued fraction produced by this sequence is $[\overline{n}]$ and the
expression for $b_{(n,1)}$ becomes\footnote{$[\overline{n}]$ is the 
positive solution to the quadratic equation $x^2+nx-1=0$ and it 
represents the quadratic surd $\half(-n+\sqrt{n^2+4})$.}
\be
b_{(n,1)} =  \left( [\overline{n}] \right)^{2\beta}
\label{eq:coeffbn1}
\ee

The behaviour of $b_{(n,1)}$ as $n$ approaches infinity is also clear.
\be
b_{(n,1)} \sim \left(\frac{1}{n}\right)^{2\beta} 
\ \ \mbox{as} \ \ n \rightarrow \infty
\label{eq:coeffbn1asy}
\ee

In order to calculate the quantity $b_{(n,n)}$, note that
there is only one ordered partition of $n$
into $n$ parts.  So, as for the previous partition, only one sequence is 
summed over, this time being $\{\overbrace{1,1,\ldots,1}^{n \ \mbox{1's}}\}$.
The corresponding continued fraction is the reciprocal of the golden ratio\footnote{
The golden ratio is, in some sense, the most essential periodic continued fraction.
It satisfies the quadratic relation, $x^2-x-1=0$ and may be expressed in
quadratic surd form as $\frac{1+\sqrt{5}}{2}$.  From the quadratic equation, it
follows that $\rho_g$ satisfies several other nice relationships such as
$\rho_g^2=\rho_g+1$ and $\rho_g=1+\frac{1}{\rho_g}$.  There is also a
fundamental connection to the Fibonacci series which 
will be utilised later on.}, $\rho_g=\frac{1+\sqrt{5}}{2}$.
This sequence of ones is unchanged by cyclic permutation, so the product 
over $l=1$ to $n$ produces the number $(\rho_g-1)^n=\rho^{-n}$.  All this implies
\be
b_{(n,n)} = \frac{1}{n} \rho_g^{-2n\beta}
\ee

As a lower bound for $a_n$, consider the term $b_{(n,1)}$ and the following:
\begin{eqnarray}
a_n & = & \sum_{m=1}^n b_{n,m} >  b_{n,1} = ([\overline{n}])^{2\beta}
\label{eq:anlowbound1}
\end{eqnarray}
Note that $\frac{1}{n} > [\overline{n}] = \frac{1}{n+[\overline{n}]} > \frac{1}{2n}$
and therefore for $\beta > 0$
\be
\lim_{n\rightarrow\infty}|a_n|^{\frac{1}{n}} \geq 
\lim_{n\rightarrow\infty}(2n)^{\frac{-2\beta}{n}}
= 1
\label{eq:anlowbound3}
\ee
and similarly for $\beta<0$,
\be
\lim_{n\rightarrow\infty}|a_n|^{\frac{1}{n}} \geq 
\lim_{n\rightarrow\infty}(n)^{\frac{-2\beta}{n}}
= 1
\label{eq:anlowbound4}
\ee
Therefore a lower bound on $\frac{1}{\rho_c}$ is 1.
The first part of this section is then complete as this implies for 
the pressure function that $-\Fb \geq \log 1 = 0$ for all $\beta$.

Consider the form of the continued fractions in the expression for
the induced zeta function (equation~(\ref{eq:fzetaind5})). They are all
strictly less than one and raised to a power $2\beta$.  Thus, as
$\beta$ increases, each of these terms decrease.  Hence, the $b_{(n,m)}$
and finally the coefficients $a_n$ must also decrease with increasing
$\beta$.  This means that the radius of convergence of the induced zeta
function may not decrease and conversely the pressure function
may not increase with $\beta$.  

It is clear then that the pressure function is a monotonically
decreasing function of $\beta$.  Since the pressure function is always
greater than or equal to 0, if it can be shown to be 0 for any finite
$\beta$ then it must be 0 for all $\beta$ to the right of this point as
well.  Consider then the case for $\beta=1$.  Here the transfer operator
reduces to the Perron-Frobenius operator which was mentioned in the introduction.  
The invariant density $\psi_I$ of a mapping
$f$ is known to be a solution to the Perron Frobenius equation which is
defined as
\be
\psi_I(x)=\sum_{f^n y=x}\frac{\psi_I(y)}{|f'(y)|}
\label{eq:perfrob}
\ee
The existence of $\psi_I$ would show that 1 is an eigenvalue of the transfer
operator when $\beta=1$.  However, for the Farey map, $\psi_I(x)=1/x$ which
cannot be normalised and is therefore not an eigenfunction of $\LB$.
However, it is still observed that $r(\L^{(1)})=1$; see, for example,
\cite{feig:transfer} and \cite{prelslawn:interm}.  
Since the spectral radius of 
the transfer operator corresponds to $\exp -\Fb$, \cite{ruelle:therm},
it can then be inferred that $\left. -\Fb \right|_{\beta=0}=0$.  Therefore, 
using the reasoning above, the pressure function is equal to 0 for all
$\beta \geq 1$.  This does not preclude that values of $\beta$ to the left
of 1 might also be 0 and this will be discussed later on. 

\hfill
$\Box$

\begin{figure}[htbp!]
  \includegraphics[width=\textwidth]{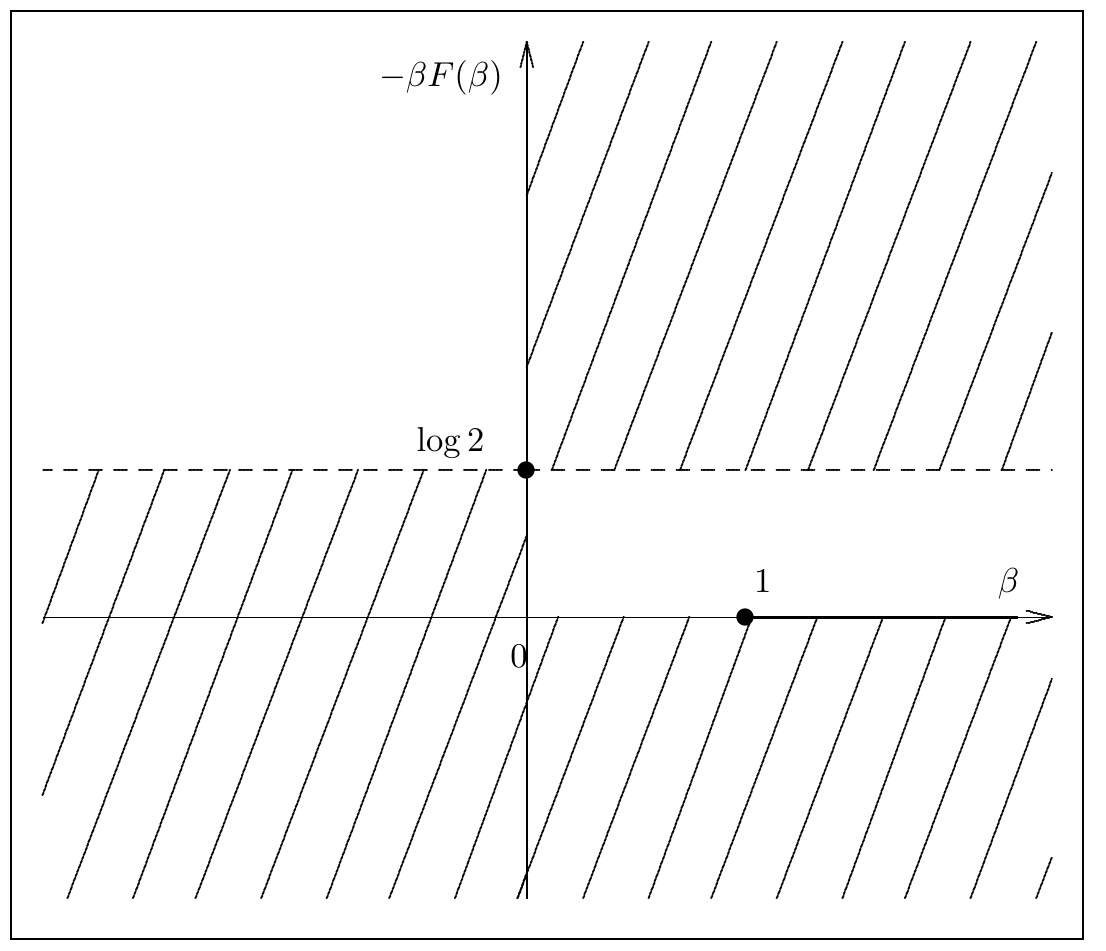}
  \caption{Preliminary bounds on the pressure function}
  \label{fig:bound1}
\end{figure}

Graphically, the constraints found so far are represented in 
figure~(\ref{fig:bound1}).  The clear regions represent the possible
region where the pressure function may lie.  The thick line and the
two dots are the actual values of -\Fb\ .  Since the pressure function
is decreasing, it cannot be smaller than $\log2$ for $\beta<0$ and 
conversely, it cannot be larger that $\log2$ for $\beta>0$. ( Note that
bounds found later on will be displayed separately and then all
together in a final figure. )

\begin{corollary}
The pressure function for the Farey map exhibits a phase transition
for $\beta$ somewhere in $(0,1]$.
\label{cor:phasetransexists}
\end{corollary}
\Proof. \ \
The analytic continuation of a constant function is itself the same
constant function.  Since the pressure function is not 0 at 
least when $\beta=0$ by proposition~(\ref{prop:beta=0}),
\Fb\ cannot be analytic everywhere on the real line and, in particular, must express
non-analytic behaviour between $\beta=0$ and $\beta=1$.
The preceding work is therefore 
enough to provide conclusive evidence for the existence of a phase transition 
in the pressure function.\footnote{The appearance of non-analytic behaviour
in some quantity with respect to a given parameter is equivalent to the physical
notion of the existence of a
phase transition.  For a formal definition of a phase transition, 
see Huang~\cite{huang:therm}.}
\hfill 
$\Box$

\subsection{Some general bounds on the pressure function}
\label{subsec:pfgenbound}
Some more precise bounds are developed in this section, further 
constraining the shape of the pressure function.
Recall that $b_{(n,n)} = \frac{1}{n} \rho_g^{-2n\beta}$ was the final
term in the sum for the $a_n$.  In fact, 
$\rho_g^n$ is a bound on all the other possible periodic continued 
fractions for a given $n$, i.e. all of those periodic continued fractions whose
elements in their repeating sequence of entries sum to $n$.
This is proven in the following lemma:
\begin{lemma}
\be
\frac{1}{\rho_g^n}
<\prod_{k=1}^m [\overline{i_k,\ldots,i_m,i_1,\ldots,i_{k+1}}] \ \
\forall \ \ \mbox{sequences} \ \ \{i_k\}_1^m \ \ \mbox{such that}
\ \ \sum_{k=1}^m i_k = n 
\nonumber
\ee
for all $m = 1,2,\ldots,n-1$.  Trivially, for $m=n$, the inequality becomes
an equality.
\label{lemma:goldmeanbound}
\end{lemma}
\Proof. \ \
Consider a periodic continued fraction whose entries of the generating
sequence sum to $n$ written as $\mu=\mu_0=[\overline{i_1,\ldots,i_m}]$.
The overall idea of the proof is to show something along the lines of the 
statement:
\be
\frac{1}{\rho_g^{i_1}} < [\overline{i_1,\ldots,i_m}]
\label{eq:pridea}
\ee
To do this, it is helpful to obtain an estimate of the integer part of $\rho_g^k$.
An examination of the first few $k$ with the help of the relation 
$\rho_g=1+\frac{1}{\rho_g}$ shows that
\begin{eqnarray}
\rho_g   & = & 1+\frac{1}{\rho_g}  \nonumber \\
\rho_g^2 & = & \rho_g\left(1+\frac{1}{\rho_g}\right)  
 = \rho_g+1 = 1+\frac{1}{\rho_g}+1 = 2+\frac{1}{\rho_g}   \nonumber \\
\rho_g^3 & = & \rho_g\left(2+\frac{1}{\rho_g}\right)  
 = 2\rho_g+1 = 2\left(1+\frac{1}{\rho_g}\right) + 1 = 3+\frac{2}{\rho_g}
\label{eq:rho^k egs}
\end{eqnarray}
It is clear from the above that $\rho_g^k$ may be expressed in the form
$a_k+\frac{b_k}{\rho_g}$.  Therefore:
\begin{eqnarray}
\rho_g^{k+1} & = & \rho_g \cdot \rho_g^k
 = \rho_g \left( a_k+\frac{b_k}{\rho_g} \right)
 = a_k\rho_g+b_k=a_k\left(1+\frac{1}{\rho_g}\right)+b_k \nonumber \\
 & = & a_k+b_k+\frac{a_k}{\rho_g} = a_{k+1}+\frac{b_{k+1}}{\rho_g}
\label{eq:rhon->n+1}
\end{eqnarray}
The recursion relations are then $a_{k+1}=a_k+b_k$ and $b_{k+1}=a_k$.
The combining of these two expressions leads to the familiar Fibonacci formula:
$a_{k+1}=a_k+a_{k-1}$.  Noting also that $a_0=a_1=1$, it is clear that
the $a_k$ sequence is indeed the Fibonacci series $1,1,2,3,5,8,13,\ldots$,
the elements of which will be denoted by $f_k$, $k=0,1,\ldots$. 
Therefore, $\rho_g^{i_1}$ may be expressed in the following form:
\be
\rho_g^{i_1}=f_{i_1} + \frac{f_{(i_1-1)}}{\rho_g}
\label{eq:rhoik1}
\ee
Now, $f_n > n $ for all $n>3$ while $f_n=n$ for $n=1,2$ and 3. 
Note that because $\rho_g = \frac{1+\sqrt{5}}{2} \approx 1.618$, the second term in
relation~(\ref{eq:rhoik1}) is strictly greater than one for all $i_1 \geq 3$.
Returning then to the conjecture of equation~(\ref{eq:pridea}), it is 
seen to be true for $i_1 \geq 3$ since, using relation~(\ref{eq:rhoik1})
and the above inequalities:
\begin{eqnarray}
\rho_g^{i_1} & = & f_{i_1} + \frac{f_{(i_1-1)}}{\rho_g} 
\geq i_1 + \frac{f_{(i_1-1)}}{\rho_g} > i_1 + 1  \nonumber \\
 & > & i_1 + [\overline{i_2,\ldots,i_m,i_1}] = [\overline{i_1,\ldots,i_m}]^{-1}
\label{eq:rhoik2}
\end{eqnarray}
where the fact has been used that the periodic continued fractions being
considered are all strictly less than 1.
Equation~(\ref{eq:pridea}) is therefore true for $i_k \geq 3$.
The two cases left will be treated separately.

\noindent
\fbox{$i_k=1$}: assume that $\frac{1}{\rho_g} < [\overline{1,i_2,\ldots,i_m}]$.  
Writing the continued fraction in the form
\be
[\overline{1,i_2,\ldots,i_m}]=\cfrac{1}{1+\cfrac{1}{\mu_2}}
\label{eq:ik=1a}
\ee
where $\mu_k$ represents the $k^{\mbox{th}}$-complete quotient and
$\mu_2=[\overline{i_2;i_3,\ldots,i_m,1}]$.  It follows that
\be
\cfrac{1}{\rho_g} < \cfrac{1}{1+\cfrac{1}{i_2+\mu_3^{-1}}}
\Rightarrow \rho_g < \mu_2 = i_2+\cfrac{1}{\mu_3}
\label{eq:ik=1b}
\ee
This is always true if $i_2 \geq 2$ since $\rho_g < 2$ but
it may break down if $i_2=1$.  However, all is not lost.
Assume that indeed 
\be
\frac{1}{\rho_g}  >  [\overline{1,1,i_3,\ldots,i_m}]
\label{eq:ik=1c}
\ee
It is simple to show that this implies 
\be
\rho_g  <  [\overline{i_3;\ldots,i_m,1,1}] = \mu_3 
\label{eq:ik=1d}
\ee
Then the following inequality still holds:
\be
\frac{1}{\rho_g^2} <  
[\overline{1,1,i_3,\ldots,i_m}] \cdot [\overline{1,i_3,\ldots,i_m,1}]
\label{eq:ik=1e}
\ee
This is proven as follows:
\begin{eqnarray}
[\overline{1,1,i_3,\ldots,i_m}] \cdot [\overline{1,i_3,\ldots,i_m,1}] 
& = & \cfrac{1}{1+\cfrac{1}{1+\cfrac{1}{\mu_3}}}
\cdot \cfrac{1}{1+\cfrac{1}{\mu_3}}  
 = \frac{\mu_3}{1+2\mu_3}  \cr
 & = & \cfrac{1}{\cfrac{1}{\mu_3} + 2} 
 > \cfrac{1}{\cfrac{1}{\rho_g} + 2} = \frac{1}{\rho_g^2}
\label{eq:ik=1f}
\end{eqnarray}
where equation~(\ref{eq:ik=1d}) has been used to bring about
the inequality.  So while the idea of equation~(\ref{eq:pridea}) is not exactly
true for all $i_1$, an inequality still exists if more terms of the 
product are involved. The important fact is that the power of the inverse
of the golden mean (i.e., 2) is the sum of the leading entries in the two continued
fractions (both 1).

\noindent
\fbox{$i_k=2$}

A similar approach to the above is employed in the case $i_1=2$.  
Assume that $\frac{1}{\rho_g^2} < [\overline{2,i_2,\ldots,i_m}]$.
The implication of this is given as follows:
\be
\frac{1}{\rho_g^2} < \cfrac{1}{2+\cfrac{1}{\mu_2}}
\Rightarrow  \rho_g < \mu_2 = i_2+\cfrac{1}{\mu_3}
\label{eq:ik=2a}
\ee
which is always true if $i_2 \geq 2$ since $\rho_g < 2$.
Once again, the case $i_2=1$ must be considered further.  Assume
that the desired result is not true, i.e. 
\be
\frac{1}{\rho_g^2} > [\overline{2,1,i_3\ldots,i_m}] =
\cfrac{1}{2+\cfrac{1}{1+\cfrac{1}{\mu_3}}}
\label{eq:ik=2b}
\ee
It easily follows that 
\be
\rho_g < \mu_3 = i_3+\mu_4^{-1}
\label{eq:ik=2c}
\ee
Then the following inequality still holds:
\be
\frac{1}{\rho_g^3} <  
[\overline{2,1,i_3,\ldots,i_m}] \cdot [\overline{1,i_3,\ldots,i_m,2}]
\label{eq:ik=2d}
\ee
since
\begin{eqnarray}
[\overline{2,1,i_3,\ldots,i_m}] \cdot [\overline{1,i_3,\ldots,i_m,2}]
 & = &\cfrac{1}{2+\cfrac{1}{1+\cfrac{1}{\mu_3}}}
\cdot \cfrac{1}{1+\cfrac{1}{\mu_3}}  \cr
 & = & \cfrac{1}{\cfrac{2}{\mu_3} + 3} 
 > \cfrac{1}{\cfrac{2}{\rho_g} + 3} = \cdots =\frac{1}{\rho_g^3}
\label{eq:ik=2e}
\end{eqnarray}
Once again, an inequality is found when the power of the inverse of the
golden ratio (which is 3) is equal to the sum of the leading entries in the 
two continued fractions (2 and 1 respectively).  Finally, the question may
be asked as to what power $N$ of $\rho_g^{-1}$ is required for the following
inequality to be true: 
\be
\frac{1}{\rho_g^N}
< \prod_{k=1}^m [\overline{i_k,\ldots,i_m,i_1,\ldots,i_{k+1}}]
\ee
Since for all continued
fractions with a leading entry of $i_1 \geq 3$ the least power $\rho_g^{-1}$
required for the inequality to hold is precisely $i_1$, and the special case
inequalities are satisfied when the sum of the leading entries equals the power of 
$\rho_g^{-1}$, the inequality
for the entire product holds if $\rho_g^{-1}$ is raised to the sum of 
all the leading entries which, of course, is $n$.  This rather long sentence
completes the proof.

\hfill $\Box$

The result of this lemma gives a nice bound on the pressure function.
Firstly, note the following remark:
\begin{rem}
Lemma~(\ref{lemma:goldmeanbound}) immediately implies that 
\be
\left(\frac{1}{\rho_g^n}\right)^{2\beta} <
\prod_{k=1}^m [\overline{i_k,\ldots,i_m,i_1,\ldots,i_{k+1}}]^{2\beta}
\ \ \mbox{if} \ \ \beta > 0
\label{eq:lemmacor1}
\ee
and
\be
\left(\frac{1}{\rho_g^n}\right)^{2\beta} >
\prod_{k=1}^m [\overline{i_k,\ldots,i_m,i_1,\ldots,i_{k+1}}]^{2\beta}
\ \ \mbox{if} \ \ \beta < 0
\label{eq:lemmacor2}
\ee
\label{rem:rhonbound}
\end{rem}
\begin{prop}
The pressure function for the Farey map is bounded in the following way:
\bdm 
-\Fb \geq \log2-2\beta \log \rho_g\mbox{ when }\beta>0
\edm
and
\bdm
-\Fb \leq \log2-2\beta \log \rho_g\mbox{ when }\beta<0
\edm
\label{prop:genbound1}
\end{prop}
\Proof. \ \
It will be useful to invoke the following theorem due to 
Alfred Pringsheim~\cite{hille:anal}.
\begin{theorem}
If the coefficients of a power series $\sum_{n=0}^{\infty}a_n z^n$ are
all positive then the power series has a singularity on the positive 
real axis lying on its circle of convergence.  I.e., the power series
has a singularity at $z=\rho_c$.
\label{theorem:pringsheim}
\end{theorem}
It can thus be assumed that $z$ is positive and
real in the search for the radius of convergence of the induced zeta function
since finding bounds on where its first singularity lies on the positive $z$-axis
is equivalent to finding bounds on the radius of convergence.
Therefore, for $\beta>0$, the induced zeta function, equation~(\ref{eq:fzetaind5})
can be estimated as follows using remark~(\ref{rem:rhonbound}):
\begin{eqnarray}
\zeta_{\mbox{ind}}(z,\beta) & = &
\exp \sum_{n=1}^\infty z^n \sum_{m=1}^{n} \frac{1}{m} 
\sum_{ \{i_k\}_1^m; \sum_{k=1}^m i_k=n} \prod_{l=1}^{m}
\left| [\overline{i_l,i_{l+1},\ldots,i_m,i_1,\ldots,i_{l-1}}] \right|^{2\beta}
\nonumber \\
 & > & \exp \sum_{n=1}^\infty z^n \sum_{m=1}^{n} \frac{1}{m} 
{n-1 \choose m-1} \rho_g^{2n\beta}
\nonumber \\
 & = & \exp \sum_{n=1}^\infty z^n \frac{1}{n}(2^n-1)\rho_g^{-2n\beta}
\nonumber \\
 & = & \exp \sum_{n=1}^\infty z^n \frac{(2\rho_g^{-2\beta})^n}{n}
-\sum_{n=1}^\infty z^n \frac{\rho_g^{-2n\beta}}{n}
\nonumber \\
 & = & \frac{1-z\rho_g^{-2\beta}}{1-2z\rho^{-2\beta}}
\label{eq:est1}
\end{eqnarray}
Clearly, the radius of convergence for this function is $\rho_c=\half\rho^{2\beta}$.
This implies that the radius of convergence of the induced zeta function must be 
less than or equal to $\half\rho^{2\beta}$.
Therefore, the above demonstrates that 
$-\Fb \geq \log2-2\beta \log \rho_g$ for $\beta>0$.
A similar argument shows also that $-\Fb \leq \log2-2\beta \log \rho_g$ for $\beta<0$.
Note that this line of constraint passes through $\log2$ when $\beta=0$ as 
would be expected.  
\hfill
$\Box$

A second general bound on the pressure function is presented in the
following proposition:
\begin{prop}
\bdm
-\Fb \geq -2\beta\log\rho_g \ \ \forall \ \ \beta
\edm
\label{prop:genbound2}
\end{prop}
\Proof. \ \
Recall that the coefficients of the power series for 
$\log \zeta_{\mbox{ind}}(z,\beta)$ were given by $a_n=\sum_{m=1}^n b_{(n,m)}$.
Clearly then
\be
a_n > b_{(n,n)} = \frac{1}{n} \rho_g^{-2n\beta}
\label{eq:lowboundrho1}
\ee
This implies for real and positive $z$ that
\be
\zeta_{\mbox{ind}}(z,\beta) = \exp\sum_{n=1}^\infty a_n z^n 
 > \exp \sum_{n=1}^\infty \frac{1}{n} \rho_g^{-2n\beta} z^n 
 = \frac{1}{1-z \rho_g^{-2\beta}}
\label{eq:lowboundrho2}
\ee
The function on the final line of the above has a simple pole at 
$z = \rho_g^{2\beta}$ which is therefore its radius of convergence.
Since the induced zeta function is greater than this function it must have
a radius of convergence less than or equal to $\rho_c=\rho_g^{2\beta}$.
\hfill
$\Box$

\begin{figure}[htb]
  \includegraphics[width=\textwidth]{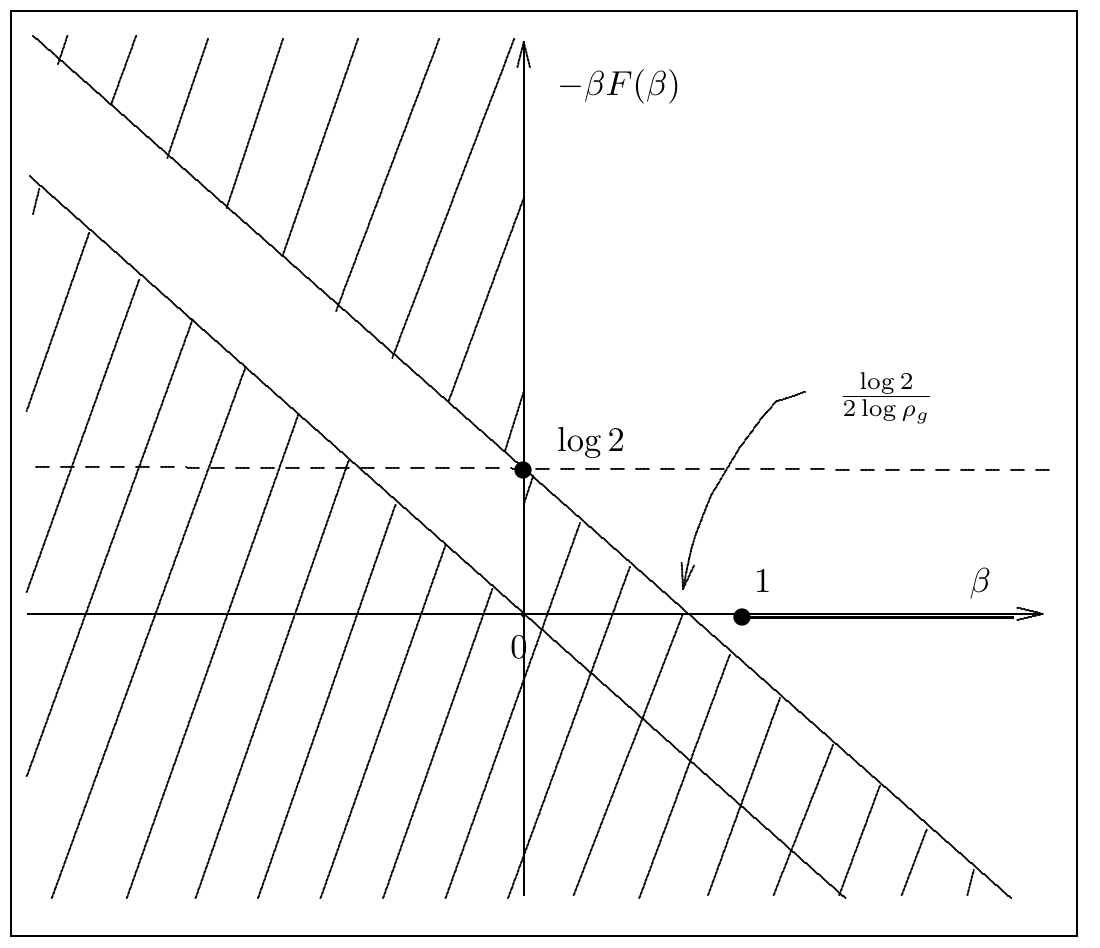}
  \caption{Some more bounds on the pressure function}
  \label{fig:bound2}
\end{figure}

The information provided by both of these bounds
is displayed in figure~(\ref{fig:bound2}).

One more bound on the pressure function is presented in this section.
\begin{prop}
For $\beta>0$, a upper bound on the pressure function for the Farey map is given by
\bdm
-\Fb \leq \log\left( 1+ \left(\frac{2}{3}\right)^{2\beta} \right)
\edm
and for $\beta<0$, the corresponding lower bound is
\bdm
-\Fb \geq \log\left( 1+ \left(\frac{2}{3}\right)^{2\beta} \right)
\edm
\label{prop:genbound3}
\end{prop}
\Proof. \ \
The proof requires an examination of the following representation of the 
induced zeta function obtained from examination of the representations
given in equations~(\ref{eq:fzetaind3}), (\ref{eq:fzetaind4}) and (\ref{eq:fzetaind5}):
\be
\log \zeta_{\mbox{ind}}(z,\beta)=
\sum_{n=1}^{\infty} z^n \sum_{m=1}^n  \frac{1}{m}
\sum_{ \{i_k\}_1^m; \sum_{k=1}^m i_k=n} 
\prod_{l=1}^{m} 
|G_{l}'([1,\overline{i_{l+1},i_{l+2},\ldots,i_m,i_1,\ldots,i_l}])|^{\beta}
\label{eq:fzetaind6}
\ee
Now, from equation~(\ref{eq:Gn'def}), $G_{n}'(x)=\frac{-1}{(1+nx)^2}$.
Since $x \in [\half,1]$ it follows that a bound on the maximum possible
value of $|G_{n}'(x)|$ would be its value for $n=1$ and $x=\half$.
Therefore, $|G_{n}'(x)| < \frac{1}{(1+1.\half)^2}=\left(\frac{2}{3}\right)^2$.
So, for real and positive $z$ and for $\beta>0$, it follows that
\begin{eqnarray}
\zeta_{\mbox{ind}}(z,\beta) & < & \exp \sum_{n=1}^{\infty} z^n \sum_{m=1}^n \frac{1}{m}
\sum_{ \{i_k\}_1^m; \sum_{k=1}^m i_k=n} 
\prod_{l=1}^{m} \left(\frac{2}{3}\right)^{2m\beta}
\nonumber \\
 & = & \exp \sum_{n=1}^{\infty} z^n \sum_{m=1}^n \frac{1}{m}
\left(\frac{2}{3}\right)^{2m\beta}
\sum_{ \{i_k\}_1^m; \sum_{k=1}^m i_k=n} 1 \nonumber \\
 & = & \exp \sum_{n=1}^{\infty} z^n \sum_{m=1}^n  \frac{1}{m}
\left(\frac{2}{3}\right)^{2m\beta} {n-1 \choose m-1} \nonumber \\
 & = & \exp \sum_{n=1}^{\infty} \frac{z^n}{n} \sum_{m=1}^n
{n \choose m} \left(\left(\frac{2}{3}\right)^{2\beta}\right)^{m} \nonumber \\
 & = & \exp \sum_{n=1}^{\infty} \frac{z^n}{n} 
\left(\left( 1+ \left(\frac{2}{3}\right)^{2\beta} \right)^n - 1 \right)
\nonumber \\
 & = &  \frac{1-z}
{1-z\left( 1+ \left(\frac{2}{3}\right)^{2\beta} \right)}  
\label{eq:bound3}
\end{eqnarray}
The final expression shows a function with a radius of convergence 
$\rho_c = \left( 1+ \left(\frac{2}{3}\right)^{2\beta} \right)^{-1}$.
Since the induced zeta function is smaller than this function its radius of 
convergence must be at least $\rho_c$.  As the pressure function is the logarithm
of the inverse of the radius of convergence of the zeta function, it is 
bounded in the following way for $\beta>0$:
\be
-\Fb \leq \log\left( 1+ \left(\frac{2}{3}\right)^{2\beta} \right)
\label{eq:bound3res1}
\ee
A similar argument for $\beta<0$ shows that there
\be
-\Fb \geq \log\left( 1+ \left(\frac{2}{3}\right)^{2\beta} \right)
\label{eq:bound3res2}
\ee
\hfill
$\Box$

\begin{figure}[htbp!]
  \includegraphics[width=\textwidth]{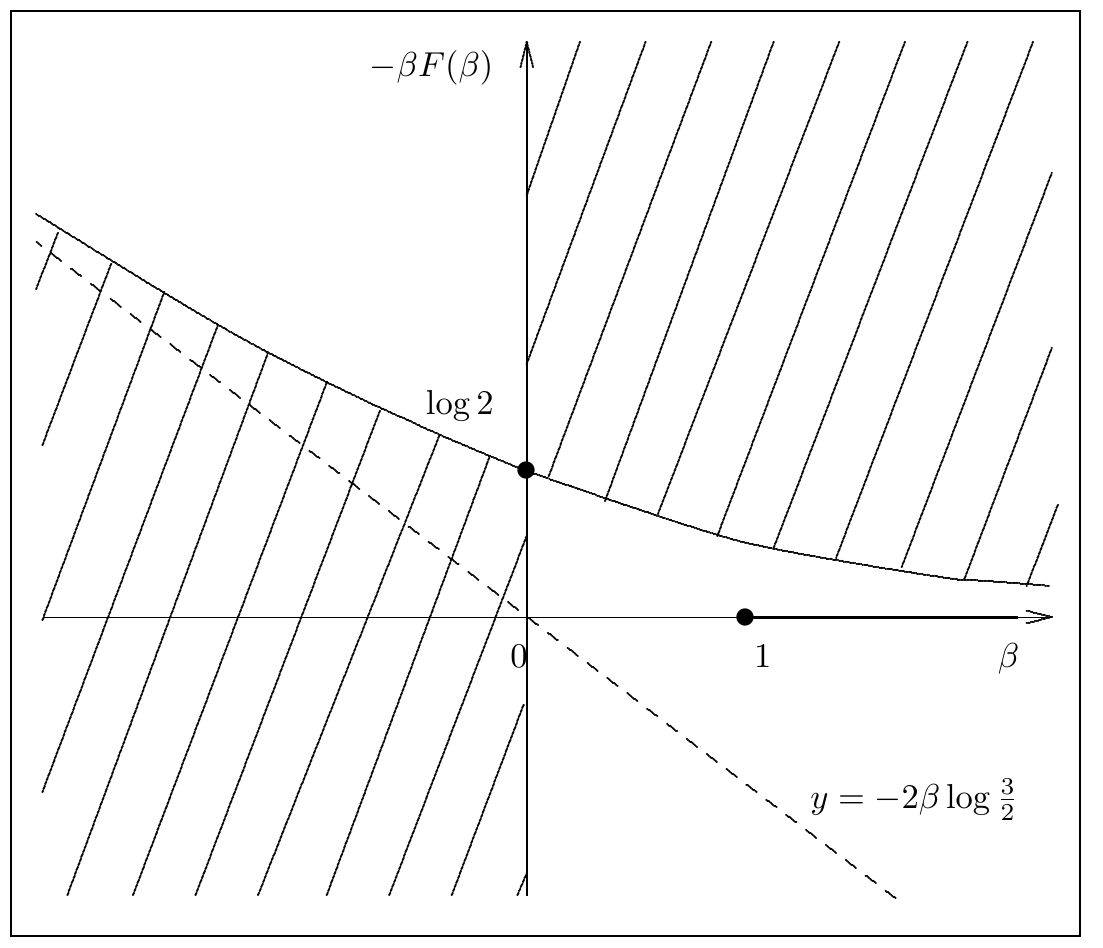}
  \caption{
    \ldots and another bound on the pressure function.
  }
  \label{fig:bound3}
\end{figure}

Note that, as expected, $\left. \log\left( 1+ \left(\frac{2}{3}\right)^{2\beta} 
\right)\right|_{\beta=0}=\log2$.  Also, this function asymptotes to the 
the $\beta$ axis as $\beta \rightarrow \infty$ and, on the other hand, 
it asymptotes to $-2\beta\log\frac{3}{2}$ as $\beta \rightarrow -\infty$.
A plot of this bound is given separately in figure~\ref{fig:bound3}.

\subsection{The pressure function for $\beta \rightarrow -\infty$}
\label{subsec:pfbeta-infty}

The previous section has provided some strong constraints on the 
behaviour of the pressure function for negative $\beta$.  In fact, it has
been shown that $-\Fb=-2\beta\log\rho_g + O(1)$.   The 
work here will improve on these bounds with the main result
being the following theorem regarding the asymptotic behaviour of the 
pressure function for large negative $\beta$:
\begin{theorem}
For every $\delta > 0$, there exists a $\beta(\delta)$ such that for
all $\beta < \beta(\delta)$
\bdm
-\Fb - -2\beta\log\rho_g < \delta
\edm
\label{theorem:pfbeta-infty}
\end{theorem}
\Proof. \ \ 

\subsubsection{Motivation}
Recall that the coefficients of the induced zeta function, $a_n$,
are ultimately a sum over the continued fraction terms
$\frac{1}{m}\prod_{l=1}^{m}
([\overline{i_l,i_{l+1},\ldots,i_m,i_1,\ldots,i_{l-1}}])^{2\beta}$,
see equation~(\ref{eq:fzetaindcoeff}).
In the limit $\beta$ approaches negative
infinity, the second part of remark~(\ref{rem:rhonbound}) appears
to indicate that $\frac{1}{n}\rho_g^{-2n\beta}=b_{(n,n)}$ will be the 
dominant term for any given $n$.  However, this is not necessarily true
since the factor $\frac{1}{m}$ has not been considered. Also, and much more
importantly, in the limit
$n \rightarrow \infty$, terms containing continued fractions which approximate the
golden mean may be of the order of $b_{(n,n)}$.  

In reference to this last
observation, consider $b_{(n,n-1)}$ for $n>2$.  This term will be evaluated 
explicitly as a motivation for the final proof of the main theorem.  
\begin{prop}
\be
b_{(n,n-1)}=(f_{n-1}+\sqrt{f_{n}f_{n-2}})^{-2\beta}
\nonumber
\ee
\label{prop:bnn-1}
where $f_n$ is the $n^{\mbox{th}}$ Fibonacci number.
\end{prop}
\Proof. \ \
From lemma~(\ref{lemma:numpart}),
it is known that there are ${n-1 \choose n-2}=n-1$ 
ordered partitions of $n$ into $n-1$ parts.  Clearly, these partitions are
those that consist of 1 `two' and $n-2$ `ones' and may be represented by
the $n-1$ sequences:
\be
\{\overbrace{1,1,\ldots,1}^{k \ \mbox{\scriptsize ones}}
,2,\overbrace{1,\ldots,1,1}^{n-k-2 \ \mbox{\scriptsize ones}}\}
\label{eq:seqnn-1}
\ee
where $0 \leq k \leq n-2$.  Note that these sequences have no periodic structures
of length less than $n-1$ and are merely cyclic rotations of each other.
Therefore, they each generate the same product of continued fractions
and the factor $\frac{1}{m}$ is cancelled.  That is, 
\be
b_{(n,n-1)} = \prod_{k=0}^{n-2}
([\overline{\underbrace{1,1,\ldots,1}_{k \ \mbox{\scriptsize ones}}
,2,\underbrace{1,\ldots,1,1}_{n-k-2 \ \mbox{\scriptsize ones}}}])^{2\beta}
\label{eq:bn-1def}
\ee
More generally, if a sequence with $m$ elements contains no smaller periodic blocks
than itself then the $m-1$ cyclic rotations of it will also be summed over in the
expression for the induced zeta function~(\ref{eq:fzetaind5}) thereby
cancelling the factor of $\frac{1}{m}$.
\begin{lemma}
Any continued fraction $C(n,x)$ of the form
\be
[\underbrace{1,1,\ldots,1}_{n \ \mbox{\scriptsize ones}},x]
\nonumber
\ee
where $x$ is any real number and $n=0,1,2,\ldots$,
is equal to the fraction
\be
\cfrac{f_{n-1} x + f_{n-2}}{f_{n} x + f_{n-1}}
\nonumber
\ee
where $f_0,f_1,f_2,f_3,f_4\ldots = 1,1,2,3,5,\ldots$ is the Fibonacci series
which satisfies the recursion relation $f_{n+1}=f_n+f_{n-1}$.
\label{lemma:fibonexp}
\end{lemma}
\Proof. \ \
Assume the assertion of the lemma is true for $k$ some positive integer.
That is,
\be
C(k,x)=[\overbrace{1,1,\ldots,1}_{k \ \mbox{\scriptsize ones}},x]
=\cfrac{f_{k-1} x +f_{k-2}}{f_{k} x + f_{k-1}}
\label{eq:fibonexp1}
\ee
Then,
\begin{eqnarray}
C(k+1,x) & = & [\overbrace{1,1,\ldots,1}_{k+1 \ \mbox{\scriptsize ones}},x] 
 = \cfrac{1}{1+C(k,x)} \cr
 & = & \cfrac{1}{1+\cfrac{f_{k-1} x +f_{k-2}}{f_{k} x + f_{k-1}}}
 = \cfrac{f_{k} x + f_{k-1}}{f_{k} x + f_{k-1}+f_{k-1}x+f_{k-2}}
 = \cfrac{f_{k} x + f_{k-1}}{f_{k+1}x + f_{k}}
\label{eq:fibonexp2}
\end{eqnarray}
Therefore, if the assertion is true for $n=k$ then it is true for $n=k+1$.
Now consider the case $n=0$. The Fibonacci series may be extended to negative
values of $n$ by application of the recursion relation rewritten as
$f_n=f_{n+2}-f_{n+1}$. This gives $f_{-1}=0,f_{-2}=1,f_{-3}=-1$ and so on.
Using this when there are no 1's preceding $x$ in the continued fraction:
\be
\frac{1}{x}=\frac{0x+1}{x+0}=\frac{f_{-1}x+f_{-2}}{f_0x+f_{-1}}
\label{eq:fibonexp3}
\ee
Thus the statement is true for the case $n=0$ and by induction must be 
therefore be true for all $n=0,1,2,\ldots$.
\hfill $\Box$

Two results needed to calculate $b_{(n,n-1)}$ now follow from this lemma.
\begin{corollary}
Let $t(n)=[\overline{2,\underbrace{1,1,\ldots,1}_{n-2 \ \mbox{\scriptsize ones}}}]$.
Then $b_{(n,n-1)}$ simplifies to the following:
\be
b_{n,n-1}=\cfrac{t(n)}{f_{n-2}+f_{n-3}t(n)}
\nonumber
\ee
\label{cor:fibonbnn-1:1}
\end{corollary}
\Proof. \ \
Returning to the definition of $b_{(n,n-1)}$ in equation~(\ref{eq:bn-1def}),
it can be rewritten using $t(n)$ as follows
\be
b_{(n,n-1)} = \prod_{k=0}^{n-2}
([\underbrace{1,1,\ldots,1}_{k \ \mbox{\scriptsize ones}},t(n)^{-1}])^{2\beta}
\label{eq:bn-1defwitht(n)1}
\ee
Using lemma~(\ref{lemma:fibonexp}) this simplifies as follows:
\begin{eqnarray}
\lefteqn{ b_{(n,n-1)} = \prod_{k=0}^{n-2}
\left( \cfrac{f_{k-1}t(n)^{-1}+f_{k-2}}{f_{k}t(n)^{-1}+f_{k-1}} \right)^{2\beta}} \cr
 & = & \left(
\cfrac{f_{-1}t(n)^{-1}+f_{-2}}{f_{0}t(n)^{-1}+f_{-1}} \cdot
\cfrac{f_{0}t(n)^{-1}+f_{-1}}{f_{1}t(n)^{-1}+f_{0}} 
\cdots
\cfrac{f_{n-4}t(n)^{-1}+f_{n-3}}{f_{n-3}t(n)^{-1}+f_{n-4}} \cdot
\cfrac{f_{n-3}t(n)^{-1}+f_{n-4}}{f_{n-2}t(n)^{-1}+f_{n-3}}
\right)^{2\beta}  \cr
 & = & \left( \cfrac{f_{-1}t(n)^{-1}+f_{-2}}{f_{n-2}t(n)^{-1}+f_{n-3}} \right)^{2\beta}
 = \left( \cfrac{1}{f_{n-2}t(n)^{-1}+f_{n-3}} \right)^{2\beta}
 = \left( \cfrac{t(n)}{f_{n-2}+f_{n-3}t(n)} \right)^{2\beta}
\label{eq:bn-1defwitht(n)2}
\end{eqnarray}
\hfill $\Box$

\begin{corollary}
The periodic continued fraction $t(n)$ defined in corollary~(\ref{cor:fibonbnn-1:1})
is given in quadratic surd form by
\be
t(n) = \cfrac{-f_{n-2}+\sqrt{f_nf_{n-2}}}{f_{n-1}}
\nonumber
\ee
\label{cor:fibonbnn-1:2}
\end{corollary}
\Proof. \ \
Since $t(n)$ is a periodic continued fraction it can be written in terms of
itself as follows:
\be
t(n)=[2,\underbrace{1,1,\ldots,1}_{n-2 \ \mbox{\scriptsize ones}},t(n)^{-1}]
\label{eq:t(n)recursive}
\ee

Using lemma~(\ref{lemma:fibonexp}) to simplify this:
\begin{eqnarray}
t(n) & = & \cfrac{1}{2+\cfrac{f_{n-3}t(n)^{-1}+f_{n-4}}{f_{n-2}t(n)^{-1}+f_{n-3}}}
 = \cfrac{f_{n-2}+f_{n-3}t(n)}
{2(f_{n-2}+f_{n-3}t(n))+f_{n-3}+f_{n-4}t(n)} \cr
 & = &
\cfrac{f_{n-2}+f_{n-3}t(n)}
{2f_{n-2}+f_{n-3}+(2f_{n-3}+f_{n-4})t(n)} 
 = \cfrac{f_{n-2}+f_{n-3}t(n)}{f_{n}+f_{n-1}t(n)} 
\label{eq:t(n)simp}
\end{eqnarray}
where the fact that $2f_{k-1}+f_{k-2}=f_{k-1}+(f_{k-1}+f_{k-2})
=f_{k-1}+f_{k}=f_{k+1}$ has been used to obtain the final expression.
A rearrangement of equation~(\ref{eq:t(n)simp}) yields the appropriate
quadratic equation of which $t(n)$ is a solution:
\be
f_{n-1}t(n)^2+(f_{n}-f_{n-3})t(n)-f_{n-2}=0
\label{eq:t(n)quad}
\ee
Note that $f_{n}-f_{n-3}=(f_{n-1}+f_{n-2})+(f_{n-2}-f_{n-1})=2f_{n-2}$.
Equation~(\ref{eq:t(n)quad}) has only one positive solution which must be $t(n)$:
\be
t(n) = \cfrac{-f_{n-2}+\sqrt{f_{n-2}^2+f_{n-1}f_{n-2}}}{f_{n-1}}
= \cfrac{-f_{n-2}+\sqrt{f_{n-2}f_{n}}}{f_{n-1}}
\label{eq:t(n)soln}
\ee

\hfill $\Box$

Substituting the result of corollary~(\ref{cor:fibonbnn-1:2}) into that of
corollary~(\ref{cor:fibonbnn-1:1}) and proceeding through much algebra, 
the result for $b_{(n,n-1)}$ is obtained as follows:
\begin{eqnarray}
\lefteqn{b_{(n,n-1)} = \left( \cfrac{t(n)}{f_{n-2}+f_{n-3}t(n)} \right)^{2\beta}}
\nonumber \\
 & = & \left(  \cfrac{-f_{n-2}+\sqrt{f_{n-2}f_{n}}}
{f_{n-2}f_{n-1}-f_{n-3}f_{n-2}+f_{n-3}\sqrt{f_{n-2}f_{n}}}  \right)^{2\beta}
\nonumber \\
 & = & \left(  \cfrac{-f_{n-2}+\sqrt{f_{n-2}f_{n}}}
{f_{n-2}f_{n-1}-f_{n-3}f_{n-2}+f_{n-3}\sqrt{f_{n-2}f_{n}}} 
\cdot \cfrac{f_{n-2}+\sqrt{f_{n-2}f_{n}}}
{f_{n-2}+\sqrt{f_{n-2}f_{n}}} \right)^{2\beta}
\nonumber \\
 & = &  \left(  \cfrac{f_{n-2}f_{n-1}}
{f_{n-2}f_{n-1}^2 + f_{n-2}f_{n-1}\sqrt{f_{n-2}f_{n}}} \right)^{2\beta}
\nonumber \\
 & = &  \left(  \cfrac{f_{n-2}f_{n-1}}
{f_{n-2}f_{n-1}(f_{n-1}+\sqrt{f_{n-2}f_{n}})} \right)^{2\beta}
\nonumber \\
 & = & \left(f_{n-1}+\sqrt{f_{n-2}f_{n}}\right)^{-2\beta}
\label{eq:bnn-1proof}
\end{eqnarray}
This completes the proof of proposition~(\ref{prop:bnn-1}).
\hfill $\Box$

It is now of interest to examine the behaviour of $b_{(n,n-1)}$ as
$n$ approaches infinity.  This requires knowledge of how Fibonacci
numbers behave for large $n$.  The appendices contain a derivation
of the following expression for $f_n$:
\be
f_n=\frac{1}{1+\rho_g^2} \left(\rho_g^{n+2}+\frac{(-1)^n}{\rho_g^n} \right)
\label{eq:fibform}
\ee
Since $\rho_g>1$ it follows that
\be
f_n \sim \frac{\rho_g^{n+2}}{1+\rho_g^2} \ \ \mbox{as} \ \ n \rightarrow \infty
\label{eq:fibasy}
\ee
So, using equation~(\ref{eq:fibform}) and proposition~(\ref{prop:bnn-1}),
the asymptotic behaviour of $b_{(n,n-1)}$ is found as follows:
\begin{eqnarray}
b_{(n,n-1)} & = & \left(f_{n-1}+\sqrt{f_{n-2}f_{n}}\right)^{-2\beta}
\nonumber \\
 & \sim & \left(\frac{\rho_g^{n+1}}{1+\rho_g^2}
+\sqrt{\frac{\rho_g^{n}}{1+\rho_g^2}\frac{\rho_g^{n+2}}{1+\rho_g^2}}\right)^{-2\beta}
\nonumber \\
 & = & \left(2\frac{\rho_g^{n+1}}{1+\rho_g^2}\right)^{-2\beta}
\ \ \mbox{as} \ \ n \rightarrow \infty
\label{eq:bnn-1asy}
\end{eqnarray}
Recalling that $b_{(n,n)}=\frac{1}{n}\rho_g^{-2n\beta}$ it is now observed that
\be
b_{(n,n-1)} \sim  n \left( \frac{1+\rho_g^2}{2\rho_g}\right)^{2\beta} b_{(n,n)} 
\label{eq:bcomp}
\ee
in the limit of large $n$.
Thus, as was mooted earlier, $b_{(n,n)}$ is not the sole dominant term of $a_n$ in
the limit $n \rightarrow \infty$.  
Indeed, equation~(\ref{eq:bcomp}) shows that it is negligible in comparison with
$b_{(n,n-1)}$.  The insight gained here is that any sequence with enough
`ones' will contribute to $a_n$ in the limit. 
  
\subsubsection{The Main Proof}

It is useful to divide positive integer sequences whose entries sum to $n$ 
into two main groups.  These categories
will be referred to as Type I and Type II.  Type I sequences are those
which have at least $\cne$ 
entries being equal to 1 where $0 < \epsilon < 1$ and the expression
$[i]$ denotes the integer part of $i$.  Type II sequences
will be all the rest.  Note that $n \leq \cne+\en+1 \leq n+1$.

Each $a_n$ will then be estimated by considering upper bounds on the number of each
type of sequence, $N(I)$ and $N(II)$ respectively, 
and upper bounds on the continued fraction products corresponding to
these sequences.

Firstly, consider type I sequences and assume that $\beta <0$.
Lemma~(\ref{lemma:goldmeanbound}) and 
remark~(\ref{rem:rhonbound}) show that 
the simplest bound on each Type I continued fraction product is $\rho_g^{2n|\beta|}$.
A sufficiently useful bound on the number of these sequences is not so trivial
and will require some work.  A typical Type I 
sequence may be represented as follows\footnote{Note that there is no loss 
of generality caused by assuming that the first
entry is $i_1$ as all sequences starting with a 1 may be cyclically permuted
to obtain a first entry greater than 1.  This equivalence means the two
cycles give the same continued fraction product term.}
\be
\{ i_1,\underbrace{1,1,\ldots,1}_{N_1},
i_2,\underbrace{1,1,\ldots,1}_{N_2},
i_3,1,\ldots,1,i_k,\underbrace{1,1,\ldots,1}_{N_k}\}
\label{eq:seqexample}
\ee
where $\sum_{j=1}^{k} N_k = \cne$ and $\sum_{j=1}^{k} i_k = n-\cne$.
Using lemma~(\ref{lemma:numpart}), there are ${n-\cne-1 \choose k-1}$ different ways of
partitioning the $\{i_j\}$.  Since the $N_i$ represent lengths of strings of 1's,
they may be equal to 0.  A simple extension of lemma~(\ref{lemma:numpart}) 
shows that the number of different partitions of $\cne$ into the $N_j$ is 
${\cne+k-1 \choose k-1}$.  Note that overcounting has occurred as cyclically
equivalent sequences have been separately counted in this process.  
Also, the number of entries of each sequence must be at least $\cne+1$ 
and at most $n$ when all are equal to 1, i.e. $ 1 \leq k \leq n-\cne$.
This leads to the following estimate:
\begin{eqnarray}
N(I) & < & \sum_{k=1}^{n-\cne} {\cne+k-1 \choose k-1} {n-\cne-1 \choose k-1} \cr
 & < & \sum_{k=1}^{n-\cne} {\cne+k-1 \choose k-1}
\cdot \sum_{k=1}^{n-\cne} {n-\cne-1 \choose k-1}
\label{eq:numtypeIa}
\end{eqnarray}
Relabeling the index $k \rightarrow k+1$, using $n-\cne \leq \en+1$ and
noting that ${n+j-\cne \choose j-1}<{n+k-\cne \choose k-1}$ for 
all $j<k$, it follows from equation~(\ref{eq:numtypeIa}) that 
\begin{eqnarray}
N(I) & < & (\en+1){n-1 \choose n-\cne-1} 
\cdot \sum_{k=0}^{n-\cne} {n-\cne \choose k} \cr
 & = & (\en+1){n-1 \choose n-\cne-1} 2^{n-\cne}
\label{eq:numtypeIb}
\end{eqnarray}
For large $\epsilon n$, $\en$ may be approximated  by $\epsilon n$ 
( note that $\epsilon$ is 
fixed).  Using Stirling's approximation
that $n! \sim n^n e^{-n}$, a final estimate is obtained:
\be
N(I) \lapprox  \frac{\epsilon n 2^{\epsilon n}}
{\epsilon^{\epsilon n} (1-\epsilon)^{(1-\epsilon)n}}  
\mbox{ for large $n$. }
\label{eq:numtypeIc}
\ee

The number of Type II sequences will be estimated by the total number of 
sequences, $2^n$.  The main work of this section is therefore to calculate a
suitable upper bound for Type II orbit products.  This will be done by showing
that the orbit product of any sequences with less that $\cne$ `ones' is bounded by 
the orbit product of some sequence containing $\cne$ `ones' and $\enb$ `twos'. 
An appropriate bound on the orbit product of any sequence of the latter type 
will then be found.

Firstly, the orbit product
of a purely periodic continued fraction of period $m$ is found to have the following 
representation. This is developed in the appendices in section~(\ref{Asec:pcfracs}).
\begin{lemma}
The orbit product of a purely periodic continued fraction
\bdm
C = [\overline{i_0;i_1,\ldots,i_{m-1}}]
\edm
is given by
\bdm
\prod_{j=0}^{m-1} \mu_j 
= \frac{(A_{m-1}+B_{m-2})+\sqrt{(A_{m-1}+B_{m-2})+4(-1)^{m-1}}}{2}
\edm
\label{lemma:orbitcalc}
\end{lemma}
\Proof. \ \
See lemma~(\ref{Alemma:orbitprod}) and corollary~(\ref{Acor:orbitprod}) in
the appendices. 
\hfill
$\Box$

Note that by derivation the $A_j$ and $B_j$ are calculated from $\mu_0$.
The first step is to show a form of ordering for these orbit products:
\begin{lemma}
Consider two sequences $\{i_j\}_{k=0}^{m-1}$ and $\{i_j'\}_{k=0}^{m}$
where $i_j=i_j'$ for all $k=0,\ldots,m-2$ and $i_{m-1}=i_{m-1}'+i_{m}'$.
Note that all entries must be greater than or equal to 1 so $i_{m-1}$ must 
be at least 2.
Then
\bdm
\prod_{j=0}^{m-1} [\overline{i_{j};i_{j+1},\ldots,i_{m-1},i_{0},\ldots,i_{j-1}}] 
  < 
\prod_{j=0}^{m} [\overline{i_{j}';i_{j+1}',\ldots,i_{m}',i_{0}',\ldots,i_{j-1}'}] 
\edm
i.e., 
\bdm
\prod_{j=0}^{m-1} \mu_j < \prod_{j=0}^{m} \mu_j'
\edm
\label{lemma:ordering}
\end{lemma}
\Proof.
Let $i_m'=k$ and $i_{m-1}'=i_{m-1}-k$ and so $1 \leq k \leq i_{m-1}-1$.
The aim of the proof is to show that the quantity
$p(k)=\prod_{j=0}^{m} \mu_j'-\prod_{k=0}^{m-1} \mu_j$ is strictly greater than 0.
Since the first $m-2$ entries for the continued fractions $\mu_0$ and $\mu_0'$ 
are the same, it is clear that
$B_k=B_k'$ and $A_k=A_k'$ for $0 \leq k \leq m-2$.
Using the recursion relations for the $A_k$ and $B_k$ it is then seen
that $B_{m-1}'=(i_{m-1}-k)B_{m-2}+B_{m-3}$, $A_{m-1}'=(i_{m-1}-k)A_{m-2}+A_{m-3}$
and $A_{m}'=(k(i_{m-1}-k)+1)A_{m-2}+kA_{m-3}$.
Therefore,
\begin{eqnarray}
A_m'+B_{m-1}' & = & (a_{m-1}-k)B_{m-2}+B_{m-3}+ (k(i_{m-1}-k)+1)A_{m-2}+kA_{m-3} \cr
 & = & B_{m-1}+A_{m-2}+k(A_{m-1}-B_{m-2})-k^2 A_{m-2} \equiv q(k)
\label{eq:order1}
\end{eqnarray}
These observations together with 
lemma~(\ref{lemma:orbitcalc}) provide the following expression
\begin{eqnarray}
\lefteqn{2p(k)=2\prod_{j=0}^{m} \mu_j'-2\prod_{k=0}^{m-1} \mu_j} \cr
 & = & q(k)-(A_{m-1}+B_{m-2}) \cr
 & + & \sqrt{q(k)^2+4(-1)^{m}}-\sqrt{(A_{m-1}+B_{m-2})+4(-1)^{m-1}}
\label{eq:order2}
\end{eqnarray}
Now consider $p(x)$ where $x \in [1,a_{m-1}-1]$.  Note that since $p(1)$ and
$p(a_{m-1}-1)$ must exist with a non-zero surd term 
and $q(k)$ has a single maximum, the square root
term must be well defined and non-zero for $x$ in this range.  
The derivative of $p(x)$ is given by
\be
2p'(x)=q'(x)\left( 1+\frac{q(x)}{\sqrt{q(x)^2+4(-1)^{m}}} \right)
\label{eq:order3}
\ee
where $q'(x)=A_{m-1}-B_{m-2}-2k A_{m-2}$.
The term in brackets can never equal zero and $q'(x)$ has one zero
at $x=\frac{A_{m-1}-B_{m-2}}{2A_{m-2}}$.  This corresponds to a sole 
maximum which may or may not be in the interval $[1,a_{m-1}-1]$.  Whatever
the case, the minimum value of $p(x)$ on $[1,a_{m-1}-1]$ must occur at 
one or both of the end points of the interval.  To complete the proof, it remains then 
to show that $p(1)$ and $p(a_{m-1}-1)$ are strictly greater than zero.

Now, $q(1)=B_{m-1}-B_{m-2}+A_{m-1}$, so
\begin{eqnarray}
2p(1) & = & B_{m-1}-2B_{m-2} \\
 & + & \sqrt{(B_{m-1}-B_{m-2}+A_{m-1})^2+4(-1)^{m}} 
- \sqrt{(A_{m-1}+B_{m-2})^2+4(-1)^{m-1}}  \nonumber
\label{eq:order4}
\end{eqnarray}
The cases $m=1$ and $m \geq 2$ will be examined separately.
Consider the first term in the above.
Since $B_{m-1}=i_{m-1}B_{m-2}+B_{m-3}$ for all $m \geq 2$ 
and $i_{m-1} \geq 2$ it follows that $B_{m-1}-2B_{m-2} \geq B_{m-3} \geq 0$.  

For $m\geq 2$ the difference of the square roots in equation~(\ref{eq:order4}) can be
shown to be greater than zero by examining the difference of the arguments
of these square roots. This difference simplifies to
\be
d=(2A_{m-1}+B_{m-1})(B_{m-1}-2B_{m-2})+8(-1)^{m}
\label{eq:order5}
\ee
Now, $B_{m-1}-2B_{m-2} \geq 0$ with equality only holding when $m=2$ and
$i_1=2$.  Clearly, $d \geq 8 > 0$ when $m$ is even.  For $m$ odd, note first that the 
minimum value of $(2A_{m-1}+B_{m-1})(B_{m-1}-2B_{m-2})$ increases with $m$.
Thus, the case $m=3$ will provide enough evidence for all odd $m$.  For $m=3$, 
equation~(\ref{eq:order5}) becomes explicitly
\be
d=(2i_2+2i_0+2i_0i_1i_2+i_1i_2+1)(i_1(i_2-1)+1)-8
\label{eq:order6}
\ee
Noting that $i_2\geq 2$ and $i_0,i_1 \geq 1$, it is clear that $d>0$.

Finally, for $m=1$, equation~(\ref{eq:order4}) reduces to
\begin{eqnarray}
2p(1) & = & B_{0}-2B_{-1}+\sqrt{(B_{0}-B_{-1}+A_{0})^2-4} \cr
 & - & \sqrt{(A_{0}+B_{-1})^2+4}=1+\sqrt{(1+i_0)^2-4}-
\sqrt{i_0^2+4}
\label{eq:order7}
\end{eqnarray}
By inspection, the minimum of $2p(1)$ for $m=1$ must occur when
$i_0$ is smallest, i.e. when $i_0=2$.  This gives $2p(1)=1+\sqrt{5}-\sqrt{8}>0$.
Therefore $p(1)$ is always strictly greater than zero.  A similar line of reasoning
shows the same is true for $p(i_{m-1}-1)$ completing the proof. 
\hfill
$\Box$

An immediate generalization is the following
\begin{corollary}
Given the orbit product corresponding to an ordered partition of $n$, any further
partitioning will give an orbit product strictly larger than the original.
I.e., given a partition (sequence) $\{i_0,i_2,\ldots,i_{m-1}\}$ and a finer
partition 
\bdm
\{i_{01},i_{02},\ldots,i_{0n_1},i_{11},i_{12},\ldots,i_{1n_2},
\ldots,i_{m-11},i_{m-12},\ldots,i_{m-1n_{m-1}}  \}
\edm
where $\sum_{l=1}^{n_j} i_{jl}=i_j$, then the latter has
a strictly larger orbit product.
\label{cor:ordering}
\end{corollary}
Note that this corollary is a much stronger result than previously shown as it gives
lemma~(\ref{lemma:goldmeanbound}) straight away.  Corollary~(\ref{cor:ordering})
also provides that the orbit product of every Type II sequence is bounded
by the orbit product of a sequence that contains $2\cneb$ `ones' and $\enb$ `twos'.
A bound on the latter quantity is now determined.
\begin{lemma}
Given $\{i_j\}_{j=0}^{m-1}$,
a sequence of $m_1$ 1's and $m_2$ 2's in any order with $m_1,m_2>0$
and $m_1+m_2=m$,
then the following holds:
\bdm
\prod_{j=0}^{m-1} \mu_j=
\prod_{j=0}^{m-1} [\overline{i_j;\ldots,i_{m-1},i_0,\ldots,i_{j-1}}]
< \rho_g^{m_1}\left( \frac{5}{2} \right)^{m_2}
\edm
where $\rho_g = [\overline{1};] = [\overline{1}]^{-1}$ is the golden ratio.
\label{lemma:lessthan1n2k}
\end{lemma}
\Proof.
The proof proceeds along similar lines to that of lemma~(\ref{lemma:goldmeanbound});
basic building blocks of inequalities are found which can then be pieced together 
to provide the overall inequality.  Consider any orbit product based on
a sequence of 1's and 2's.  The indices of this product may be cyclically
permuted so that $\mu_0$ has a 1 as its first entry and a 2 as its last entry
in the periodic block.  The sequence corresponding to this $\mu_0$ can
then be broken down into $M$ blocks containing a string of 1's followed by a
string of 2's.  I.e.
\be
\mu_0=[\overline{\underbrace{1;\ldots,1,2,\ldots,2}_{\mbox{\scriptsize block 1}},
\underbrace{1,\ldots,1,2,\ldots,2}_{\mbox{\scriptsize block 2}},\ldots,
\underbrace{1,\ldots,1,2,\ldots,2}_{\mbox{\scriptsize block $M$}}}]
\ee
Consider now the continued fraction
\be
t = [\underbrace{1;1,\ldots,1}_{\mbox{\scriptsize $N_1$ `ones'}},
\underbrace{2,2,\ldots,2}_{\mbox{\scriptsize $N_2$ `twos'}},\mu_N]
\ee
where $N=N_1+N_2$ and the first entry of $\mu_N$ is a 1 and 
also the partial orbit product
\begin{eqnarray}
\lefteqn{\prod_{j=0}^{N-1} \mu_j = \prod_{j=1}^{N_1} 
[\underbrace{1;1,\ldots,1}_{\mbox{\scriptsize $N_1-j$ `ones'}},\mu_{N_1}]} \cr
 & \times & \prod_{j=1}^{N_2} 
[\underbrace{2;2,\ldots,2}_{\mbox{\scriptsize $N_2-j$ `twos'}},\mu_N]
\end{eqnarray}
Now, a similar result to lemma~(\ref{lemma:fibonexp}) shows
that $\mu_{N_1}=\frac{g_{N_2}\mu_N+g_{N_2-1}}{g_{N_2-1}\mu_N+g_{N_2-2}}$.
The $g_n$ are related to the `silver mean', $\rho_s$, which
is equal to $\sqrt{2}+1=[\overline{2};]$ and
satisfies the equation $x=2+\frac{1}{x}$.  The $g_n$ also
follow the recursion relation $g_n=2g_{n-1}+g_{n-2}$ and the first few
are given by $g_0,g_1,g_2,g_3,\ldots=1,2,5,12,\ldots$.

In a similar fashion to the proof of corollary~(\ref{cor:fibonbnn-1:1})
the product simplifies to
\be
\prod_{j=0}^{N-1} \mu_j
=f_{N_1}g_{N_2}+f_{N_1-1}g_{N_2-1} + \frac{f_{N_1}g_{N_2-1}+f_{N_1-1}g_{N_2-2}}{\mu_N}
\ee
Note that the smallest continued fraction that can be made out of 1's and
2's is $[\overline{1,2}]=\frac{1+\sqrt{3}}{2}$.  So, $\mu_N \geq
\frac{1+\sqrt{3}}{2}$ and therefore $\frac{1}{\mu_N} \leq \sqrt{3}-1$.
The product is then bounded in the following way
\be
\prod_{j=0}^{N-1} \mu_j \leq
f_{N_1}g_{N_2}+f_{N_1-1}g_{N_2-1} + (\sqrt{3}-1)(f_{N_1}g_{N_2-1}+f_{N_1-1}g_{N_2-2})
\label{eq:nastyguy}
\ee
The proof now proceeds by the method of induction.  Assume that
\be
\prod_{j=0}^{N-1} \mu_j < \rho_g^{N_1} \left( \frac{5}{2} \right)^{N_2}
\label{eq:nastyguy2}
\ee
holds for $N_1=k$ and $N_1=k-1$.  Then for $N_1=k+1$, it follows easily 
using the relationship $f_{k+1}=f_{k}+f_{k-1}$ that 
\begin{eqnarray}
\lefteqn{f_{k+1}g_{N_2}+f_{k}g_{N_2-1} + 
(\sqrt{3}-1)(f_{k+1}g_{N_2-1}+f_{k}g_{N_2-2})} \cr
 & < & \rho_g^{k} \left( \frac{5}{2} \right)^{N_2} + \rho_g^{k-1} 
\left( \frac{5}{2} \right)^{N_2} 
 = \rho_g^{k+1} \left( \frac{5}{2} \right)^{N_2}
\end{eqnarray}
The same procedure also shows that if the assertion is true for $N_2=k$ and
$N_2=k-1$, it is true for $N_2=k+1$.
\begin{eqnarray}
\lefteqn{f_{N_1}g_{k+1}+f_{N_1-1}g_{k} + 
(\sqrt{3}-1)(f_{N_1}g_{k}+f_{N_1-1}g_{k-1})} \cr
 & < & 2\rho_g^{N_1} \left( \frac{5}{2} \right)^{k} + \rho_g^{N_1} 
\left( \frac{5}{2} \right)^{k-1} \cr
 & = & \rho_g^{N_1} \left( \frac{5}{2} \right)^{k}\left( 2+\frac{1}{2.5} \right)
 < \rho_g^{N_1} \left( \frac{5}{2} \right)^{k}\left(2+\frac{1}{\rho_s} \right) \cr
 & = & \rho_g^{N_1} \left( \frac{5}{2} \right)^{k}\rho_s <\rho_g^{N_1} 
\left( \frac{5}{2} \right)^{k+1} 
\end{eqnarray}
Finally, it remains to check that the claim is true for $(N_1,N_2)=
(1,1),(1,2),(2,1)$ and $(2,2)$.  Substituting the values for the $f_n$
and the $g_n$ into equation~(\ref{eq:nastyguy}) when $(N_1,N_2)=(1,1)$:
\bdm
\prod_{j=0}^{N-1} \mu_j \leq
3+(\sqrt{3}-1)(1)=\sqrt{3}+2 \approx 3.73 < \rho_g^1 
\left( \frac{5}{2} \right)^1 \approx 3.91
\edm
Similarly, $3\sqrt{3}+4 \approx 9.20 < \rho_g^1 
\left( \frac{5}{2} \right)^2 \approx 9.43$,
$2\sqrt{3}+3 \approx 6.46 < \rho_g^2 \left( \frac{5}{2} \right)^1 \approx 6.54$ and
$3\sqrt{3}+4 \approx 15.7 < \rho_g^2 \left( \frac{5}{2} \right)^2 \approx 16.4$.
So, by induction, the inequality in equation~(\ref{eq:nastyguy2}) is 
true for all $N_1,N_2 \geq 1$.  By joining the inequalities for each of the
blocks together, the lemma is seen to be true.
\hfill
$\Box$

A corollary to lemmas~(\ref{lemma:ordering}) and~(\ref{lemma:lessthan1n2k})
is the following:
\begin{corollary}
All Type II periodic continued fraction products are bounded above by the 
quantity
\bdm
\rho_g^{2 \cneb |\beta|}\left( \frac{5}{2} \right)^{2 \enb |\beta|}
\edm
\label{cor:alllessthan1n2k}
\end{corollary}

For fixed $\epsilon$ and $\beta$, the $a_n$ may now be estimated by 
adding the bounds on the number of Type I  and Type II sequences 
multiplied by the bounds on Type I and Type II orbit products respectively.
I.e.,
\begin{eqnarray}
a_n & < & N(I) \rho_g^{2n\beta} + N(II) \rho_g^{2 \cneb \beta}
\left( \frac{5}{2} \right)^{2 \enb \beta} \cr
 & = & \frac{\epsilon n}
{\epsilon^{\epsilon n} (1-\epsilon)^{(1-\epsilon)n}}  2^{\epsilon n}
\rho_g^{2n|\beta|} + 2^n \rho_g^{4 \cneb |\beta|}
\left( \frac{5}{2} \right)^{2 \enb |\beta|}
\label{eq:final1}
\end{eqnarray}
Taking the $n^{\th}$ root of this estimate and using the fact that
$|a+b|^{\frac{1}{n}}<|a|^{\frac{1}{n}}+|b|^{\frac{1}{n}}$,
it is observed that
\begin{eqnarray}
|a_n|^{\frac{1}{n}} & < & \left( \frac{\epsilon n}
{\epsilon^{\epsilon n} (1-\epsilon)^{(1-\epsilon)n}}  2^{\epsilon n}
\rho_g^{2n|\beta|} \right)^{\frac{1}{n}} 
 + \left( 2^n \rho_g^{4 \cneb |\beta|}\left( \frac{5}{2} \right)^{2 \enb |\beta|} 
\right)^{\frac{1}{n}}  \cr
 & \rightarrow & \frac{2^{\epsilon }}
{\epsilon^{\epsilon}(1-\epsilon)^{(1-\epsilon)}}  
\rho_g^{2|\beta|} + 2 \rho_g^{2 (1-\epsilon) |\beta|}
\left( \frac{5}{2} \right)^{\epsilon |\beta|}
\mbox{ as $n \rightarrow \infty$}
\label{eq:final2}
\end{eqnarray}
Thus, a lower bound for the radius of convergence of the induced zeta function
has been obtained.  Finally, the logarithm of the above gives the following
bound on the pressure function:
\be
-\Fb - 2|\beta| \log \rho_g < \log \left( \frac{2^{\epsilon }}
{\epsilon^{\epsilon}(1-\epsilon)^{(1-\epsilon)}}  
+\left(\frac{5}{2\rho_g^2} \right)^{\epsilon |\beta|}
\right)
\label{eq:final3}
\ee

Let $r(\epsilon)=\frac{2^{\epsilon }}{\epsilon^{\epsilon}(1-\epsilon)^{(1-\epsilon)}}$.
Note that $r(\epsilon) \rightarrow 1$ as $\epsilon \rightarrow 0$. In
particular, $r(\epsilon)>1$  for all $\epsilon \in (0,1]$.  Therefore, for any
$\delta>0$, there exists a value $\epsilon(\delta)$ such that for all
$\epsilon < \epsilon(\delta)$, $r(\epsilon) < 1+\delta/2$.  Now set
$\epsilon=\epsilon(\delta)/2$.  Consider now the other term in the logarithm,
$\left(\frac{5}{2\rho_g^2} \right)^{\epsilon |\beta|}$.  Since 
$\frac{5}{2\rho_g^2} < 1$, this term will approach zero for fixed $\epsilon$ and
$\beta \rightarrow -\infty$.
Clearly then, there exists a 
$\beta(\delta,\epsilon(\delta))=\beta(\delta)$ such that for all
$\beta<\beta(\delta)$
\bdm
\left(\frac{5}{2\rho_g^2} \right)^{\frac{\epsilon(\delta)}{2} |\beta|} 
< \frac{\delta}{2}
\edm 
Returning to equation~(\ref{eq:final3}), these observations show that
for all $\beta<\beta(\delta)$
\be
-\Fb - 2|\beta| \log \rho_g < 
\log \left( 1+\frac{\delta}{2}+\frac{\delta}{2} \right)
=\log ( 1+\delta ) < \delta
\label{eq:final4}
\ee
This completes the proof of theorem~(\ref{theorem:pfbeta-infty}).
\hfill $\Box$

\begin{figure}[tp!]
  \includegraphics[width=\textwidth]{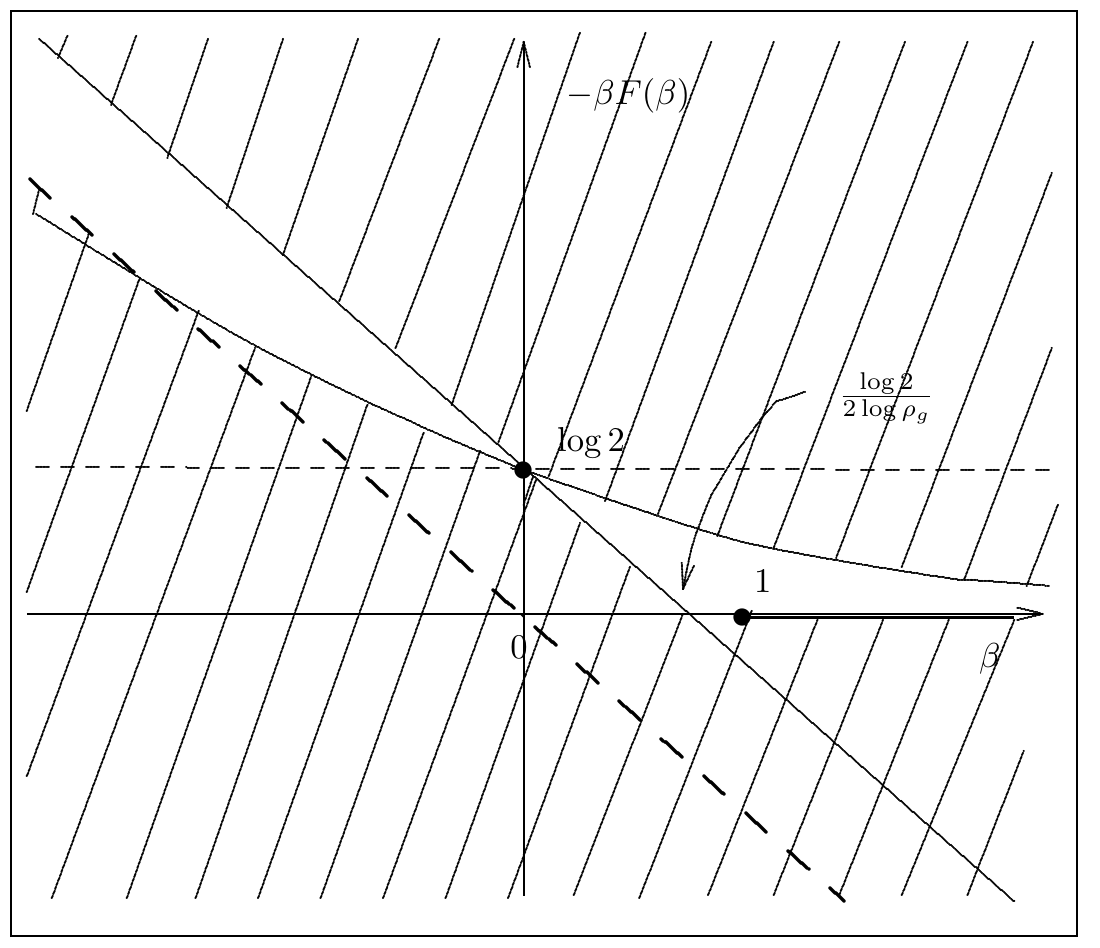}
  \caption{All bounds on the pressure function}
  \label{fig:bound4}
\end{figure}

All of the information obtained regarding the pressure function thus far,
see figures~(\ref{fig:bound1}),(\ref{fig:bound2}) and (\ref{fig:bound3}),
is displayed in figure~(\ref{fig:bound4}).
The bold dashed line indicates the asymptote $y=-2\beta\log 2$.

\clearpage

\subsection{The phase transition at $\beta=1$}
\label{subsec:pfphaset}
It has been demonstrated that a phase transition exists at some value of $\beta$
in the interval $(0,1]$.  It has in fact been shown for a
certain class of maps of the interval in~\cite{prelslawn:interm} that
the critical value of $\beta$ is actually 1.  The proof uses the induced transfer
operator and the fact that this has a simple leading eigenvalue.  Perturbation
theory of simple eigenvalues, see~\cite{kato:pert}, 
may then be applied to show that the pressure function
is analytic for $0 \leq \beta < 1$.  This implies that the non-analytic point
must occur at $\beta=1$.  The asymptotic behaviour of the pressure function
is also found for this class of maps 
and is reproduced here for the special case of the Farey map: 
\be
-\Fb \sim \frac{1-\beta}{-\log(1-\beta)} \mbox{ as } \beta \rightarrow 1^{-}
\label{eq:scaling}
\ee

\section{Concluding Remarks}
\label{sec:conclusion2}

\begin{figure}[tp!]
  \includegraphics[width=\textwidth]{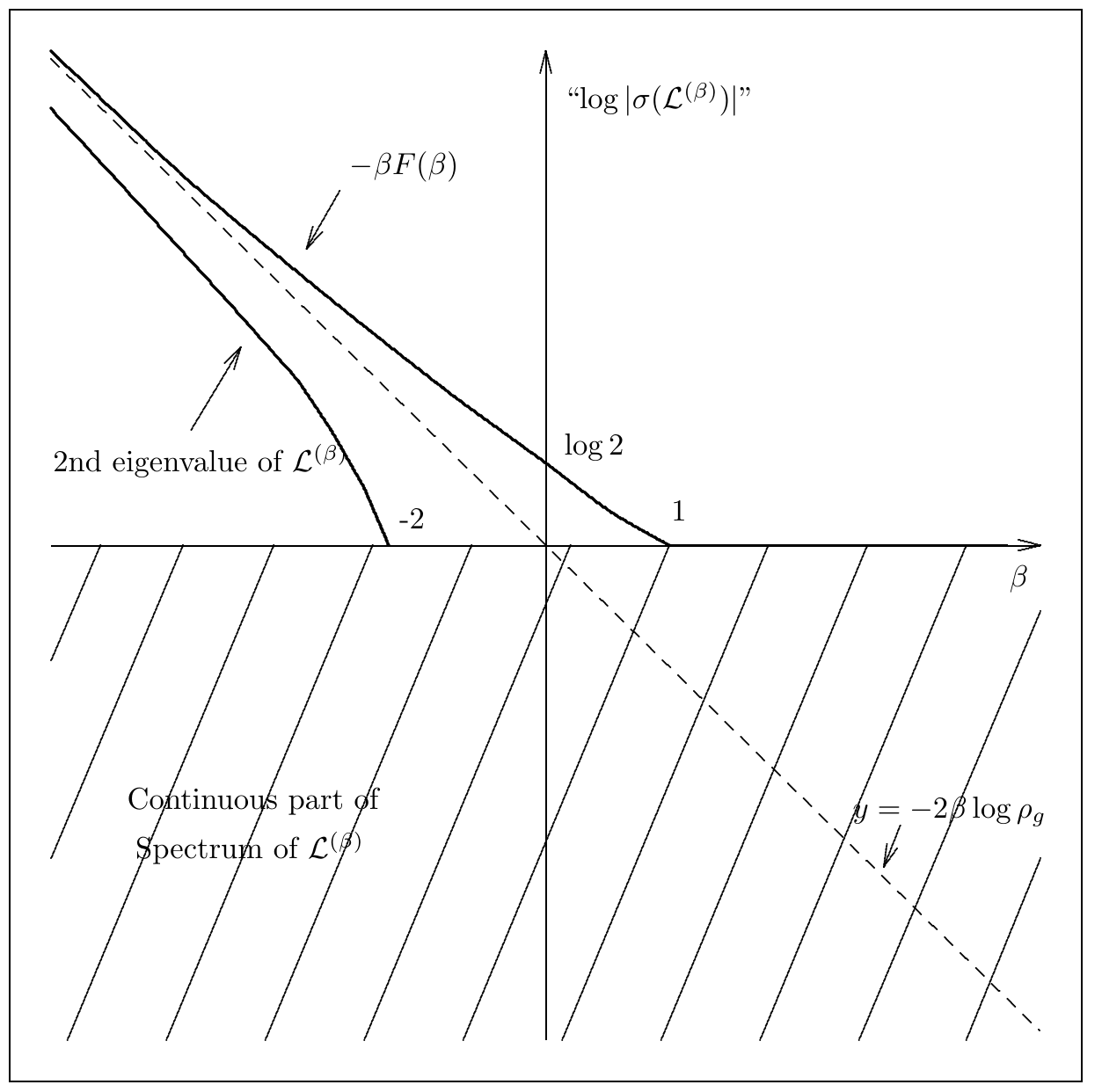}
  \caption{The pressure function obtained numerically}
  \label{fig:pressure}
\end{figure}

The constraints found on the pressure function are enough to give a very
solid picture of its shape.  In particular, the scaling behaviour at the 
phase transition and the asymptotic behaviour for 
large negative $\beta$ have been demonstrated analytically. 
Moreover, the findings are seen to agree with
previous numerical calculations of the pressure function, (\cite{dodds:honsthesis}).
Numerical observations
have identified that for the Farey map there are
no other eigenvalues with a magnitude greater than 1 for all $-2 \leq \beta < 1$.
For $-7 < \beta < -2$, it has been observed that a single secondary eigenvalue appears.
This structure is qualitatively represented in figure~(\ref{fig:pressure}). (Note
that the figure displays the logarithm of the magnitude of each element of the
spectrum of the transfer operator, the largest such value corresponding to the 
pressure function).

It was also shown analytically in~\cite{prelslawn:interm} that there exists 
a spectral gap in the spectrum of the transfer operator for all $0\leq\beta<1$.
The correlation length of the system may be obtained from the distance of the
leading eigenvalue (i.e., the exponential of the pressure function) to the 
second largest eigenvalue.
So while there is knowledge of a spectral gap, it is the secondary spectra of the
transfer operator which will provide exact details of correlations in the system.
This structure has proved to be very resistant to analytic investigation.
Note that in the positive temperature range, 
i.e. $\beta > 0$, it is observed that the leading 
eigenvalue of the transfer operator is the only eigenvalue outside of the
unit disc of continuous spectrum (i.e. the essential radius).

The spin system created through the symbolic dynamics of the Farey map is seen
to freeze into one state when $\beta \geq 1$ or $T = 1/\beta \leq 1$.  This 
state is just the one with all spins being `up' which corresponds to an
infinite string of `ones'.  In terms of the map, this says that with probability
1, a typical orbit of the Farey map will stay near the indifferent 
fixed point forever.  In terms of intermittent behaviour, the `signal' has become
completely smooth and regular.  So in this toy model of intermittency, it is
observed that there is a sudden transition to laminar behaviour, reflected in
the singular nature of the invariant density of the original Farey map.

Finally, it is remarked that the counting and ordering 
exercises involved are interesting in their own right.
In particular, the reader's attention is drawn to conjecture~(\ref{Acon:pcfineq})
in the first appendix.  This certainly stands by itself as a worthy result
away from the main content of the thesis
and it is hoped that a proof may be found in the near future.

\newpage
\appendix

\phantomsection

\addcontentsline{toc}{chapter}{Appendices}

\chapter{Continued Fractions}

The theory presented here is standard work and is taken primarily
from the book of Rockett and Sz\"{u}sz,~\cite{rockett:contfracs}.

\section{General definitions}
\label{Asec:cfdefs}
In general, a continued fraction $C$ is the following entity:
\be
C=i_0 + \cfrac{1}{i_1+\cfrac{1}{i_2+\cfrac{1}{i_3+\cfrac{1}{\cdots+\cfrac{1}{i_n}}}}}
=[i_0;i_1,i_2,\ldots,i_n]
\ee
where the $i_i$ are positive integers, $i_0$ may also be 0, and
$n \in N \bigcup \{\infty\}$.
A continued fraction with $i_0=0$ will be abbreviated as
$C = [i_1,i_2,\ldots,i_n] \ (=[0;i_1,i_2,\ldots,i_n])$.
Irrational numbers have an infinite continued fraction expansion while
the expansion for any rational terminates at some finite $n$.
The `convergents' of a continued fraction, $C_k$, 
naturally arise as the rational number given by the truncated
continued fraction
\be
C_k=\frac{A_k}{B_k}=
[i_0;i_1,i_2,\ldots,i_k]
\ee
where $k \leq n$, the number of terms in the expansion, and
the sequences $\{A_k\}$ and $\{B_k\}$ are generated by the
recursion relations
\be
\left[
\begin{array}{l}
A_{k+1}=i_{k+1}A_k+A_{k-1} \\
B_{k+1}=i_{k+1}B_k+B_{k-1}
\end{array}
\right.  \ \ \mbox{for} \ k=0,1,2,\ldots,n-1
\label{Aeq:cfrec1}
\ee
with the seed values $A_{-1}=1,A_0=i_0$ and $B_{-1}=0,B_0=1$.
It can be easily shown using these relations that
\be
A_{m}B_{m-1} - A_{m-1}B_{m} = (-1)^{m+1}
\label{Aeq:convcomm}
\ee

The $\mbox{k}^{\mbox{th}}$ complete quotient $\mu_k$ of a continued 
fraction $C=[i_0;i_1,i_2,\ldots]$ is defined by:
\be
\mu_k=[i_k;i_{k+1},i_{k+2},\ldots]
\label{Aeq:cfquot}
\ee
A simple proof by induction shows that 
\be
C=\mu_0=\frac{A_{k}\mu_{k+1}+A_{k-1}}{B_{k}\mu_{k+1}+B_{k-1}} \ \ \mbox{for $k \geq 0$}
\label{Aeq:Cmukform}
\ee

Finally, a small result used in the text is that the
terms of the sequence $\{\alpha_k\}=\{A_k+B_k\}$
are always positive and increasing.  To show this, note that
$\alpha_{-1}=1$ and $\alpha_0=1+i_0$ and that $\{\alpha_k\}$ obeys the same
recursion relation as $\{A_k\}$ and $\{B_k\}$:
\be
A_{k+1}+B_{k+1}=i_{k+1}(A_k+B_k)+(A_{k-1}+B_{k-1})
\label{Aeq:cfrec2}
\ee
Thus, for $k \geq 0$, $\alpha_{k+1}-\alpha_{k}=(i_{k+1}-1)\alpha_k + \alpha_{k-1}$ for
$k=0,1,\ldots,n-1$.  Since $i_{k+1}\geq 1$ for all $k \geq 0$ the first term
on the righthand side is $ \geq 0$.  By inspection, $\alpha_k$ is always
$> 0$ so the second term is always $>0$. Thus, the
terms must increase with the possible exception of the initial terms $\alpha_{-1}$
and $\alpha_0$ when $i_0=0$.

\section{Periodic Continued Fractions}
\label{Asec:pcfracs}

The fixed points of the induced Farey map are periodic
continued fractions.  These are simply those continued
fractions whose entries are periodic.  In particular, the
sequence of entries representing a periodic continued fraction,
$\{i_i\}$, is made up of a finite initial sequence
$\{i_0,i_1,\ldots,i_{n-1} \}$ followed by a
repeating sequence $\{i_n,i_{n+1},\ldots,i_{n+m-1}\}$ where
$i_{n+j+km}=i_{n+j}$ for all $0 \leq j \leq m-1$ and
$k=1,2,\ldots$.  If $n=0$, i.e. there is no preliminary sequence,
the continued fraction is said to be purely periodic.  Most of the
main work involves purely periodic continued fractions and only 
these will be considered here.

The major result regarding periodic continued fractions is a bijective map
between them and quadratic surds.  A quadratic surd is a solution
to a quadratic equation such that the discriminant $(b^2-4ac)$ is not
an perfect square.  This result is quoted as a theorem in the main body of the
text and a proof can be found in Rockett.

A useful result discovered by the author and subsequently found to be
a lemma on p.54 of ~\cite{rockett:contfracs}(!) is the following:
\begin{lemma}
Let $C=[\overline{i_0;i_1,\ldots,i_{m-1}}]$.  Then
\bdm
\prod_{k=0}^{m-1}\mu_k=\mu_0 \cdot \mu_1 \cdot \cdots \cdot \mu_{m-1}=B_{m-1}C+B_{m-2}
\edm
This product is referred to as an `orbit product'.
\label{Alemma:orbitprod}
\end{lemma}
\Proof. \ \
Note firstly that $C=\mu_0=\mu_m$. Therefore
\be
\prod_{k=0}^{m-1}\mu_k=\mu_0 \cdot \mu_1 \cdot \cdots \cdot \mu_{m-1}
= \mu_1 \cdot \mu_2 \cdot \cdots \cdot \mu_{m-1} \cdot \mu_m
\ee
In general, if it is assumed that $\mu_k=B_{k-1}\mu_k+B_{k-2}$ then
\begin{eqnarray}
\mu_k \cdot \mu_{k+1}
 & = & \left(B_{k-1}\mu_k+B_{k-2}\right)\cdot \mu_{k+1}
   = \left( B_{k-1}(i_k+\frac{1}{\mu_{k+1}})+B_{k-2} \right) \cdot \mu_{k+1} \cr
 & = & (i_k B_{k-1}+B_{k-2})\mu_{k+1} + B_{k-1}
   =  B_k \mu_{k+1} + B_{k-1}
\end{eqnarray}
Since $\mu_1=B_0 \mu_1 + B_{-1}$ it follows by induction that
\be
\prod_{k=0}^{m-1}\mu_k=B_{m-1} \mu_{m} + B_{m-2}=B_{m-1}C+B_{m-2}
\ee
\hfill
$\Box$

Using equation~(\ref{Aeq:Cmukform}), a purely periodic continued fraction with
period $m$ can be written in terms of itself as
\be
C=\frac{A_{m-1}C+A_{m-2}}{B_{m-1}C+B_{m-2}}
\ee
Rearranging this gives a quadratic equation for $C$ with the 
positive solution necessarily being $C$.
\be
C=\frac{A_{m-1}-B_{m-2} + \sqrt{(A_{m-1}-B_{m-2})^2+4A_{m-2}B_{m-1}}}{2B_{m-1}}
\label{Aeq:cfracsoln}
\ee
This observation allows for the following corollary to 
lemma~(\ref{Alemma:orbitprod}).
\begin{corollary}
The orbit product of a purely periodic continued fraction
\bdm
C = [\overline{i_0;i_1,\ldots,i_{m-1}}]
\edm
is given by
\bdm
\prod_{k=0}^{m-1}\mu_k=
\frac{(A_{m-1}+B_{m-2})+\sqrt{(A_{m-1}+B_{m-2})^2+4(-1)^{m-1}}}{2}
\edm
\label{Acor:orbitprod}
\end{corollary}
\Proof. \ \
Relationship~(\ref{Aeq:convcomm}) shows that 
\bdm
(A_{m-1}-B_{m-2})^2+4A_{m-2}B_{m-1}=(A_{m-1}+B_{m-2})^2+4(-1)^{m-1}
\edm
The corollary is then shown by inserting this into the
solution for $C$, equation~(\ref{Aeq:cfracsoln}), and then all
this into the result of lemma~(\ref{Alemma:orbitprod}):
\begin{eqnarray}
\prod_{k=0}^{m-1}\mu_k & = &
B_{m-1}\left( \frac{A_{m-1}-B_{m-2} + 
\sqrt{(A_{m-1}+B_{m-2})^2+4(-1)^{m-1}} }{2B_{m-1}} \right) + B_{m-2} \cr
 & = & \frac{A_{m-1}-B_{m-2} + 2B_{m-2} +
\sqrt{(A_{m-1}+B_{m-2})^2+4(-1)^{m-1}}}{2} \cr
 & = & \frac{(A_{m-1}+B_{m-2})+\sqrt{(A_{m-1}+B_{m-2})^2+4(-1)^{m-1}}}{2}
\end{eqnarray}
\hfill
$\Box$

Note that the orbit product depends only on one quantity $(A_{m-1}+B_{m-2})$.
Finally, a very helpful relationship would be the following inequality which, as yet, 
has not been proven and only verified numerically for small values of $m$ and $n$.
\begin{con}
Let $\{i_0,i_1,\ldots,i_{m-1}\}$ be a positive integer sequence with 
$\sum_{k=0}^{m-1} i_k = n$.  Then
\bdm
\prod_{k=0}^{m-1}[\overline{i_k;i_{k+1},\ldots,i_{m-1},i_0,\ldots,i_{k-1}}] 
\leq \prod_{k=0}^{m-1}[\overline{i_k;}]
\edm
with the equality only holding when $i_1=i_2=\cdots=i_{m-1}$.
\label{Acon:pcfineq}
\end{con}
This would certainly simplify the proof of
theorem~(\ref{theorem:pfbeta-infty}) as well as
being an elegant result in its own right. 

\chapter{Miscellaneous}
\label{Achap:misc}

\section{The spectral radius of $\LB_1$}
\label{Asec:radL1}
The operator $\LB_1$ was defined in chapter~(\ref{ch:induce}) by the relation
$\LB_{1}\psi=\LB(\chi_{J^c}\psi)$.  From the definitions of the Farey map and the
transfer operator ( equation~(\ref{eq:fareydef}) and
definition~(\ref{defn:transferop2}) respectively ), $\LB_1$ is found to be
\be
\LB_1\psi(x)=(1+x)^{-2\beta}\psi(\frac{x}{1+x})
\label{Aeq:L1a}
\ee
There is the following result regarding the spectral radius of this operator.
\begin{lemma}
The spectral radius of the operator $\LB_{1}$, $r(\LB_{1})$, is equal to 1.
\end{lemma}
\Proof. \ \
Consider the following formula for the spectral radius of an operator $\O$ taken 
from~\cite{prelslawn:interm}:
\be
r(\O) \equiv lim_{n \rightarrow \infty} \| \O^n \|^{\frac{1}{n}}
\label{Aeq:L1b}
\ee
where the norm of $\O$ is defined as $\| \O \| = \sup_{\|\psi\|=1} \| \O \psi \|$.
It follows from the definition of $\LB_1$ that its $n^{\th}$ iterate is given
by
\be
\LB_1\psi(x)=(1+nx)^{-2\beta}\psi(\frac{x}{1+nx})
\label{Aeq:L1c}
\ee
Therefore, using expression~(\ref{Aeq:L1b})
\be
\| (\LB_1)^n\|^{\frac{1}{n}}=\sup_{\|\psi\|=1} \| (\LB_1)^n \psi \|
=\sup_{\|\psi\|=1} \sup_{x} |1+nx|^{-2\beta}
\left| \psi(\frac{x}{1+nx}) \right|
\label{Aeq:L1d}
\ee
Now, setting $\psi = 1$ shows that
$\| (\LB_1)^n\|^{\frac{1}{n}} \geq \sup_{x} |1+nx|^{-2\beta}$. On the other
hand, it follows from equation~(\ref{Aeq:L1d}) that
\be
\| (\LB_1)^n\|^{\frac{1}{n}} \leq \sup_{x} |1+nx|^{-2\beta}
\sup_{\|\psi\|=1,x} \left| \psi(\frac{x}{1+nx}) \right| \leq \sup_{x} |1+nx|^{-2\beta}
\label{Aeq:L1e}
\ee
Since $|1+nx|^{\frac{-2\beta}{n}} \rightarrow 1$ as $n \rightarrow \infty$, the
proof is finished.
\hfill
$\Box$

\section{The trace of the operator $\Rk$}
\label{Asec:tracedetail}
The work in section~\ref{sec:trace}
hinges upon the calculation or at least analysis of the single eigenvalue
of the operator $\Rk$:
\be
\lambda_k = \frac{(-1)^k z}{k!}  \phi_k^{(k)}(1)= \frac{(-1)^k z}{k!}
\frac{d^k}{d\xi^k} \left.
\xi^{-2\beta} \Phi(z,2\beta+k,\frac{1}{\xi}+1) \right|_{\xi=1}
\label{Aeq:Ropeig1}
\ee
where $k>0$.  The following lemma provides an expression for 
$\lambda_k$ from which the singularity structure may be easily obtained.
\begin{lemma}
The sole eigenvalue of $\Rk$ is given by
\begin{eqnarray*}
\lambda_k  & = &  \frac{z}{k!}
\frac{\Gamma(2\beta+k)}{\Gamma(2\beta)}\Phi(z,2\beta+k,2) \cr
  &   & \mbox{} +  \frac{z}{k!}
\sum_{l=1}^k \sum_{j=1}^l (-1)^j a_{(l,j)} {k \choose l}
\frac{\Gamma(2\beta+k-l)}{\Gamma(2\beta)}
\frac{\Gamma(2\beta+k+j)}{\Gamma(2\beta+k)}\Phi(z,2\beta+k+j,2)
\end{eqnarray*}
where each $a_{(l,j)}$ is some positive integer, $\Gamma(z)$ is the
gamma function and $\Phi(z,s,\nu)$ is the Lerch transcendent function.
For $k>0$, this eigenvalue is holomorphic in $z$ for $|z|<1$ and
$\beta \in C$ and is also holomorphic in $\beta$ for $\beta \in C$ and
$|z| \leq 1$.
\label{Alemma:tracedetail}
\end{lemma}
\Proof. \ \
Using equations~\ref{eq:Ropeig2} and~\ref{eq:diff1}, the expression for
$\lambda_k$ may be written as follows:
\begin{eqnarray}
\lambda_k  & = & \frac{(-1)^k z}{k!}  \left.
\sum_{l=0}^k {k \choose l}
(-1)^{(k-l)} 2\beta(2\beta+1)(2\beta+2)\ldots(2\beta+k-l-1) \right.
\nonumber \\
  &   &  \left.\times \frac{d^l}{d\xi^l}
\Phi(z,2\beta+k,\frac{1}{\xi}+1) \right|_{\xi=1}
\nonumber \\
  & = &  \frac{z}{k!}  \left.
\sum_{l=0}^k {k \choose l} (-1)^{l}
\frac{\Gamma(2\beta+k-l)}{\Gamma(2\beta)}
\frac{d^l}{d\xi^l} \Phi(z,2\beta+k,\frac{1}{\xi}+1) \right|_{\xi=1}
\label{Aeq:Ropeig2}
\end{eqnarray}
where $\Gamma$ is the Gamma function. $\Gamma$ satisfies
the factorial-like relationship $\Gamma(s+1)=s\Gamma(s)$, see~\cite{lerch},
and hence has been used to express $\lambda_k$ in a more compact form.
This section is mainly concerned with the term
\be
T(l,z,2\beta+k)=\left. \frac{d^l}{d\xi^l} \Phi(z,2\beta+k,\frac{1}{\xi}+1)
\right|_{\xi=1}
\label{Aeq:Ropeig3}
\ee
the righthand part of the summand in~\ref{Aeq:Ropeig2}.

Let $u=\frac{1}{\xi}$. The operator $\frac{d^l}{d\xi^l}$ then becomes:
\be
\frac{d^l}{d\xi^l}=\left( \frac{d}{d\xi} \right)^l
=\left( \frac{du}{d\xi} \frac{d}{du} \right)^l
=\left( -u^2 \frac{d}{du} \right)^l
=(-1)^l \left( u^2 \frac{d}{du} \right)^l
\label{Aeq:diffnew}
\ee
So, with this change of variable, equation~\ref{Aeq:Ropeig3} transforms to
\be
(-1)^l T(l,z,2\beta+k)=
\left. \left( u^2 \frac{d}{du} \right)^l \Phi(z,2\beta+k,u+1)
\right|_{u=1}
\label{Aeq:Ropeig4}
\ee
Now assume that for $l=L\geq 1$ the righthand side of the above ( without the
evaluation at $u=1$ ) is of the form
\be
\left( u^2 \frac{d}{du} \right)^L \Phi(z,2\beta+k,u+1)
=\sum_{j=1}^{L} a_{(L,j)} u^{L+j} \frac{d^j}{du^j} \Phi(z,2\beta+k,u+1)
\label{Aeq:Ropeig4a}
\ee
where the $a_{(L,j)}$ are positive integers and have an extended definition to
$j=0$ and $j=L+1$ where $a_{(L,L+1)}=a_{L,0}=0$.  Note that this is 
clearly true for $L=1$. Then it follows that
\begin{eqnarray}
(-1)^{L+1} T(L+1,z,2\beta+k) & = & 
\left( u^2 \frac{d}{du} \right)
\sum_{j=1}^{L} a_{(L,j)} u^{L+j} \frac{d^j}{du^j} \Phi(z,2\beta+k,u+1) \cr
& = & u^2 \sum_{j=1}^{L} a_{(L,j)} \left[  (L+j)u^{L+j-1} 
\frac{d^j}{du^j} \Phi(z,2\beta+k,u+1) \right. \cr
&   & \left. \mbox{} + u^{L+j} \frac{d^{j+1}}{du^{j+1}} 
\Phi(z,2\beta+k,u+1) \right] \cr
& = &  \sum_{j=1}^{L} a_{(L,j)} (L+j)u^{L+1+j} 
\frac{d^j}{du^j} \Phi(z,2\beta+k,u+1) \cr 
&   & \mbox{} + \sum_{j=1}^{L} a_{(L,j)} u^{L+2+j} \frac{d^{j+1}}{du^{j+1}} 
\Phi(z,2\beta+k,u+1) \cr
& = & \sum_{j=1}^{L+1} a_{(L+1,j)} 
u^{L+j+1} \frac{d^j}{du^j} \Phi(z,2\beta+k,u+1)
\end{eqnarray}
where the $a_{(L+1,j)} \equiv (L+j)a_{(L,j)}+a_{(L,j-1)}$ and must be 
positive integers since the $a_{(L,j)}$ are never negative and the sum of
consecutive values are never zero.  Therefore, by induction,
it is clear that $T(l,z,2\beta+k)$ is of the form
\be
(-1)^l T(l,z,2\beta+k)
=\left. \delta_{l,0} \Phi(z,2\beta+k,2)
+\sum_{j=1}^l a_{(l,j)} \frac{d^j}{du^j} \Phi(z,2\beta+k,u+1) \right|_{u=1}
\label{Aeq:Ropeig4z}
\ee
where the case $l=0$ has been included with the help of the 
Kronecker delta, $\delta_{m,n}$, and the understanding that the
second sum disappears when $l=0$.
So, without exactly determining the coefficients $a_{(l,j)}$ of
the differentiations of the Lerch transcendents, it can be seen
that they are positive integers and of course finite for $l$ finite.
Note that $a_{(l,1)}=l!$ and $a_{(l,l)}=1$ for $l=1,2,\ldots$.

Equation~\ref{eq:lerchdiff} shows
that $\frac{d}{du} \Phi(z,s,u+1) = -s\Phi(z,s+1,u+1)$ from which it
follows that
\begin{eqnarray}
\frac{d^j}{du^j} \Phi(z,s,u+1)
 & = & (-1)^j s(s+1)\cdots(s+j-1)\Phi(z,s+j,u+1)
\nonumber \\
 & = & (-1)^j \frac{\Gamma(s+j)}{\Gamma(s)}\Phi(z,s+j,u+1)
\label{Aeq:lerchdiffmult}
\end{eqnarray}
Inserting this expression with $u=1$ and $s=2\beta+k$
into the equation~\ref{Aeq:Ropeig4z} gives the result
\begin{eqnarray}
(-1)^l T(l,z,2\beta+k)
 & = &\delta_{l,0} \Phi(z,2\beta+k,2) \\
 &   & + \sum_{j=1}^l a_{l,j}  (-1)^j
\frac{\Gamma(2\beta+k+j)}{\Gamma(2\beta+k)}\Phi(z,2\beta+k+j,2)
\nonumber
\label{Aeq:Ropeig5}
\end{eqnarray}
where
Finally, this result for $T(l,z,2\beta+k)$
may be substituted into the expression for $\lambda_k$,
equation~\ref{Aeq:Ropeig2}:
\begin{eqnarray}
\lefteqn{ \lambda_k = \frac{z}{k!}
\sum_{l=0}^k {k \choose l} (-1)^{l}
\frac{\Gamma(2\beta+k-l)}{\Gamma(2\beta)}  } \\
  &  & \times (-1)^l \left[ \delta_{l,0} \Phi(z,2\beta+k,2)
 + \sum_{j=1}^l (-1)^j a_{(l,j)}
\frac{\Gamma(2\beta+k+j)}{\Gamma(2\beta+k)}\Phi(z,2\beta+k+j,2) \right] \cr
  & = &  \frac{z}{k!}
\frac{\Gamma(2\beta+k)}{\Gamma(2\beta)}\Phi(z,2\beta+k,2) \cr
  &   & \mbox{} + \frac{z}{k!}
\sum_{l=1}^k \sum_{j=1}^l (-1)^j a_{(l,j)} {k \choose l}
\frac{\Gamma(2\beta+k-l)}{\Gamma(2\beta)}
\frac{\Gamma(2\beta+k+j)}{\Gamma(2\beta+k)}\Phi(z,2\beta+k+j,2) \nonumber
\label{Aeq:Ropeig6}
\end{eqnarray}

The comment in the main body of the thesis that $\lambda_k$ is a
holomorphic function of $\beta$ when $z=1$ and $k>0$ is supported
by the form of equation~\ref{Aeq:Ropeig6}.   To see this, note that $\Gamma(s)$
has no zeroes on the $s$-plane and has simple poles at $s=0,-1,-2,\ldots$,
see~\cite{lerch}.  Thus the quotient $\frac{\Gamma(s+n)}{\Gamma(s)}$, where $n$
is a positive integer, has simple zeroes at $s=0,-1,-2,\ldots,-n+1$.  It has
no poles as these can only come from the simple poles of the numerator $\Gamma(s+n)$, at
$s=-n,-n-1,-n-2,\ldots$, but these are cancelled by the poles of the denominator
at those same points.

So the two quotients of Gamma functions in the second term of the last
line of equation~\ref{Aeq:Ropeig6} provide no poles since $k-l \geq 0$ and
$j>0$.  They do however have zeroes at $2\beta=0,-1,-2,\ldots,-k+l+1$ and
at $2\beta+k=0,-1,-2,\ldots,-j+1$, respectively.  All of this was clear, of
course, from the fact that these quotients were introduced to represent
finite polynomials with simple factors which are very well behaved entities.
Together, the two quotients provide zeroes at the following values of $\beta$:
\be
2\beta=0,-1,-2,\ldots,-k+l+1;-k,-k-1,-k-2,\ldots,-k-j+1
\label{Aeq:zeroes}
\ee
For $z=1$, the Lerch transcendent $\Phi(z,2\beta+k+j,2)$ reduces to the
Riemann zeta function less 1, i.e. $\zeta_R(2\beta+k+j)-1$.  This can be
analytically continued to the whole of the $\beta$-plane with a simple
pole at $2\beta+k+j=1 \Leftrightarrow \beta=\frac{1-k-j}{2}$
with residue $\half$; again see~\cite{lerch}.  This pole is cancelled by
the most negative zero of the quotients, the last value in~\ref{Aeq:zeroes}.
Thus, the second term is a holomorphic function of $\beta$ for $z=1$.

Since $k \geq 1$, the quotient of gamma functions for the first term
has at least one zero.  The zeroes in general are at
$2\beta=0,-1,-2,\ldots,1-k$.  The Lerch transcendent at $z=1$ reduces
to the function $\zeta_R(2\beta+k)-1$ which has one simple pole at
$2\beta=1-k$ which is balanced by the most negative zero of the preceding
quotient.  Thus, both terms are holomorphic in $\beta$ for all $\beta \in C$
and $z=1$.  In conclusion, for $k>0$, the following is true:
\begin{itemize}
\item The function $(z,\beta) \rightarrow \lambda_k$ is holomorphic
in $z$ for $|z|<1$ and $\beta \in C$.
\item The function $(z,\beta) \rightarrow \lambda_k$ is holomorphic
in $\beta$ for $\beta \in C$ and $|z| \leq 1$.
\end{itemize}
as is required for the completion of the proof of the lemma.
\hfill
$\Box$

\section{Generalised Fibonacci Numbers}
\label{Asec:fibnum}
This section provides a derivation of an exact expression for
generalised Fibonacci numbers $\gk_n$.  This sequence of numbers is
generated by setting $\gk_{-1}=0$ and $\gk_{0}=1$ and applying the recursion
formula
\be
\gk_{n+1}=k\gk_{n}+\gk_{n-1}
\label{Aeq:fibrec}
\ee
Note that $k \geq 1$ and for $k=1$ this produces the normal Fibonacci series.
\begin{lemma}
The generalised Fibonacci number, $\gk_n$, has an exact expression of the 
form
\bdm
\gk_n = \frac{1}{1+\rho_k^2}\left(\rho_k^{n+2}
+\left(\frac{-1}{\rho_k}\right)^n \right)
\edm
where $\rho_k=\frac{k+\sqrt{k^2+4}}{2}=[\overline{k;}]$ is the positive
solution to the quadratic equation $x^2-kx-1=0$.
Moreover, this relationship holds for all integers $n$.
\label{Alemma:fibform}
\end{lemma}
\Proof. \ \
Firstly, note that $\rho_k = k + \frac{1}{\rho_k}$.
Assume the statement is true for $n=m$ and $n=m-1$.  Then the term $\gk_{m+1}$
is given by
\begin{eqnarray}
\gk_{m+1} & = & k\gk_{m}+\gk_{m-1} 
 = \frac{1}{1+\rho_k^2}\left( k\rho_k^{m+2}
+k\left(\frac{-1}{\rho_k}\right)^{m} + \rho_k^{m+1}
+\left(\frac{-1}{\rho_k}\right)^{m-1} \right) \cr
 & = & \frac{1}{1+\rho_k^2}\left( \rho_k^{m+1}(k\rho_k+1)
+(1-\frac{k}{\rho_k})\left(\frac{-1}{\rho_k}\right)^{m-1}\right) \cr
 & = & \frac{1}{1+\rho_k^2}\left(\rho_k^{m+3}
+\left(\frac{-1}{\rho_k}\right)^{m+1} \right)
\label{Aeq:fib1}
\end{eqnarray}
since $(k\rho_k+1)=\rho_k^2$ and $(1-\frac{k}{\rho_k})=\frac{1}{\rho_k^2}
=\left(\frac{-1}{\rho_k}\right)^2$.
A similar procedure shows that this the relationship holds for $n=m-2$ as 
well.  Also,  the formula holds true for $n=0$ and $n=-1$ since
\bdm
\gk_{0}=\frac{1}{1+\rho_k^2} (\rho_k^2 + 1 )=1
\edm
and
\bdm
\gk_{-1}=
\frac{1}{1+\rho_k^2}\left(\rho_k-\left(\frac{1}{\rho_k}\right)^{-1} \right)=0
\edm
Therefore, by induction, the proof is finished.
\hfill
$\Box$

Note that $\rho_1=\rho_g=\frac{1+\sqrt{5}}{2}$ is the golden ratio.  Also,
$\rho_2=\rho_s=1+\sqrt{2}$ is sometimes called the `silver ratio'.  
Both of these numbers and their corresponding Fibonacci sequences 
are used in the final section of the thesis.

\newpage

\bibliographystyle{plain}

\begin{thebibliography}{10}

\bibitem{andrews:part}
George~E. Andrews.
\newblock {\em The Theory of Partitions}, volume~2 of {\em Encyclopedia of
  Mathematics and Its Applications}.
\newblock Addison-Wesley, 1976.

\bibitem{artin:periodic}
M.~Artin and B.~Mazur.
\newblock On periodic points.
\newblock {\em Annals of Mathematics}, 81(2):82--99, 1965.

\bibitem{beck:thermchaos}
Christian Beck and Friedrich Schl\"{o}gl.
\newblock {\em Thermodynamics of chaotic systems}.
\newblock Cambridge University Press, 1993.

\bibitem{ergsymdyn}
Tim Bedford, Michael Keane, and Caroline Series, editors.
\newblock {\em Ergodic Theory, Symbolic Dynamics and Hyperbolic Spaces}.
\newblock Oxford University Press, 1991.

\bibitem{bowen:therm}
R.~Bowen.
\newblock {\em Equilibrium States and the Ergodic Theory of Anosov
  Diffeomorphisms}.
\newblock Number 470 in Lecture Notes in Mathematics. Springer, New York, 1975.

\bibitem{dodds:honsthesis}
P.~S. Dodds.
\newblock Honours thesis: Multifractals, thermodynamic formalism and
  intermittency.
\newblock {\em The University of Melbourne}, November 1993.
\newblock (an abridged version will hopefully appear in print in the near
  future.).

\bibitem{lerch}
A.~Erd\'{e}lyi, editor.
\newblock {\em Higher Transcendental Functions}.
\newblock McGraw Hill Book Company, 1953.

\bibitem{feig:transfer}
Feigenbaum, Procaccia, and T\'{e}l.
\newblock Scaling properties of multifractals as an eigenvalue problem.
\newblock {\em Physical Review A}, 39(10):5359--5372, May 1989.

\bibitem{feig:farey}
Mitchell~J. Feigenbaum.
\newblock Presentation functions, fixed points, and a theory of scaling
  function dynamics.
\newblock {\em Journal of Statistical Physics}, 52(3/4):527--569, 1988.

\bibitem{grad:int}
I.S. Gradshteyn and I.M. Ryzhik.
\newblock {\em Table of Integrals, Series, and Products}.
\newblock Academic Press, $4^{\mbox{th}}$ edition, 1965.

\bibitem{groth:nuclear}
A.~Grothendieck.
\newblock Produits tensoriels topologiques et espaces nucl\'{e}aires.
\newblock {\em Memoirs of the American Mathematical Society}, 16, 1955.

\bibitem{hille:anal}
Einar Hille.
\newblock {\em Analytic Function Theory, Volumes 1 \& 2}.
\newblock Chelsea, $2^{\mbox{nd}}$ edition, 1982.

\bibitem{huang:therm}
Kerson Huang.
\newblock {\em Statistical Mechanics}.
\newblock Wiley, New York, $2^{\mbox{nd}}$ edition, 1987.

\bibitem{kamw:endspectra}
Herbert Kamowitz.
\newblock The spectra of endomorphisms of the disc algebra.
\newblock {\em Pacific Journal of Mathematics}, 46(2):433--440, 1973.

\bibitem{kamw:compop}
Herbert Kamowitz.
\newblock The spectra of composition operators on $\mbox{H}^{\mbox{p}}$.
\newblock {\em Journal of Functional Analysis}, 18:132--150, 1975.

\bibitem{kato:pert}
T.~Kato.
\newblock {\em Perturbation Theory for Linear Operators}.
\newblock Springer-Verlag, $2^{\mbox{nd}}$ edition, 1980.

\bibitem{lorenz:air}
Edward Lorenz.
\newblock Deterministic nonperiodic flow.
\newblock {\em Journal of Atmospheric Sciences}, 20:130--141, 1963.

\bibitem{pomman:interm}
P.~Manneville and Y.~Pomeau.
\newblock Intermittency and the lorenz model.
\newblock {\em Phys. Lett.}, 75 A(1):1--, 1979.

\bibitem{mayer:zeta}
Dieter~H. Mayer.
\newblock On a $\zeta$ function related to the continued fraction
  transformation.
\newblock {\em Bulletin de la Societ\'{e} mathematique de France},
  104:195--203, 1976.

\bibitem{mayer:transop}
Dieter~H. Mayer.
\newblock {\em Lecture Notes in Physics: The Ruelle-Araki Transfer Operator in
  Classical Statistical Mechanics}, volume 123 of {\em Lecture Notes in
  Physics}.
\newblock Springer-Verlag, 1980.

\bibitem{mayer:therm}
Dieter~H. Mayer.
\newblock On the thermodynamic formalism for the gauss map.
\newblock {\em Communications in Mathematical Physics}, 130:311--333, 1990.

\bibitem{prelslawn:interm}
Thomas Prellberg and Joseph Slawny.
\newblock Maps of intervals with indifferent fixed points: thermodynamic
  formalism and phase transitions.
\newblock {\em Journal of Statistical Physics}, 66:503--514, 1992.

\bibitem{rob:topvecsp}
A.~P. Robertson and W.~J. Robertson.
\newblock {\em Topological Vector Spaces}.
\newblock Cambridge Tracts in Mathematics and Mathematical Physics. Cambridge
  University Press, 1966.

\bibitem{rockett:contfracs}
Andrew~M. Rockett and Peter Sz\"{u}sz.
\newblock {\em Continued Fractions}.
\newblock World Scientific, 1992.

\bibitem{rudin:anal}
Walter Rudin.
\newblock {\em Real and Complex Analysis}.
\newblock McGraw-Hill Book company, $3^{\mbox{rd}}$ edition, 1990.

\bibitem{ruelle:zeta}
David Ruelle.
\newblock Zeta functions and statistical mechanics.
\newblock {\em Soc. Math. France, Asterisque}, 40:167--176, 1976.

\bibitem{ruelle:anosov}
David Ruelle.
\newblock $\zeta$ functions for expanding maps and anosov flows.
\newblock {\em Inventiones mathematicae}, 34:231--342, 1976.

\bibitem{ruelle:therm}
David Ruelle.
\newblock {\em Thermodynamic Formalism}.
\newblock Addison-Wesley, 1977.

\bibitem{schuster:detchaos}
H.G. Schuster.
\newblock {\em Deterministic Chaos}.
\newblock Physik Verlag, 1984.

\bibitem{tel:multifracs}
T\'{a}mas T\'{e}l.
\newblock Fractals,multifractals and thermodynamic formalism; an introductory
  review.
\newblock {\em Z. Naturforsch.}, 43a:1154--1174, September 1988.

\end{thebibliography}

\chapter*{Acknowledgments}

\addcontentsline{toc}{chapter}{Acknowledgments}

My supervisor, Dr.\ Thomas Prellberg, is thanked for his continuing
efforts and support of my work.  I appreciate the skills Thomas has imparted
to me and the opportunity to investigate dynamical systems in such an
interesting way.  He has always been more than ready to provide assistance and
to give direction when necessary.
I am also grateful to my `official' supervisor, Professor Colin
Thompson, for his confidence in my work.

\noindent
I am indebted to fruitful discussions with Rachel Haverfield
and Matthew Emerton.

\noindent
I am also pleased to acknowledge that 
this work has been supported by an APRA (Australian Postgraduate
Research Award).

\end{document}